\documentclass[twocolumn]{aastex631}

\shortauthors{Roberti et al.}

\usepackage{isotope}
\usepackage{amsmath}

\newcommand{\msun}{$\rm M_{\odot}$}
\newcommand{\cc}{$\rm ^{12}C+^{12}C$ }
\newcommand{\ccp}{$\rm ^{12}C(^{12}C,p)^{23}Na$}
\newcommand{\cca}{$\rm ^{12}C(^{12}C,\alpha)^{20}Ne$}
\newcommand{\ccn}{$\rm ^{12}C(^{12}C,n)^{23}Mg$}
\newcommand{\nen}{$\rm ^{22}Ne(\alpha,n)^{25}Mg$}
\newcommand{\cn}{$\rm ^{13}C(\alpha,n)^{16}O$}
\newcommand{\s}{$s-$}

\begin{document}

\title{Impact of Sub-2.5 MeV \cc Resonances on the Production of Elements from C to Pd in Core-Collapse Supernovae}

\correspondingauthor{Lorenzo Roberti}

\author[0000-0003-0390-8770]{Lorenzo Roberti}
\affiliation{Department of Physics and Geology, University of Perugia; Perugia, 06123, Italy}
\affiliation{Istituto Nazionale di Fisica Nucleare - Sezione di Perugia (INFN), via A. Pascoli s/n, I-06125 Perugia, Italy}
\affiliation{Konkoly Observatory, Research Centre for Astronomy and Earth Sciences, HUN-REN, Konkoly Thege Miklós út 15-17, Budapest, H-1121, Hungary}
\affiliation{CSFK HUN-REN, MTA Centre of Excellence, Konkoly Thege Miklós út 15-17, Budapest, H-1121, Hungary}
\affiliation{Istituto Nazionale di Astrofisica – Osservatorio Astronomico di Roma, Via Frascati 33, Monte Porzio Catone, I-00040, Italy}
\affiliation{NuGrid Collaboration, \url{http://nugridstars.org}}

\author[0000-0001-5386-8389]{Sara Palmerini}
\affiliation{Department of Physics and Geology, University of Perugia; Perugia, 06123, Italy}
\affiliation{Istituto Nazionale di Fisica Nucleare - Sezione di Perugia (INFN), via A. Pascoli s/n, I-06125 Perugia, Italy}
\affiliation{Istituto Nazionale di Astrofisica – Osservatorio Astronomico di Roma, Via Frascati 33, Monte Porzio Catone, I-00040, Italy}

\author[0009-0005-0324-0637]{Agnese Falla}
\affiliation{Dipartimento di Fisica, Sapienza Università di Roma, P.le A. Moro 5, Roma, I-00185, Italy}
\affiliation{Istituto Nazionale di Astrofisica – Osservatorio Astronomico di Roma, Via Frascati 33, Monte Porzio Catone, I-00040, Italy}

\author[0000-0002-4819-310X]{Luca Boccioli}
\affiliation{Department of Physics, University of California, Berkeley, CA 94720, USA}
\affiliation{NuGrid Collaboration, \url{http://nugridstars.org}}

\author[0009-0009-7734-7719]{Francesco Andreis}
\affiliation{Department of Physics and Geology, University of Perugia; Perugia, 06123, Italy}
\affiliation{Istituto Nazionale di Fisica Nucleare - Sezione di Perugia (INFN), via A. Pascoli s/n, I-06125 Perugia, Italy}

\author[0000-0002-3589-3203]{Alessandro Chieffi}
\affiliation{Istituto Nazionale di Astrofisica - Istituto di Astrofisica e Planetologia Spaziali (INAF - IAPS), Via Fosso del Cavaliere 100, I-00133, Roma, Italy}
\affiliation{Istituto Nazionale di Fisica Nucleare - Sezione di Perugia (INFN), via A. Pascoli s/n, I-06125 Perugia, Italy}
\affiliation{School of Physics and Astronomy, Monash University, VIC 3800, Australia}

\author[0000-0002-1819-4814]{Marco La Cognata}
\affiliation{Istituto Nazionale di Fisica Nucleare - Laboratori Nazionali del Sud (INFN - LNS), Via Santa Sofia 62, Catania, Italy}

\author[0000-0003-0636-7834]{Marco Limongi}
\affiliation{Istituto Nazionale di Astrofisica - Osservatorio Astronomico di Roma (INAF - OAR), Via Frascati 33, I-00040, Monteporzio Catone, Italy}
\affiliation{Kavli Institute for the Physics and Mathematics of the Universe, Todai Institutes for Advanced Study, University of Tokyo, Kashiwa, 277-8583 (Kavli IPMU, WPI), Japan}
\affiliation{Istituto Nazionale di Fisica Nucleare - Sezione di Perugia (INFN), via A. Pascoli s/n, I-06125 Perugia, Italy}

\author[0000-0000-0000-0000]{Marco Masci}
\affiliation{Department of Physics and Geology, University of Perugia; Perugia, 06123, Italy}

\author[0000-0002-0101-5226]{Aliya Nurmukhanbetova}
\affiliation{Dipartimento di Ingegneria e Architettura, Università degli Studi di Enna "Kore", Cittadella Universitaria, Enna, Italy}
\affiliation{Istituto Nazionale di Fisica Nucleare - Laboratori Nazionali del Sud (INFN - LNS), Via Santa Sofia 62, Catania, Italy}

\author[0000-0002-8642-4028]{Alessandro Alberto Oliva}
\affiliation{Istituto Nazionale di Fisica Nucleare - Laboratori Nazionali del Sud (INFN - LNS), Via Santa Sofia 62, Catania, Italy}

\author[0000-0002-9986-1518]{Roberta Spartà}
\affiliation{Dipartimento di Ingegneria e Architettura, Università degli Studi di Enna "Kore", Cittadella Universitaria, Enna, Italy}
\affiliation{Istituto Nazionale di Fisica Nucleare - Laboratori Nazionali del Sud (INFN - LNS), Via Santa Sofia 62, Catania, Italy}

\author[0000-0002-6953-7725]{Aurora Tumino}
\affiliation{Dipartimento di Ingegneria e Architettura, Università degli Studi di Enna "Kore", Cittadella Universitaria, Enna, Italy}
\affiliation{Istituto Nazionale di Fisica Nucleare - Laboratori Nazionali del Sud (INFN - LNS), Via Santa Sofia 62, Catania, Italy}

\begin{abstract}
    We explore the impact of a more efficient \cc reaction on the structure and nucleosynthesis of massive stars. We calculate non-rotating stellar models with initial masses of 15, 16, 18, 20, 22, 25, and 40 \msun\ and solar metallicity by means of the \verb|FRANEC| code. Furthermore, we simulate the core-collapse supernova of these models with the thermal bomb technique, using two different approaches to inject the thermal energy into the pre-supernova structure. Our results show that a more efficient \cc rate extends the duration of the central carbon burning phase, developing more massive convective cores and leading to a different and less compact pre-supernova structure with respect to models calculated with a standard \cc rate. These structural differences significantly impact nucleosynthesis. In particular, an increased rate enhances the production of elements heavier than Fe, produced by the \s process nucleosynthesis and driven by the more efficient activation of the \cn\ neutron source in the early carbon burning shells. We find that the differences in the chemical composition of the core-collapse supernova ejecta are primarily determined by these pre-supernova structural changes, which dominate over the effects of different explosion prescriptions.
\end{abstract}

\keywords{nuclear reactions, massive star, core-collapse supernova, s-process nucleosynthesis}


\section{Introduction} \label{sec:intro}

    Stars more massive than $\sim9$ \msun\ evolve through a sequence of nuclear burning stages (H, He, C, Ne, O, and Si) until they form a dense core of Fe-peak elements, which eventually collapses. The conditions at the end of central He burning, specifically the mass of the CO core and the \isotope[12]{C} abundance, govern the ignition of the subsequent, more energetic burning phases, starting with C fusion. These stages result from a complex interplay between the energy generated by nuclear reactions, the formation of convective regions, and neutrino losses, which become dominant once the temperature exceeds $\sim700$ MK \citep[e.g.,][and references therein]{chieffi:98,woosley:02,langer:12,limongi:24}.

    In this context, \cc is one of the most critical reactions determining the subsequent evolution of the star. Its efficiency controls, in fact, the maximum mass that does not ignite C \citep[$\rm M_{up}$,][]{becker:80} and the minimum mass required for a star to end as a core-collapse supernova \citep[$\rm M_{MAS}$,][]{chieffi:25}. Furthermore, it dictates the development and the evolution of the C burning shells that eventually sculpt the density profile and the resulting compactness at the pre-supernova stage \citep{chieffi:21}.

    Despite its importance, the \cc reaction rate was affected by large experimental uncertainties at energies of astrophysical interest ($\rm 0.9 < E_{{cm}} < 3.4$ MeV). Indeed, in such energy interval the cross section is dominated by a complex resonance structure that makes extrapolations from high energy data likely to introduce large systematic errors. In recent years, several experimental efforts have aimed to map the Gamow window. In addition to direct measurements, which suffer from a very low signal-to-noise ratio due to the Coulomb suppression of the cross section \citep[e.g.,][]{spillane:07,fruet:20}, indirect techniques such as the Trojan Horse Method (THM) have provided new insights. In particular, the experimental results of \cite{tumino:18} covering the whole Gamow window suggest a significant rate enhancement due to previously unpredicted resonances, leading to a more efficient and earlier carbon ignition compared to standard estimates \citep{caughlan:88}. Conversely, recent extrapolations by \cite{monpribat:22} has suggested the possibility of a hindrance effect at low energies, which would drastically reduce the reaction rate at the typical temperatures of carbon burning.

    In the past, the impact of the \cc reaction on the stellar structure and on the nucleosynthesis has been studied by several groups \citep{gasques:07,bennett:12,pignatari:13,chieffi:21,monpribat:22,dumont:25}. In this paper, we follow up on the work in \cite{chieffi:21} and study the impact of a more efficient \cc reaction rate by \cite{tumino:18} on the explosive yields of massive stars. The paper is organized as follows: in Sect. \ref{sec:methods} we describe the codes and the input used; in Sect. \ref{sec:strut} we discuss the impact of the \cc reaction on the stellar evolution and nucleosynthesis; in Sect. \ref{sec:exp} we present the explosive yields and compare the results with those calculated with the standard reaction rate; in Sect. \ref{sec:disc} we summarise our findings.


\section{Methods} \label{sec:methods}

    In this paper we use the same code adopted by \citet[][namely the FRANEC code]{chieffi:21}. Differences are discussed in detail in the following.
    
    \subsection{The \cc reaction rate}

        \begin{figure}[!t]
            \centering
            \includegraphics[width=\linewidth]{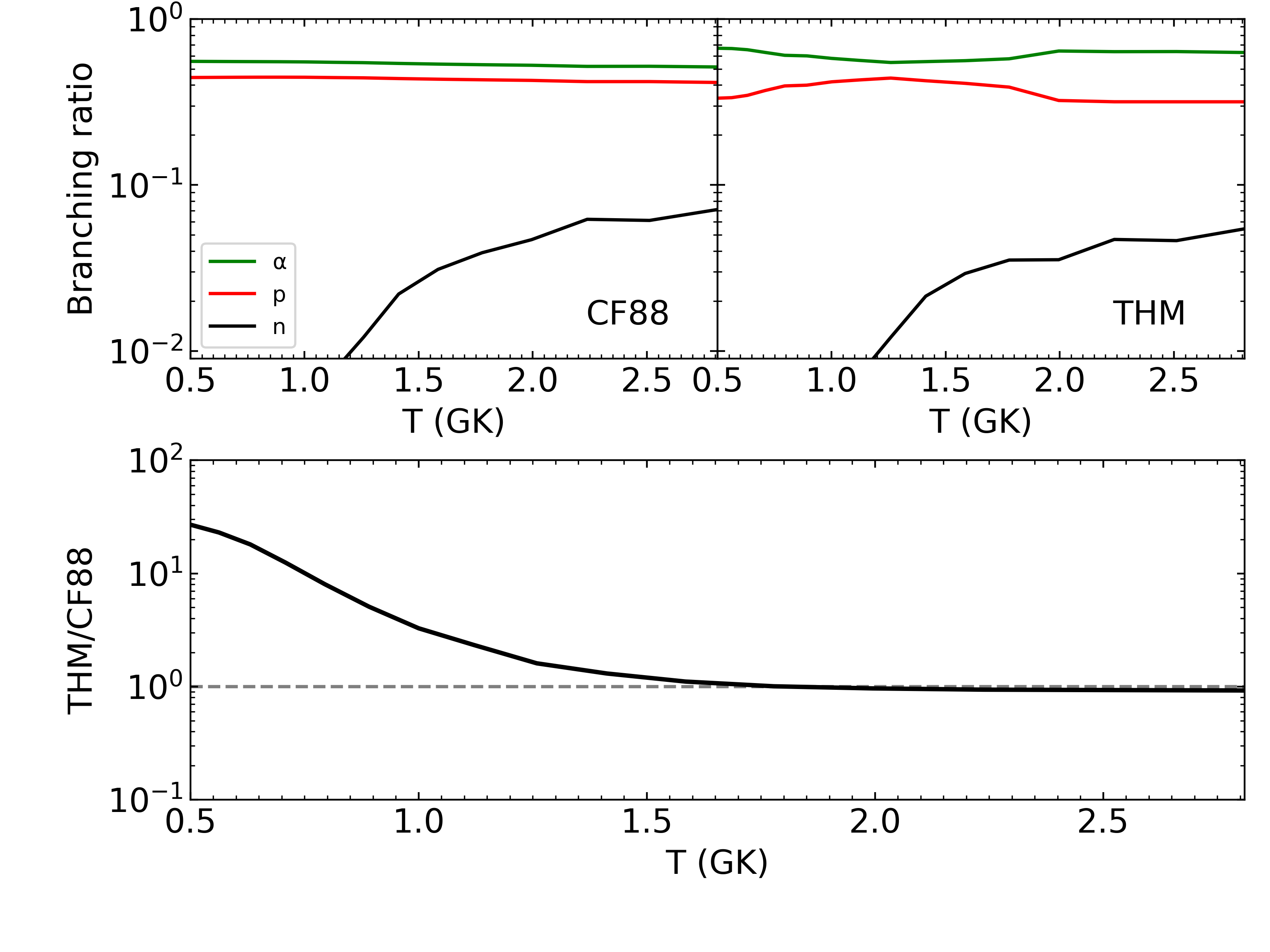}
            \caption{Upper panels: branching ratios of $\alpha$ (green), $p$ (red), and $n$ (black) exit channel of the \cc reaction rates from CF88 \citep[left][]{caughlan:88} and measured by the Trojan Horse Method\citep[THM][]{tumino:18}. Lower panel: the THM/CF88 ratio of the total rate.}
            \label{fig:reaction}
        \end{figure}

        The reference value for the \cc reaction rate used by \citet{chieffi:21}, in this paper, and commonly in the literature is the one from \citet[][hereafter CF88]{caughlan:88}, while a more efficient rate is the one reported by \citet[][hereafter THM, since it has been measured via the Trojan Horse Method]{tumino:18}. In the latter case, only the $p$ (\ccp; $r_p$) and $\alpha$ channels (\cca; $r_\alpha$) have been measured, therefore we rely on approximation to extract the $n$ channel (\ccn; $r_n$), which has a negligible contribution to the total rate ($r_t=r_p+r_\alpha+r_n$), but it may have feedback for the nucleosynthesis. In doing that, we adopted the $r_{\alpha}/r_p$ ratio ($R^{THM}_{\alpha p}$) from the THM rate, the $r_n/r_p$ ratio ($R^{CF88}_{np}$) from the CF88 rate and we estimate $r_n^{THM} = r_t^{THM}* R^{CF88}_{np}/(R^{CF88}_{np}+R^{THM}_{\alpha p}+1)$. Above 3 GK, the THM reaction rate is extrapolated by scaling the CF88 values to the THM value at 3GK. We note, however, that the impact of the rate in this temperature regime is typically negligible, as \isotope[12]{C} is rarely exposed to such conditions, even during explosive nucleosynthesis. The relative contributions to the total rate from the $p$, $\alpha$, and $n$ channels are shown in the upper panels of \figurename~\ref{fig:reaction} as a function of the temperature, in the interval of interest for the stellar evolution. The THM rate is up to a factor of 20--30 higher than that of CF88 below $\sim1.5$ GK, while differences between the two rates are negligible above that threshold (see lower panel of \figurename~\ref{fig:reaction} and also \figurename~3 from \citealt{tumino:18}). A further difference with the CF88 rate is the different $R_{\alpha p}$ ratio as a function of the temperature, that persists above the 1.5 GK threshold. The total reaction rate and $p$, $\alpha$, and $n$ channels adopted in this work are shown in \tablename~\ref{tab:rate}.

    \subsection{Stellar models}

        We computed non-rotating models at solar metallicity \citep[Z=0.01345,][]{asplund:09} for initial masses of 15, 16, 18, 20, 22, 25, and 40 \msun. The models are calculated with the \verb|FRANEC| code \citep{LC18,RLC24,falla:25}, using the exact same prescriptions for mass loss, convection and semi-convection as in \cite{chieffi:21}. In particular, we recall that the convective borders are determined according to the Ledoux criterion in the H-rich layers and according to the Schwarzschild criterion elsewhere. Core overshooting is included during H burning stage with $\rm \alpha_{ov} = 0.5Hp$, while we do not include any convective boundary mixing at the edges of convective shells.
        The nuclear network is based on that adopted in \cite{roberti:24} and it is shown in \tablename~\ref{tab:net}. It includes 356 nuclei between $n$ and Cd and more than 3000 reactions, in order to properly resolve the weak \s process peak including the short-lived radioactive (SLR) isotope \isotope[107]{Pd}. Contrary to \cite{roberti:24}, here we adopt the $3\alpha$ reaction from \cite{fynbo:05} instead of that from the NACRE database \citep{angulo:99}.
        
        The core-collapse supernova explosion is simulated as a thermal bomb, by means of the Lagrangian hydrodynamic code \verb|HYPERION| \citep{LC20} using two different setups. In the first one, the explosion is triggered by depositing the minimum amount of thermal energy, at a mass coordinate of 0.8 \msun\ in the pre-supernova model (i.e., well inside the Fe core), that leads to the full ejection of the mantle above the Fe core. The remnant mass is then determined by requiring the ejection of 0.07 \msun\ of \isotope[56]{Ni} (Set A). In the second one the explosion is triggered by depositing thermal energy at the Si/Si-O interface, since that is where shock revival occurs in more sophisticated neutrino-driven explosion simulations \citep{summa:16,vartanyan:18,bruenn:23,boccioli:23,boccioli:25a}, and typically coincides with the mass cut \citep{boccioli:24}. Furthermore, the amount of energy deposited is calculated by requiring that the final kinetic energy of the ejecta $E_{\rm exp}$ in units of $10^{51}$ erg (= 1 foe) satisfies the equation:
        \begin{equation}\label{eq:eexp}
            E_{\rm exp} = 4.32\times \xi_{2.5} - 0.032,
        \end{equation}
        where $\xi_{M_k}=M_k(M_{\odot})/R_{M_k} (1000\ \rm km)$ is the compactness of the star evaluated at $M_k=2.5$ \msun\ \citep[$\xi_{2.5}$,][]{oconnor:11}. This equation is the result of a semi-analytical fit to the explosion energies obtained in 1D+ neutrino-driven explosion simulations by \cite{BR25} in the case of \verb|FRANEC| models (Set B). For both explosion setups, the nucleosynthesis is simulated for $2.5\times10^4\ \rm s$, while the propagation of the shock is followed up to $3\times10^7\ \rm s$ to properly account for fallback, if any occurs. 

    \begin{table}
        \caption{The THM \cc reaction rates adopted in this study. All the values are $\rm log_{10}$ values. The reaction rates are in units of $\rm cm^3\ s^{-1}\ mol^{-1}$.}
        \label{tab:rate}
	    \centering
        \begin{tabular}{rrrrr}
            \hline\hline
            T(K)  & Total & $\alpha$ & $p$ & $n$ \\
            \hline
            8.00 & $-51.5575$ & $-51.7328$ & $-52.0362$ & $-90.0000$ \\
            8.05 & $-48.5925$ & $-48.7883$ & $-49.0327$ & $-90.0000$ \\
            8.10 & $-45.3358$ & $-45.6728$ & $-45.6037$ & $-89.4079$ \\
            8.15 & $-41.7909$ & $-42.1420$ & $-42.0471$ & $-88.5675$ \\
            8.20 & $-38.4925$ & $-38.8401$ & $-38.7514$ & $-83.2436$ \\
            8.25 & $-35.5575$ & $-35.9048$ & $-35.8167$ & $-73.6874$ \\
            8.30 & $-32.8924$ & $-33.2373$ & $-33.1536$ & $-65.3038$ \\
            8.35 & $-30.7684$ & $-31.0989$ & $-31.0419$ & $-58.0718$ \\
            8.40 & $-28.4363$ & $-28.7506$ & $-28.7245$ & $-51.6212$ \\
            8.45 & $-26.6438$ & $-26.9353$ & $-26.9547$ & $-46.0871$ \\
            8.50 & $-24.8762$ & $-25.1430$ & $-25.2143$ & $-41.2194$ \\
            8.55 & $-23.2257$ & $-23.4661$ & $-23.5971$ & $-36.9491$ \\
            8.60 & $-21.5873$ & $-21.7987$ & $-22.0015$ & $-33.0734$ \\
            8.65 & $-19.8297$ & $-20.0240$ & $-20.2727$ & $-29.4057$ \\
            8.70 & $-18.0865$ & $-18.2626$ & $-18.5636$ & $-26.0726$ \\
            8.75 & $-16.5500$ & $-16.7277$ & $-17.0239$ & $-23.1056$ \\
            8.80 & $-15.0210$ & $-15.2060$ & $-15.4809$ & $-20.4603$ \\
            8.85 & $-13.6023$ & $-13.8034$ & $-14.0333$ & $-18.1187$ \\
            8.90 & $-12.3089$ & $-12.5275$ & $-12.7120$ & $-16.0457$ \\
            8.95 & $-11.0878$ & $-11.3097$ & $-11.4863$ & $-14.2101$ \\
            9.00 & $ -9.9066$ & $-10.1437$ & $-10.2850$ & $-12.5326$ \\
            9.05 & $ -8.8422$ & $ -9.0915$ & $ -9.2091$ & $-10.9880$ \\
            9.10 & $ -7.6638$ & $ -7.9256$ & $ -8.0196$ & $ -9.5824$ \\
            9.15 & $ -6.5954$ & $ -6.8520$ & $ -6.9673$ & $ -8.2658$ \\
            9.20 & $ -5.5006$ & $ -5.7517$ & $ -5.8881$ & $ -7.0344$ \\
            9.25 & $ -4.4188$ & $ -4.6584$ & $ -4.8291$ & $ -5.8714$ \\
            9.30 & $ -3.3783$ & $ -3.5710$ & $ -3.8692$ & $ -4.8290$ \\
            9.35 & $ -2.4583$ & $ -2.6550$ & $ -2.9569$ & $ -3.7868$ \\
            9.40 & $ -1.4427$ & $ -1.6388$ & $ -1.9415$ & $ -2.7783$ \\
            9.45 & $ -0.5779$ & $ -0.7807$ & $ -1.0752$ & $ -1.8388$ \\
            9.50 & $  0.3225$ & $  0.1182$ & $ -0.1763$ & $ -0.9139$ \\
            9.55 & $  1.1549$ & $  0.9492$ & $  0.6547$ & $ -0.0590$ \\
            9.60 & $  1.9220$ & $  1.7139$ & $  1.4194$ & $  0.7431$ \\
            9.65 & $  2.5813$ & $  2.3674$ & $  2.0729$ & $  1.4767$ \\
            9.70 & $  3.2999$ & $  3.0850$ & $  2.7906$ & $  2.2078$ \\
            9.75 & $  3.8799$ & $  3.6590$ & $  3.3645$ & $  2.8503$ \\
            9.80 & $  4.4479$ & $  4.2229$ & $  3.9284$ & $  3.4560$ \\
            9.85 & $  4.9617$ & $  4.7326$ & $  4.4381$ & $  4.0049$ \\
            9.90 & $  5.3954$ & $  5.1631$ & $  4.8686$ & $  4.4629$ \\
            9.95 & $  5.7535$ & $  5.5208$ & $  5.2263$ & $  4.8242$ \\
           10.00 & $  6.0336$ & $  5.8058$ & $  5.5113$ & $  5.0660$ \\
            \hline
        \end{tabular}
    \end{table}

    \begin{table}
        \caption{The 356-isotope nuclear network. $\rm A_{min}$ and $\rm A_{max}$ are the minimum and maximum atomic weight of each element.}
        \label{tab:net}
	    \centering
        \begin{tabular}{lccc|lccc}
            \hline\hline
            Element & Z & $\rm A_{min}$ & $\rm A_{max}$ & Element & Z & $\rm A_{min}$ & $\rm A_{max}$\\
            \hline
            n  & 0  & 1   & 1  & Mn & 25 & 49  & 57  \\ 
            H  & 1  & 1   & 3  & Fe & 26 & 51  & 61  \\ 
            He & 2  & 3   & 4  & Co & 27 & 53  & 62  \\ 
            Li & 3  & 6   & 7  & Ni & 28 & 55  & 65  \\ 
            Be & 4  & 7   & 10 & Cu & 29 & 57  & 66  \\ 
            B  & 5  & 10  & 11 & Zn & 30 & 60  & 71  \\ 
            C  & 6  & 12  & 14 & Ga & 31 & 62  & 72  \\ 
            N  & 7  & 13  & 16 & Ge & 32 & 64  & 77  \\ 
            O  & 8  & 14  & 19 & As & 33 & 74  & 77  \\ 
            F  & 9  & 17  & 20 & Se & 34 & 74  & 83  \\ 
            Ne & 10 & 19  & 23 & Br & 35 & 79  & 85  \\ 
            Na & 11 & 21  & 24 & Kr & 36 & 78  & 89  \\ 
            Mg & 12 & 23  & 27 & Rb & 37 & 84  & 89  \\ 
            Al & 13 & 25  & 28 & Sr & 38 & 84  & 93  \\ 
            Si & 14 & 27  & 32 & Y  & 39 & 89  & 95  \\ 
            P  & 15 & 29  & 34 & Zr & 40 & 90  & 98  \\
            S  & 16 & 31  & 37 & Nb & 41 & 92  & 98  \\
            Cl & 17 & 33  & 38 & Mo & 42 & 92  & 103 \\
            Ar & 18 & 35  & 41 & Tc & 43 & 97  & 103 \\
            K  & 19 & 37  & 44 & Ru & 44 & 96  & 107 \\
            Ca & 20 & 39  & 49 & Rh & 45 & 102 & 108 \\
            Sc & 21 & 41  & 49 & Pd & 46 & 102 & 112 \\
            Ti & 22 & 44  & 51 & Ag & 47 & 107 & 112 \\
            V  & 23 & 45  & 52 & Cd & 48 & 108 & 112 \\
            Cr & 24 & 47  & 55 &    &    &     &     \\
            \hline
        \end{tabular}
    \end{table}


\section{Findings in Stellar Evolution and Nucleosynthesis}\label{sec:strut}

    \begin{figure*}[!t]
        \centering
        \includegraphics[width=0.49\linewidth]{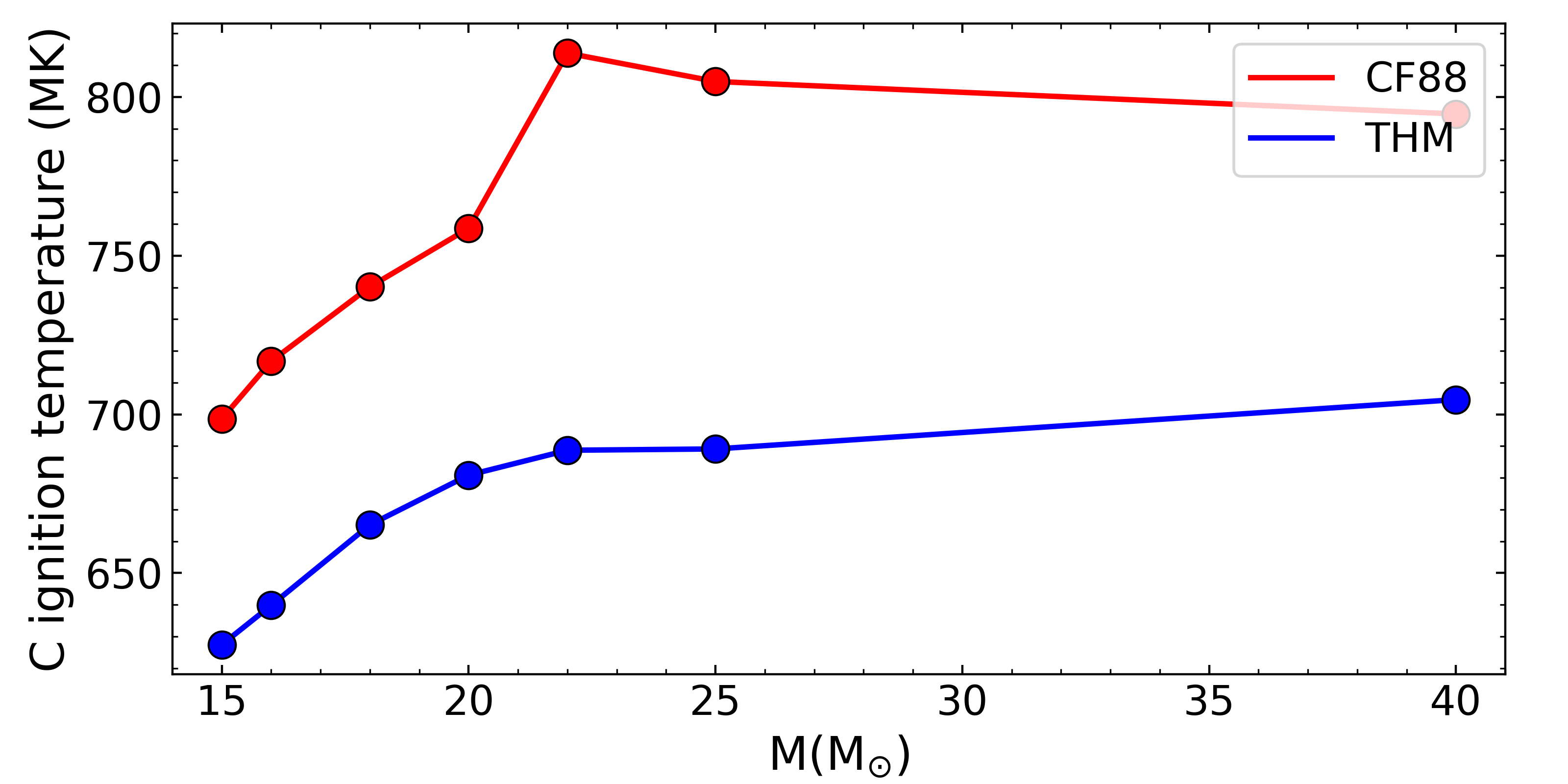}
        \includegraphics[width=0.49\linewidth]{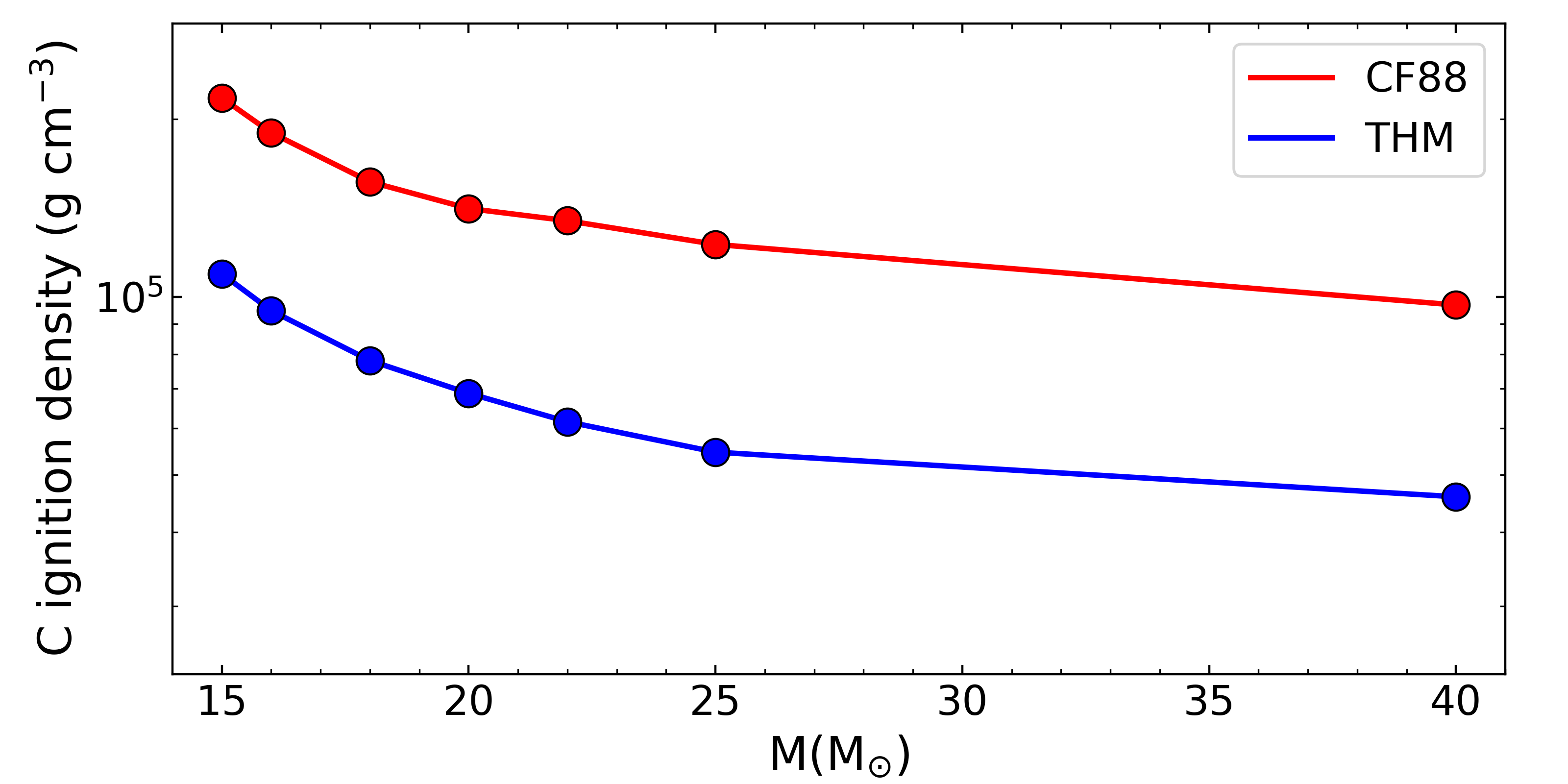}
        \includegraphics[width=0.49\linewidth]{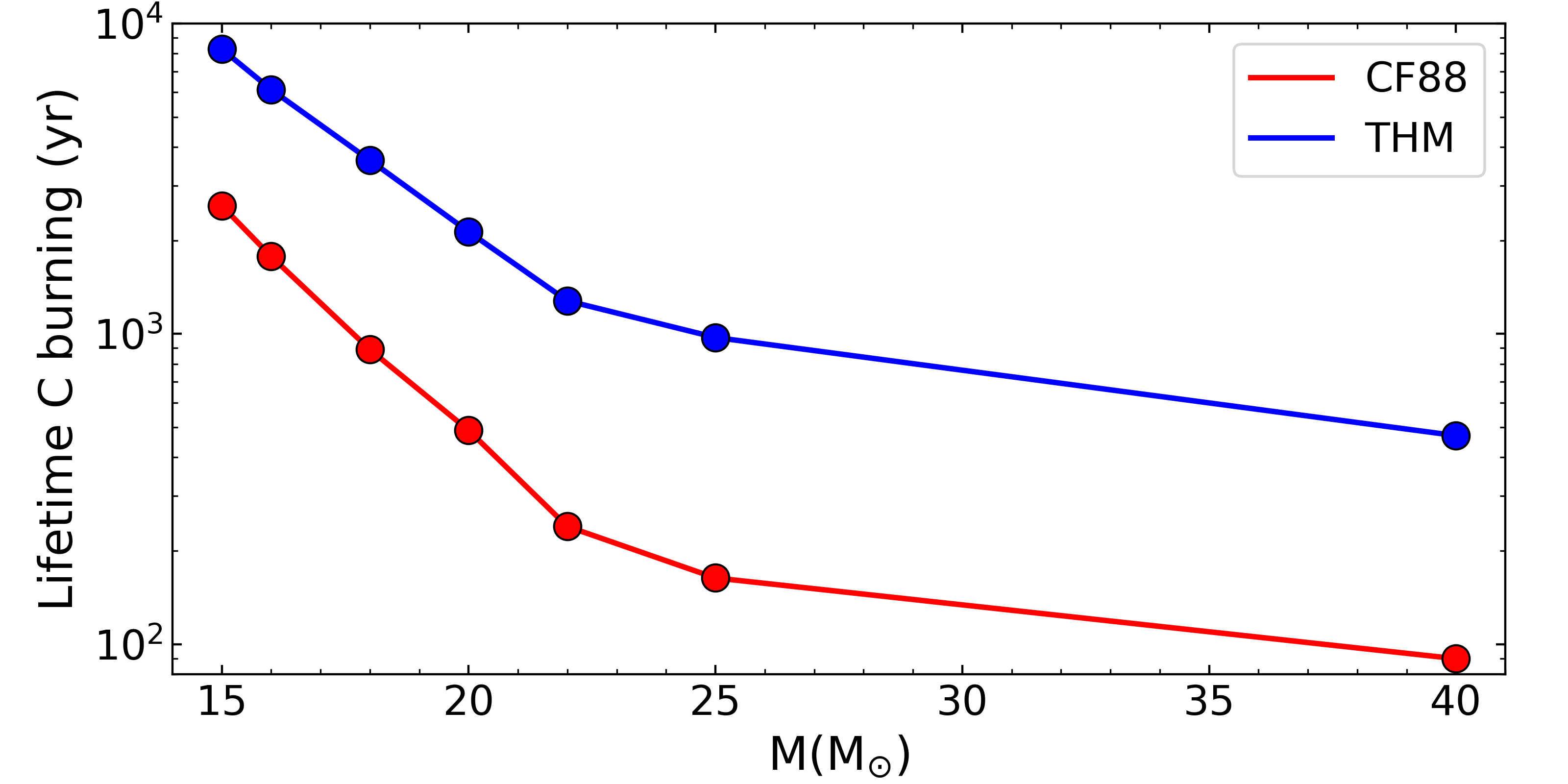}
        \includegraphics[width=0.49\linewidth]{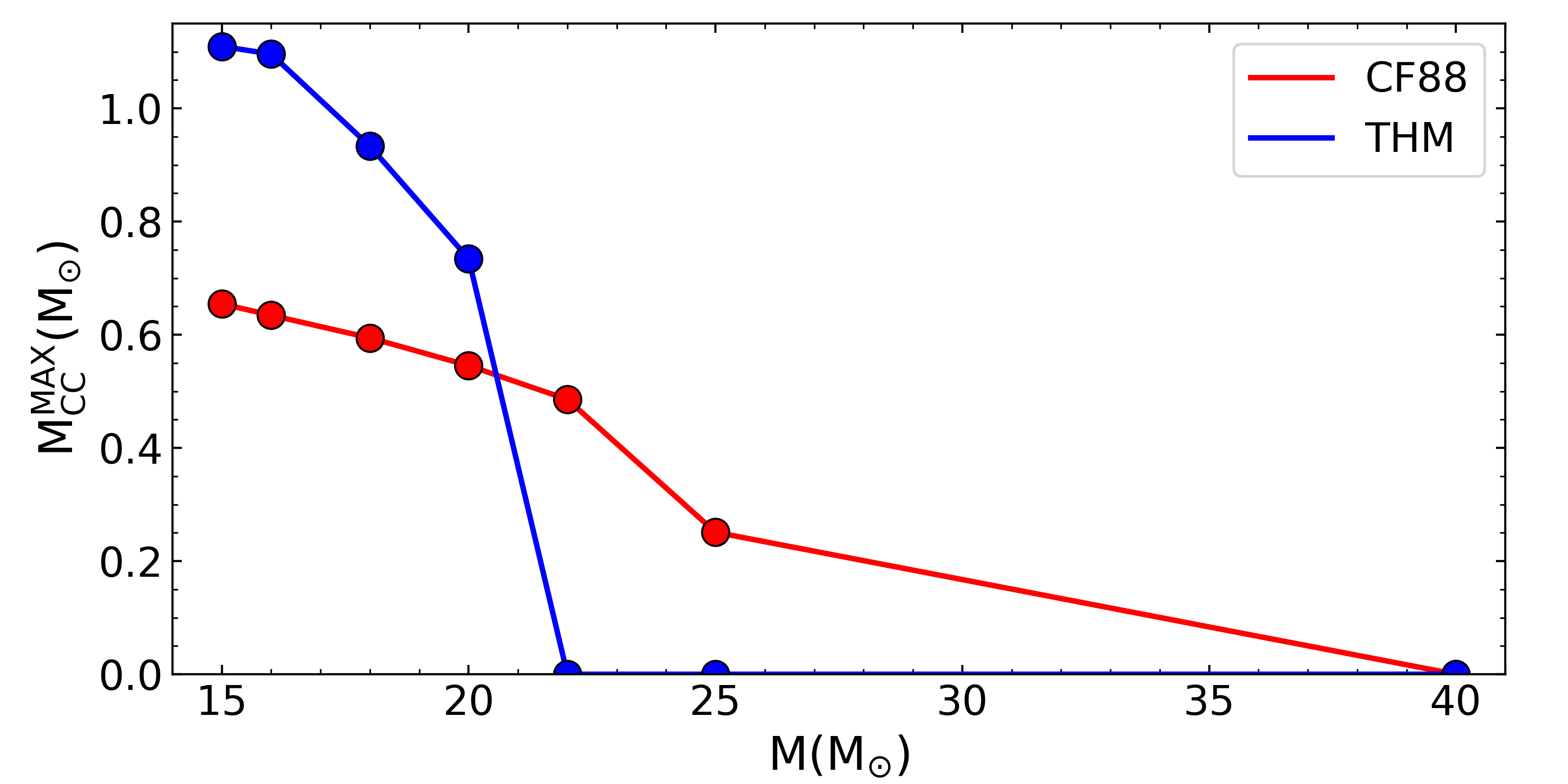}
        \caption{Central C burning ignition: temperature (upper left), density (upper right), duration (lower left), and maximum size of the convective core (lower right) for the CF88 (red) and THM (blue) models. C ignition is identified by the formation of a convective core, or when the central C mass fraction decreases by 15\% otherwise; C exhaustion is defined when the central mass fraction drops below $10^{-3}$.}
        \label{fig:cburn}
    \end{figure*}

    \begin{figure}[!t]
        \centering
        \includegraphics[width=\linewidth]{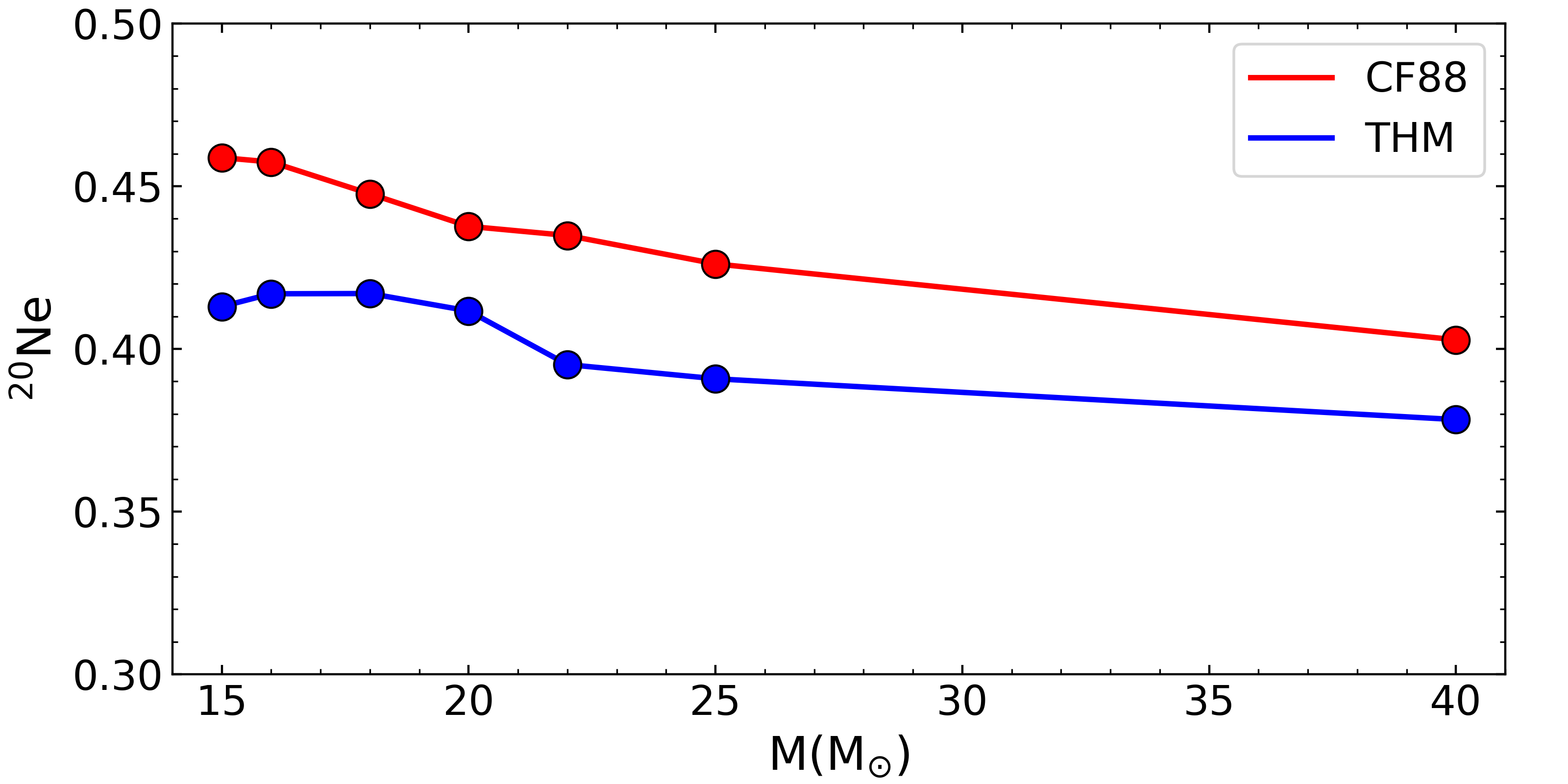}
        \includegraphics[width=\linewidth]{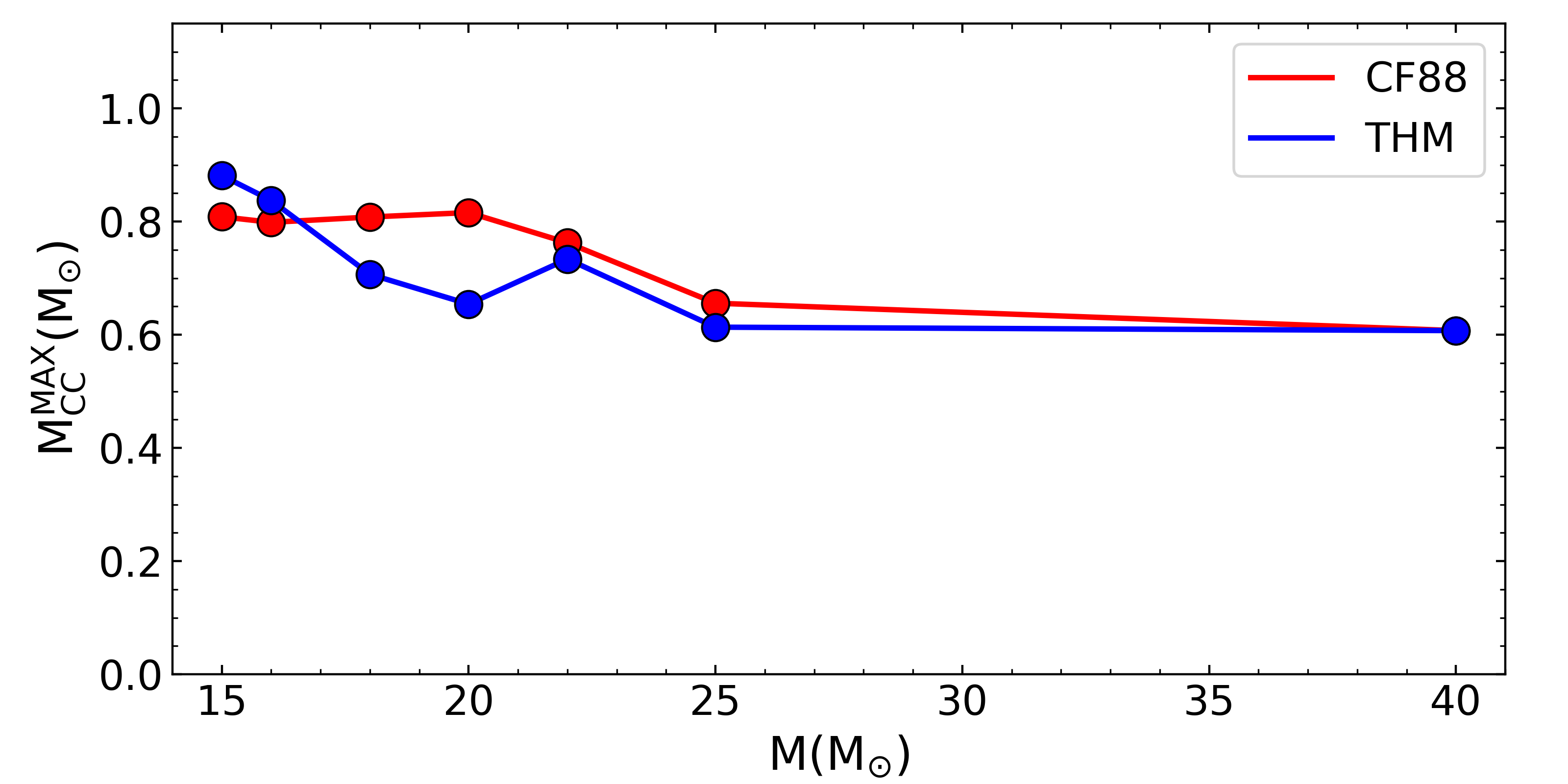}
        \caption{Upper panel: \isotope[20]{Ne} mass fraction left by central C burning for CF88 (red) and THM (blue) models. Lower panel: the maximum size of the convective core during Ne burning for CF88 (red) and THM (blue) models}
        \label{fig:ne}
    \end{figure}
    
    \begin{figure*}[!t]
        \centering
        \includegraphics[width=0.49\linewidth]{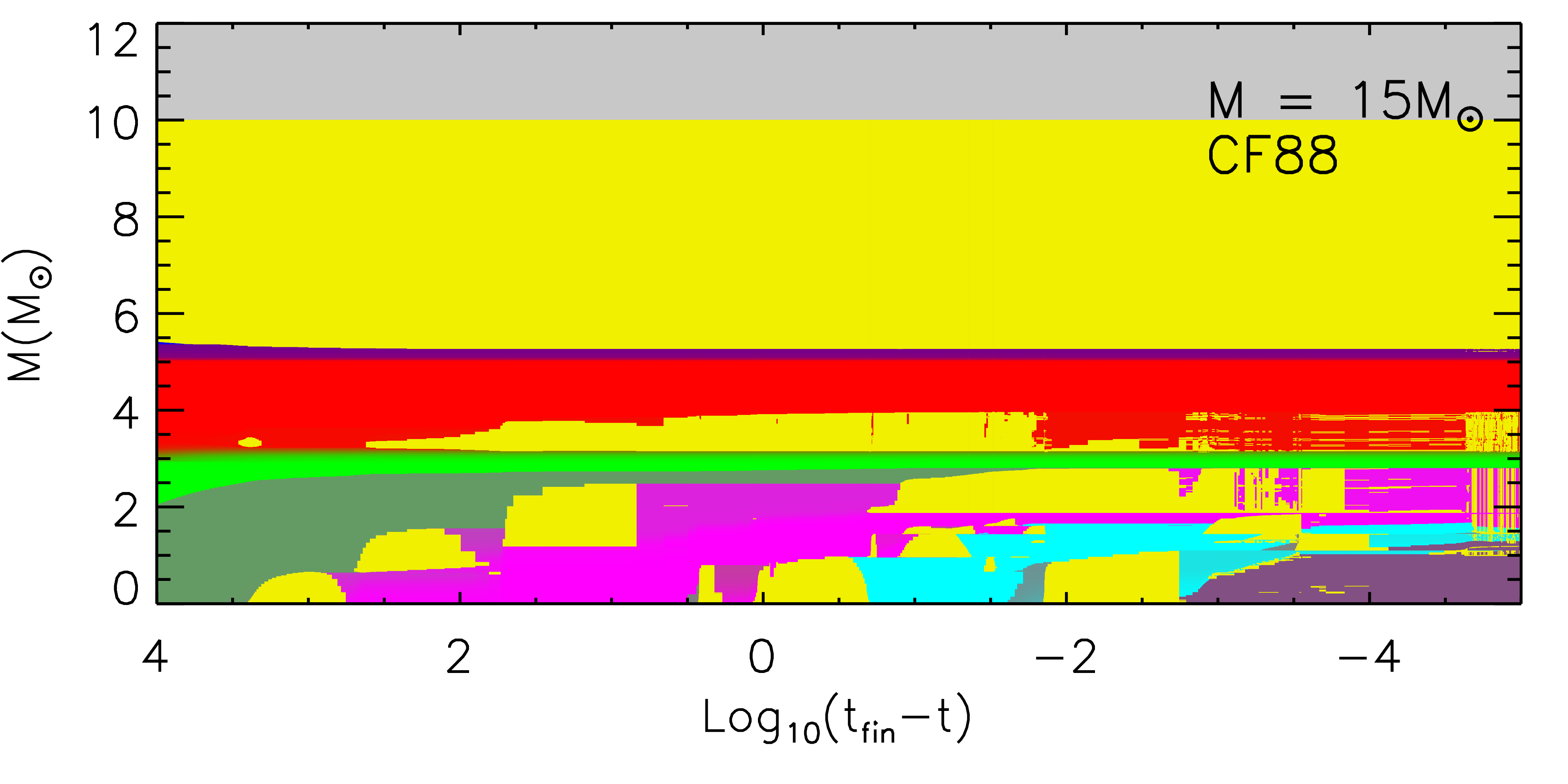}
        \includegraphics[width=0.49\linewidth]{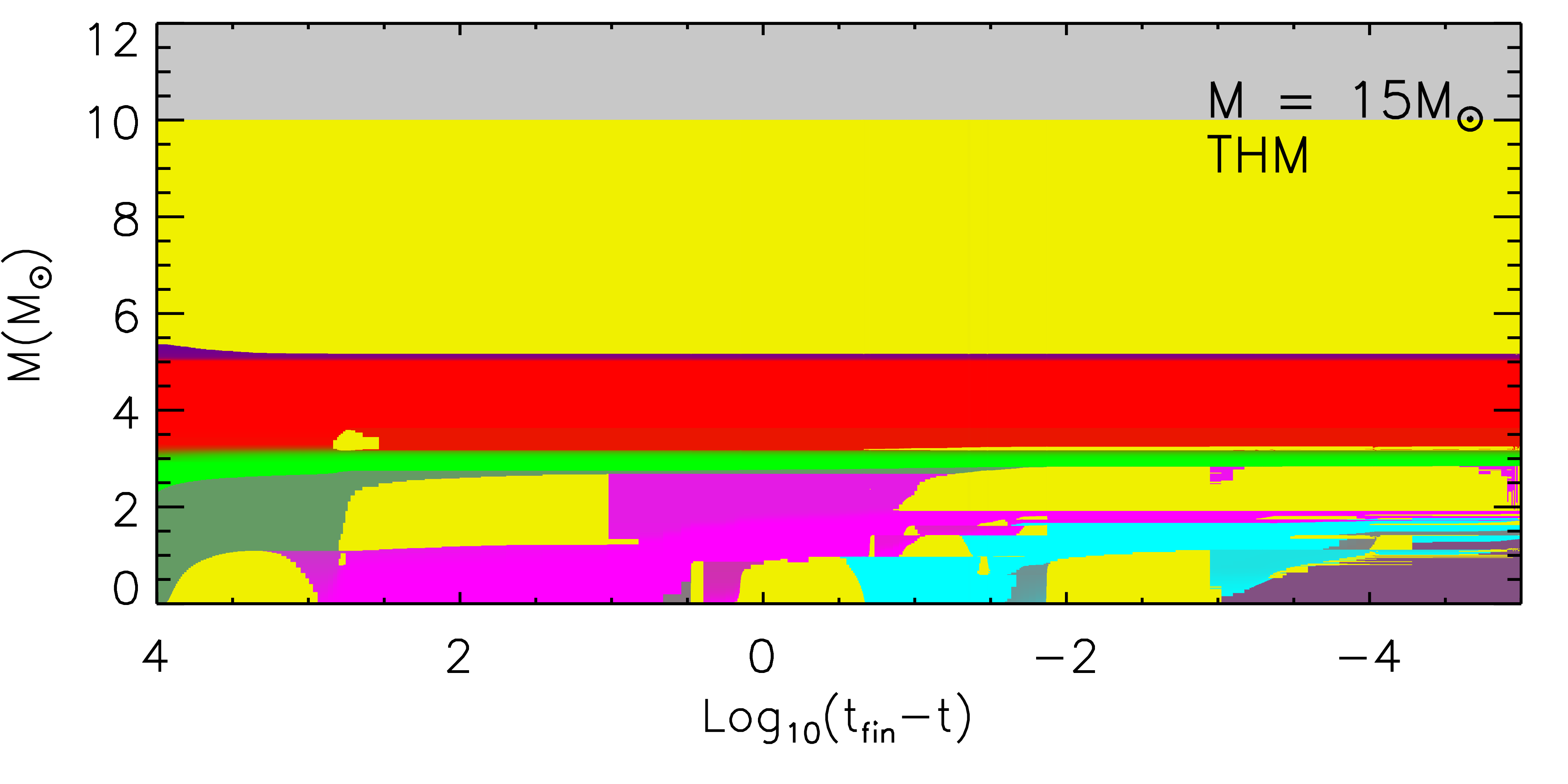}
        \includegraphics[width=0.49\linewidth]{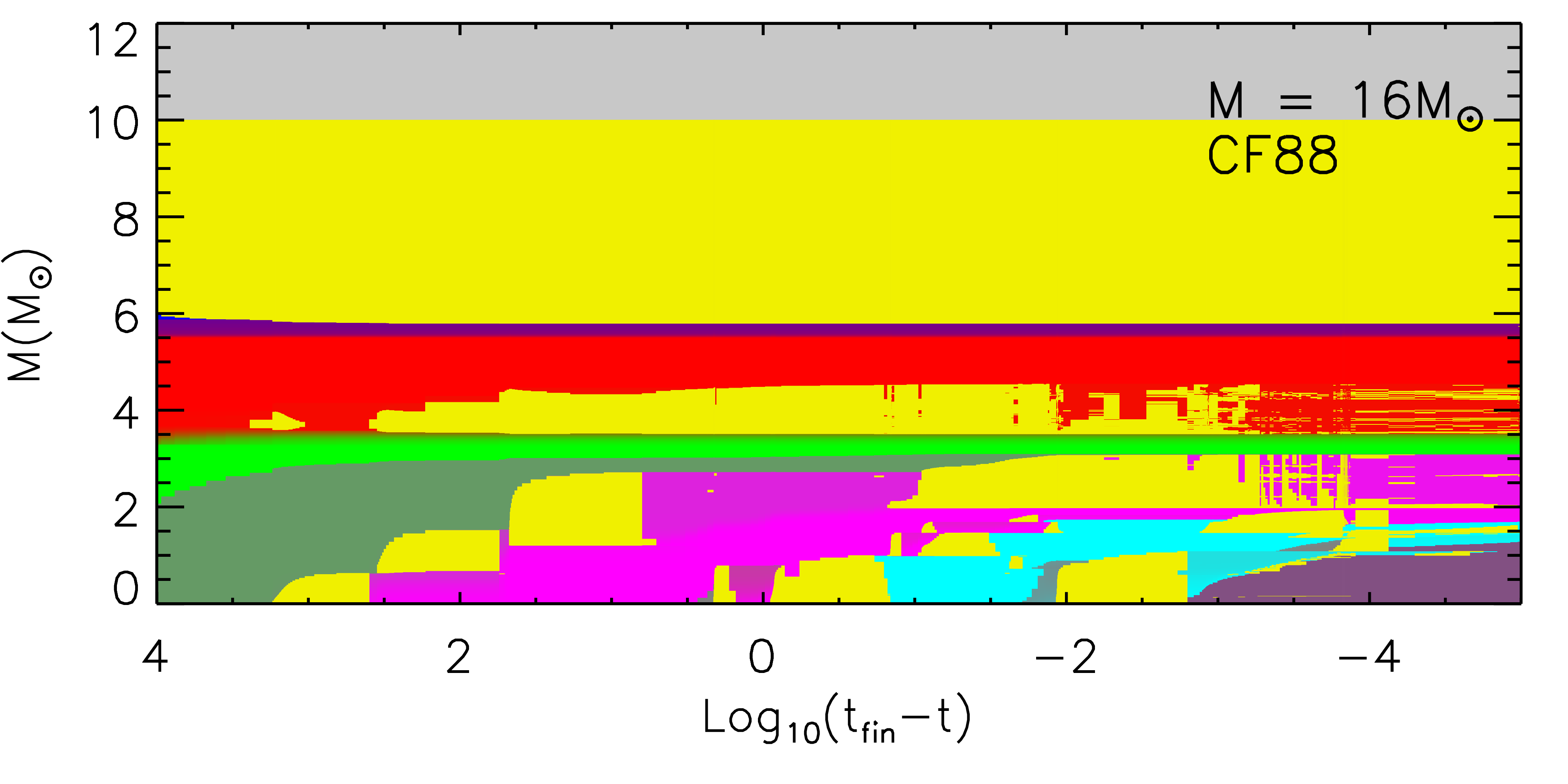}
        \includegraphics[width=0.49\linewidth]{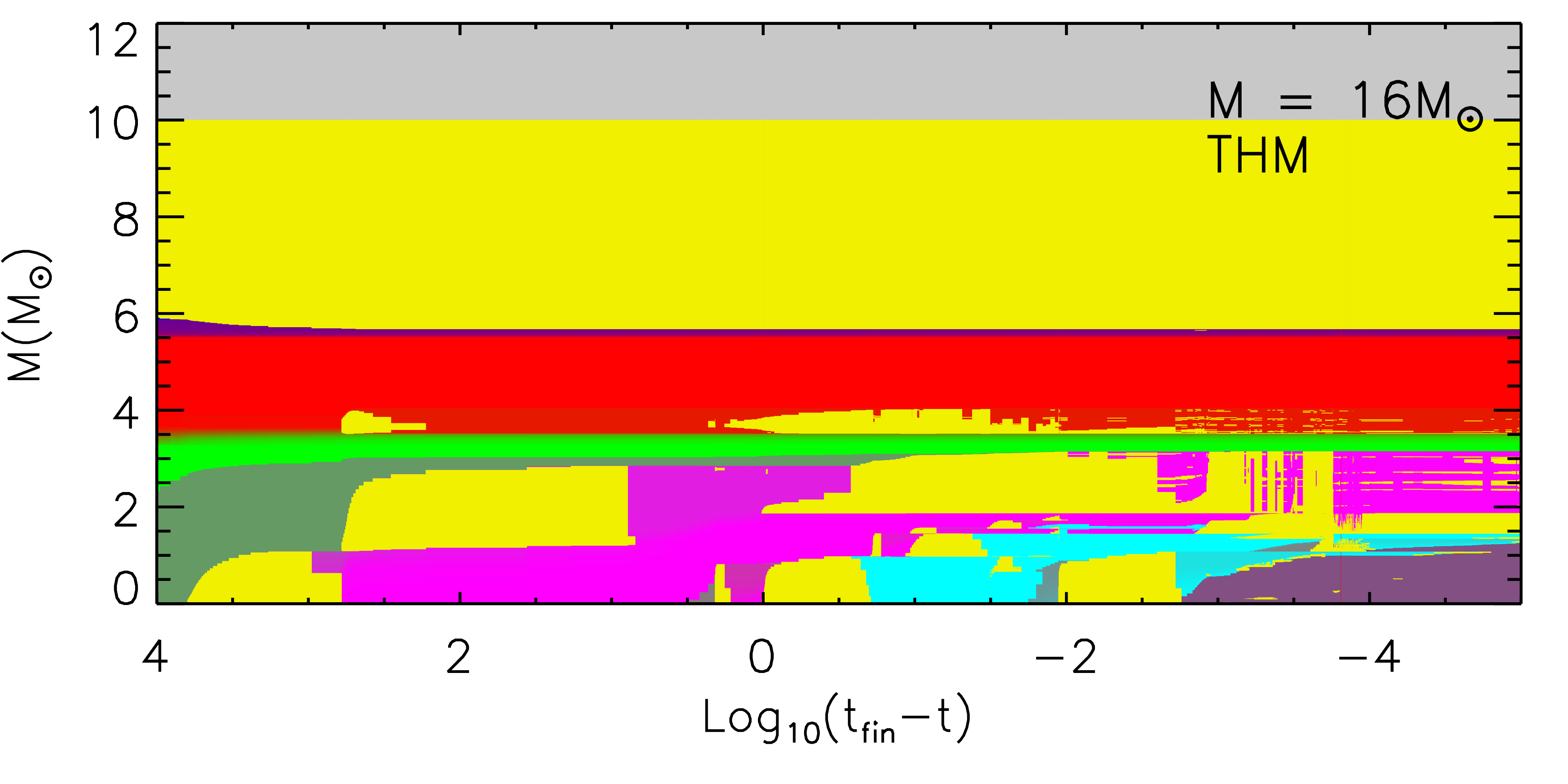}
        \includegraphics[width=0.49\linewidth]{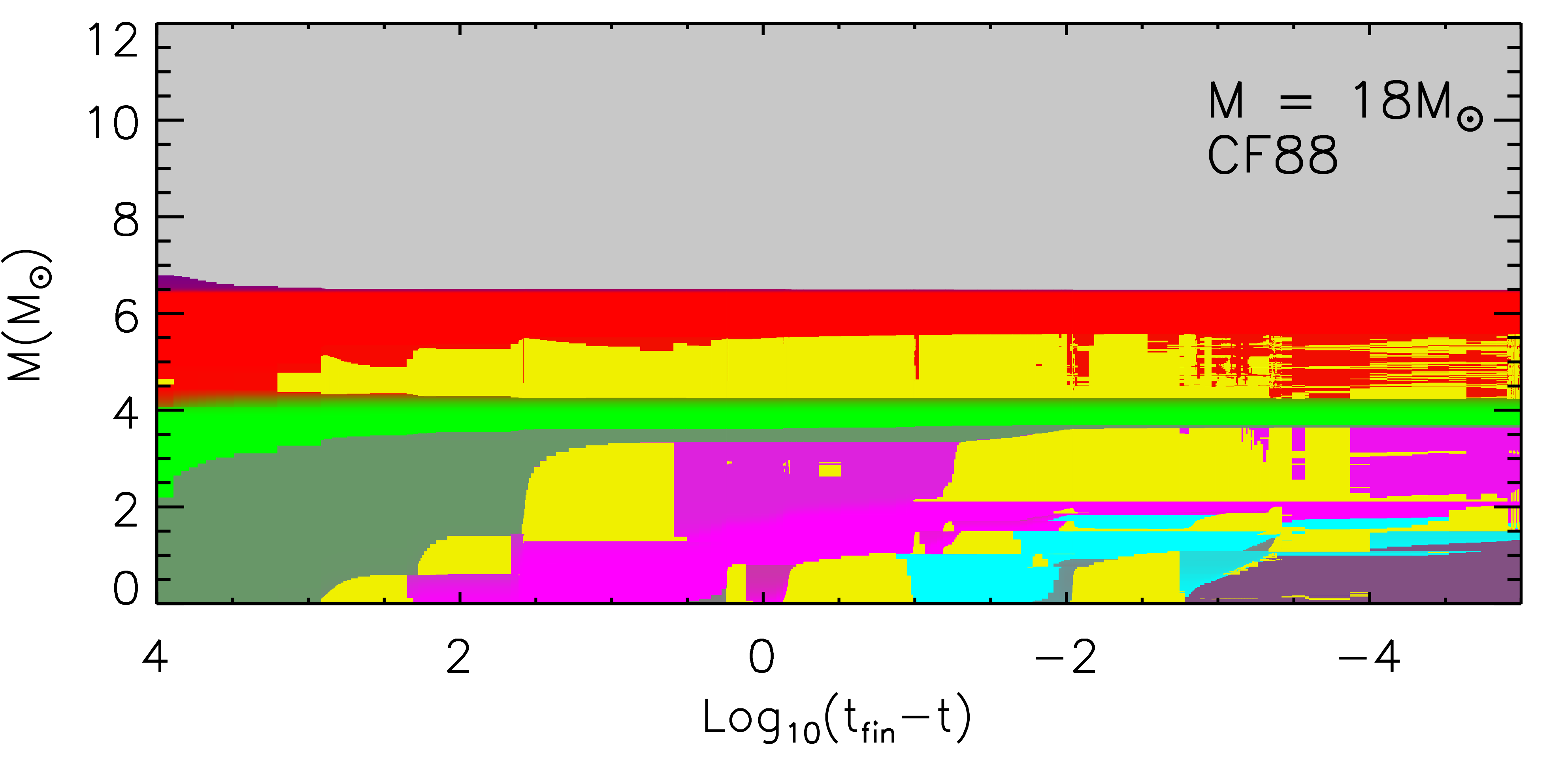}
        \includegraphics[width=0.49\linewidth]{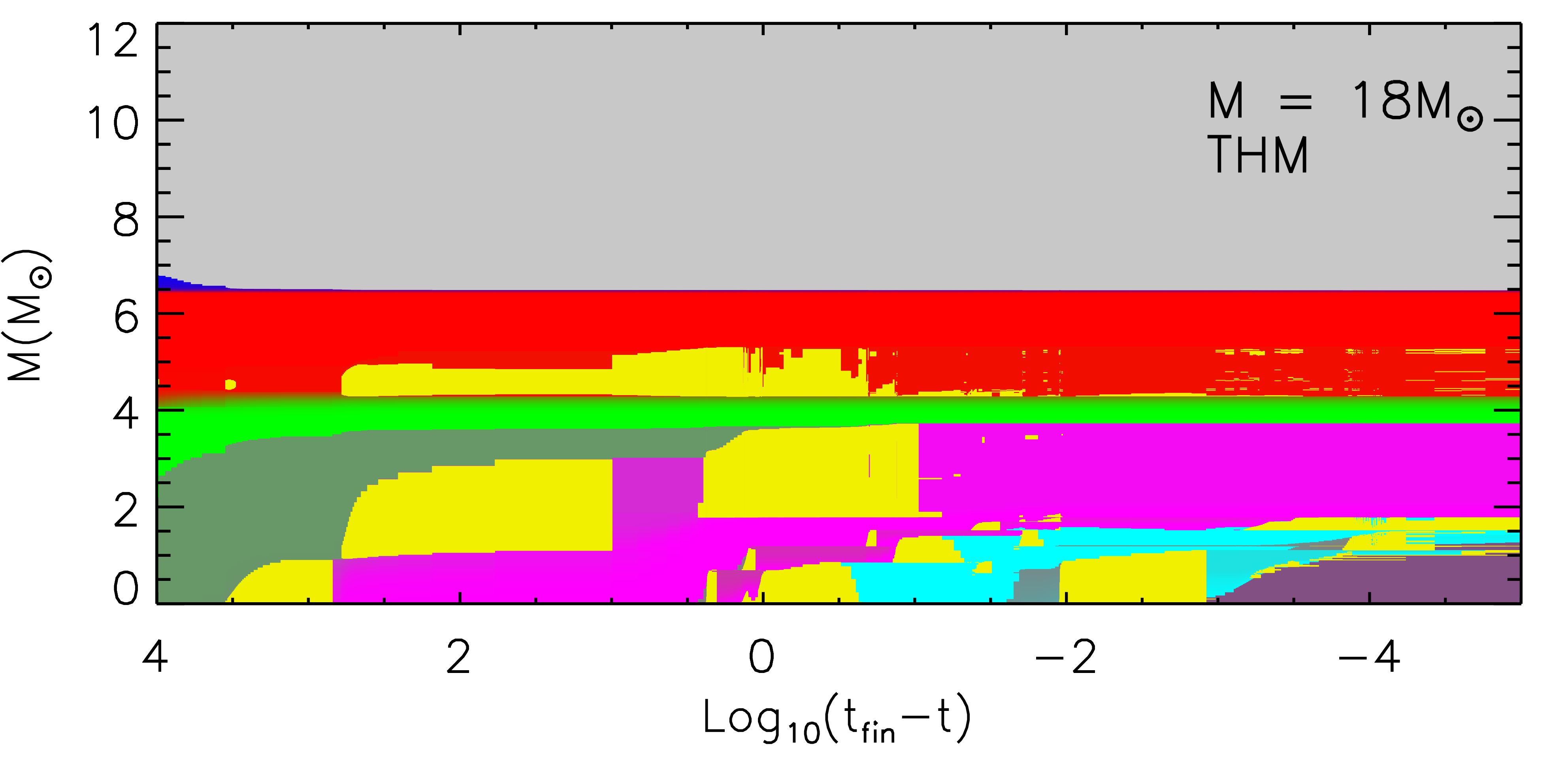}
        \includegraphics[width=0.49\linewidth]{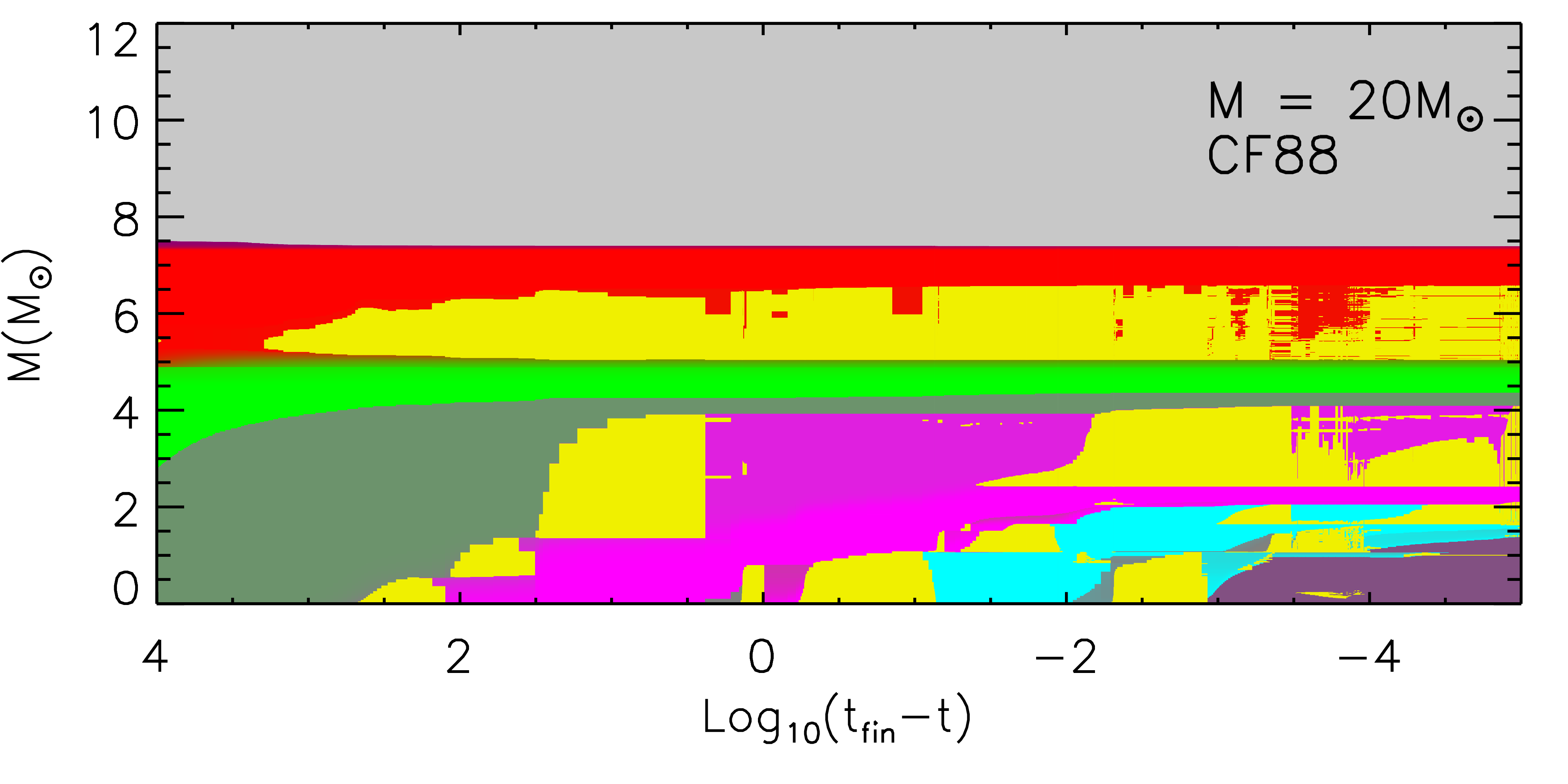}
        \includegraphics[width=0.49\linewidth]{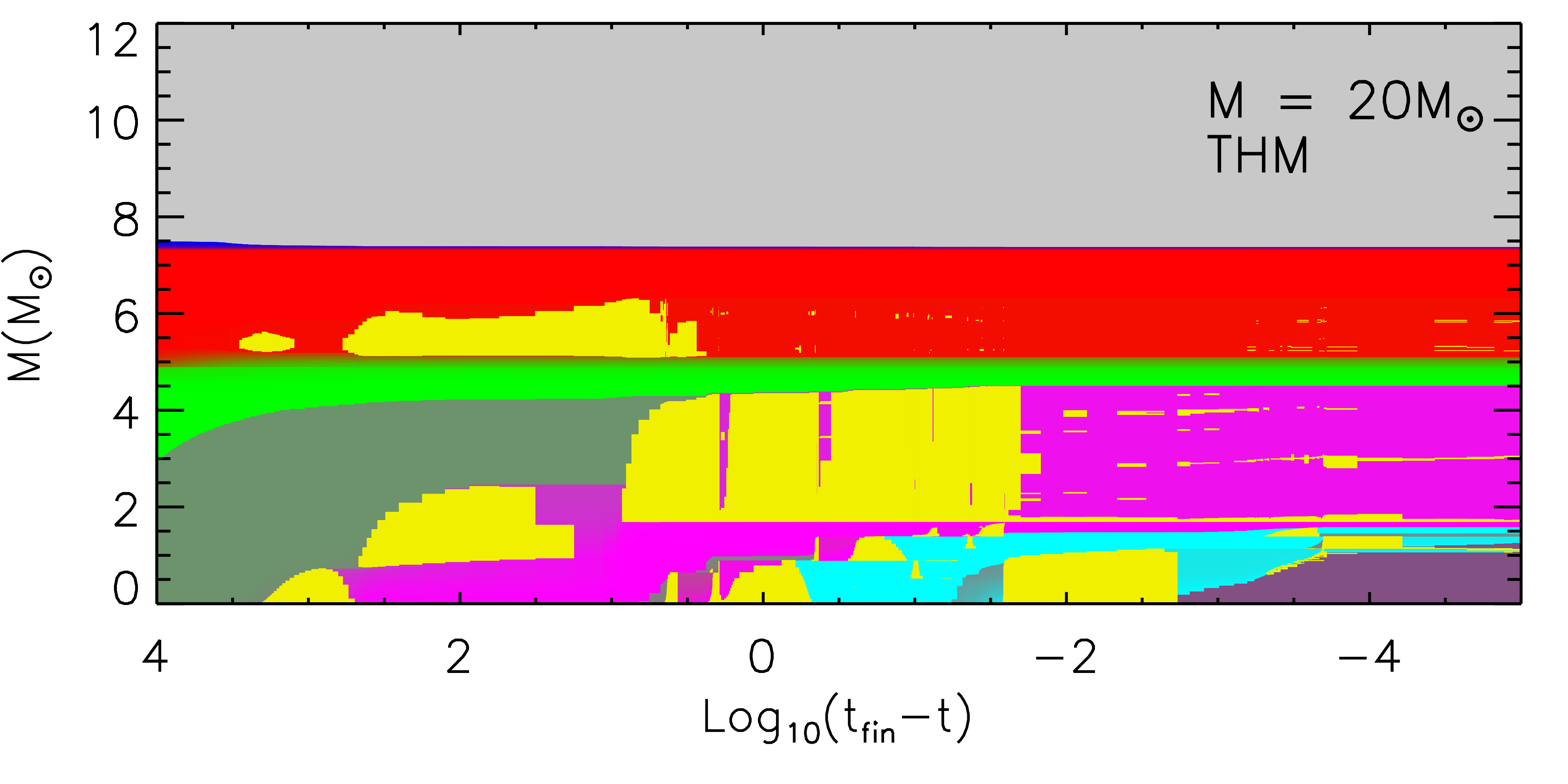}
        \caption{Kippenhahn diagrams for the CF88 (left) and THM (right) models between 15 and 20 \msun.}
        \label{fig:kip1}
    \end{figure*}

    \begin{figure*}[!t]
        \centering
        \includegraphics[width=0.49\linewidth]{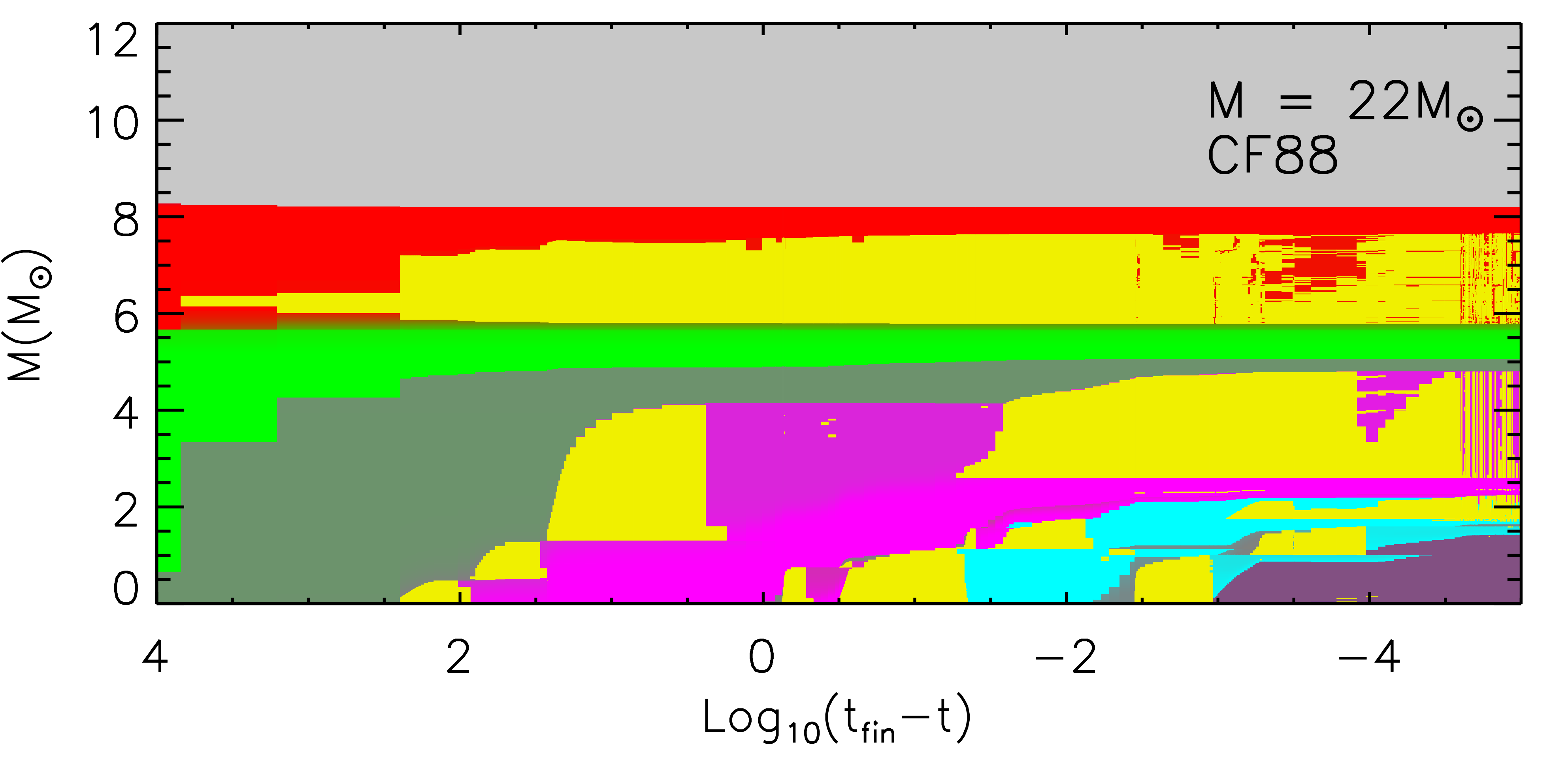}
        \includegraphics[width=0.49\linewidth]{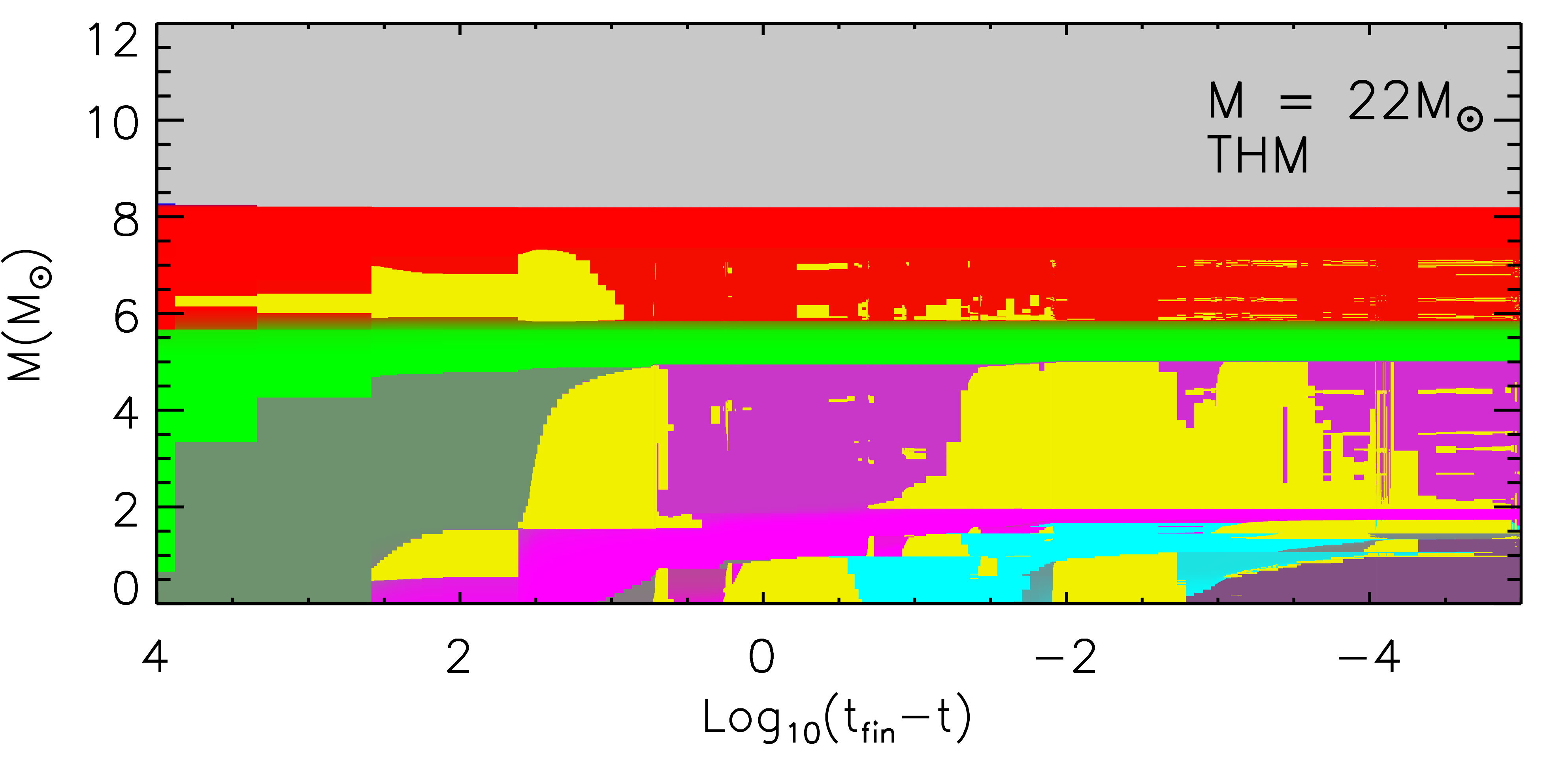}
        \includegraphics[width=0.49\linewidth]{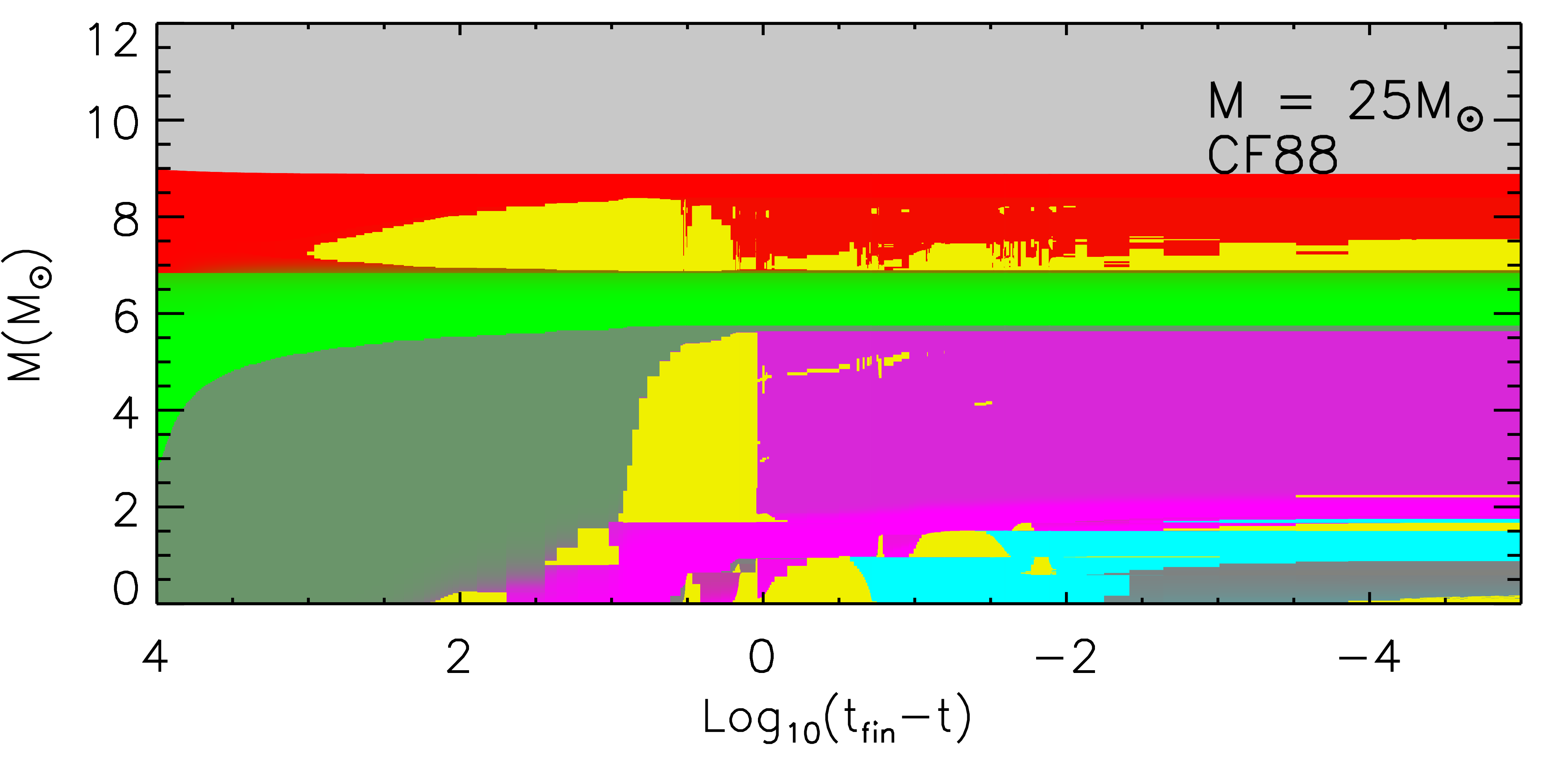}
        \includegraphics[width=0.49\linewidth]{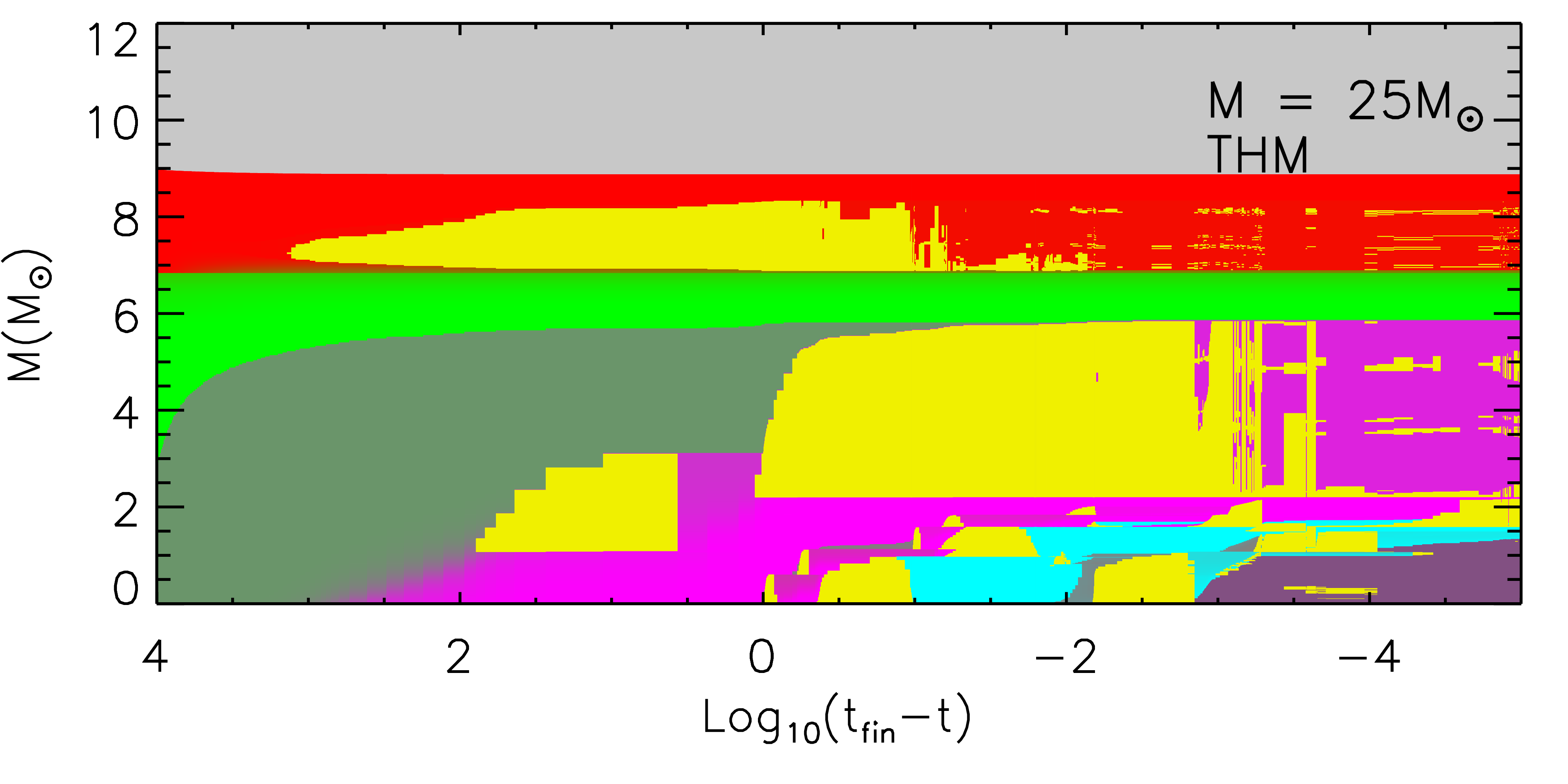}
        \includegraphics[width=0.49\linewidth]{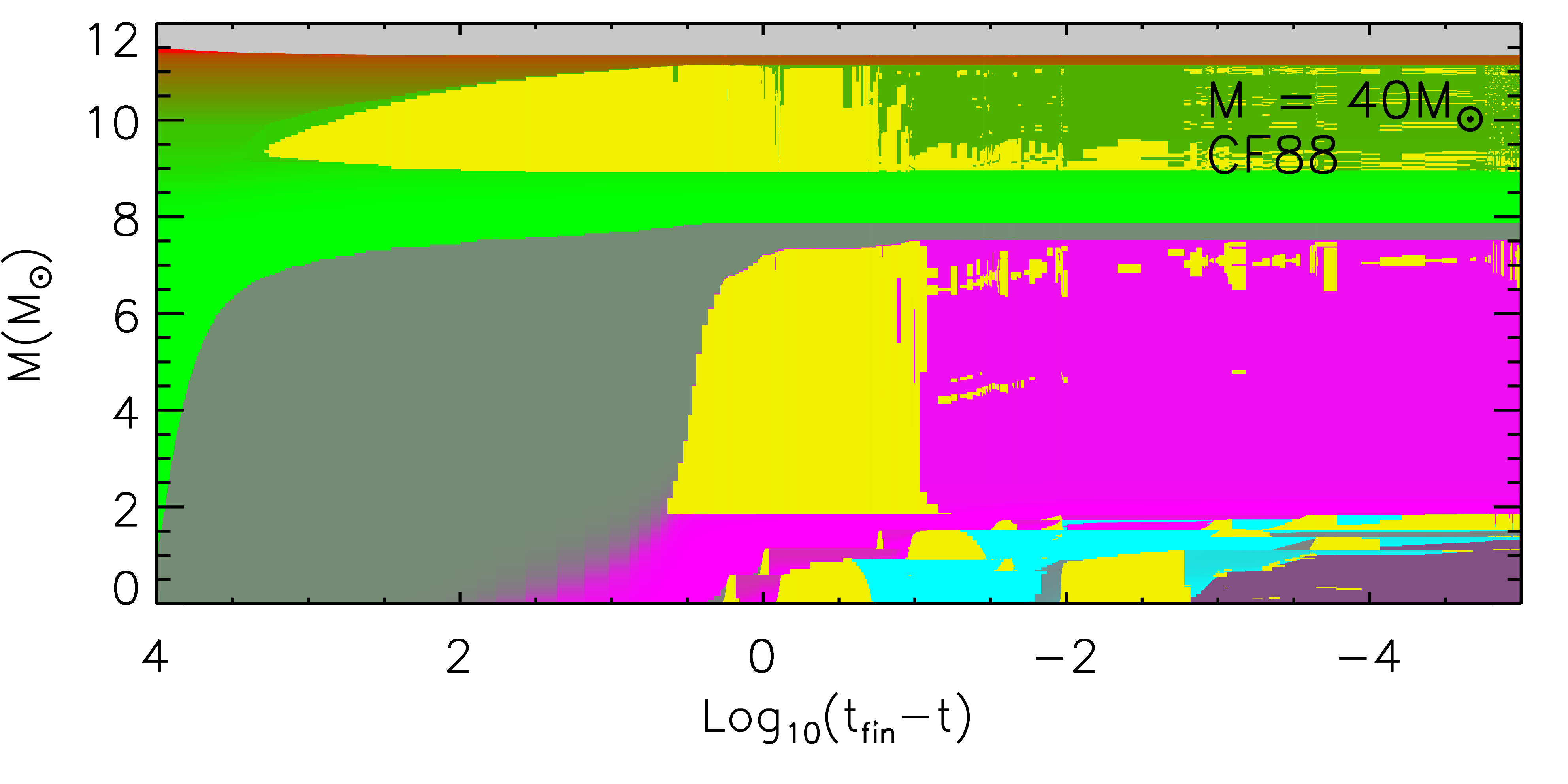}
        \includegraphics[width=0.49\linewidth]{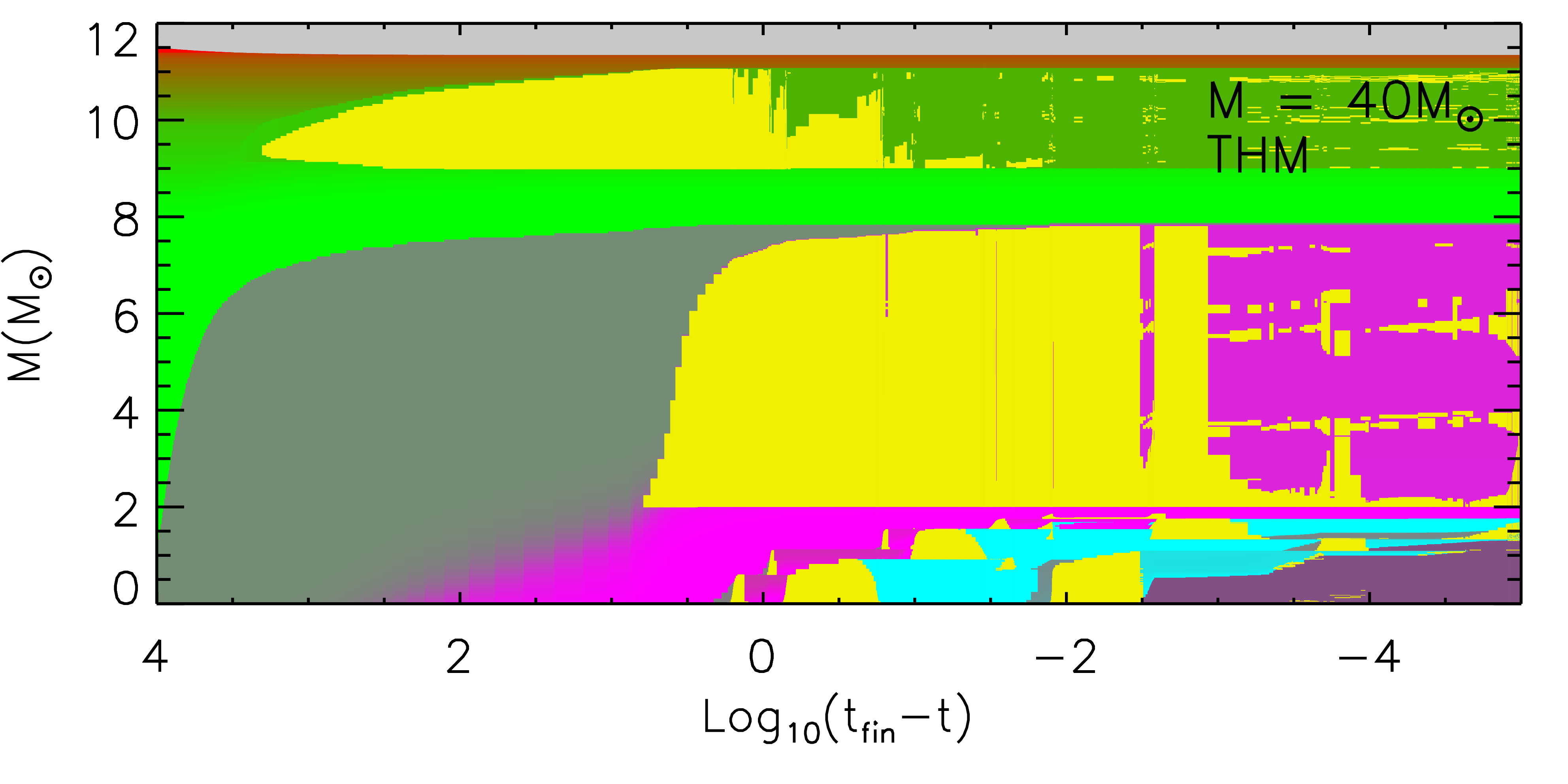}
        \caption{Same as \figurename~\ref{fig:kip1}, but for the models between 22 and 40 \msun.}
        \label{fig:kip2}
    \end{figure*}

    \begin{figure*}[!t]
        \centering
        \includegraphics[width=0.49\linewidth]{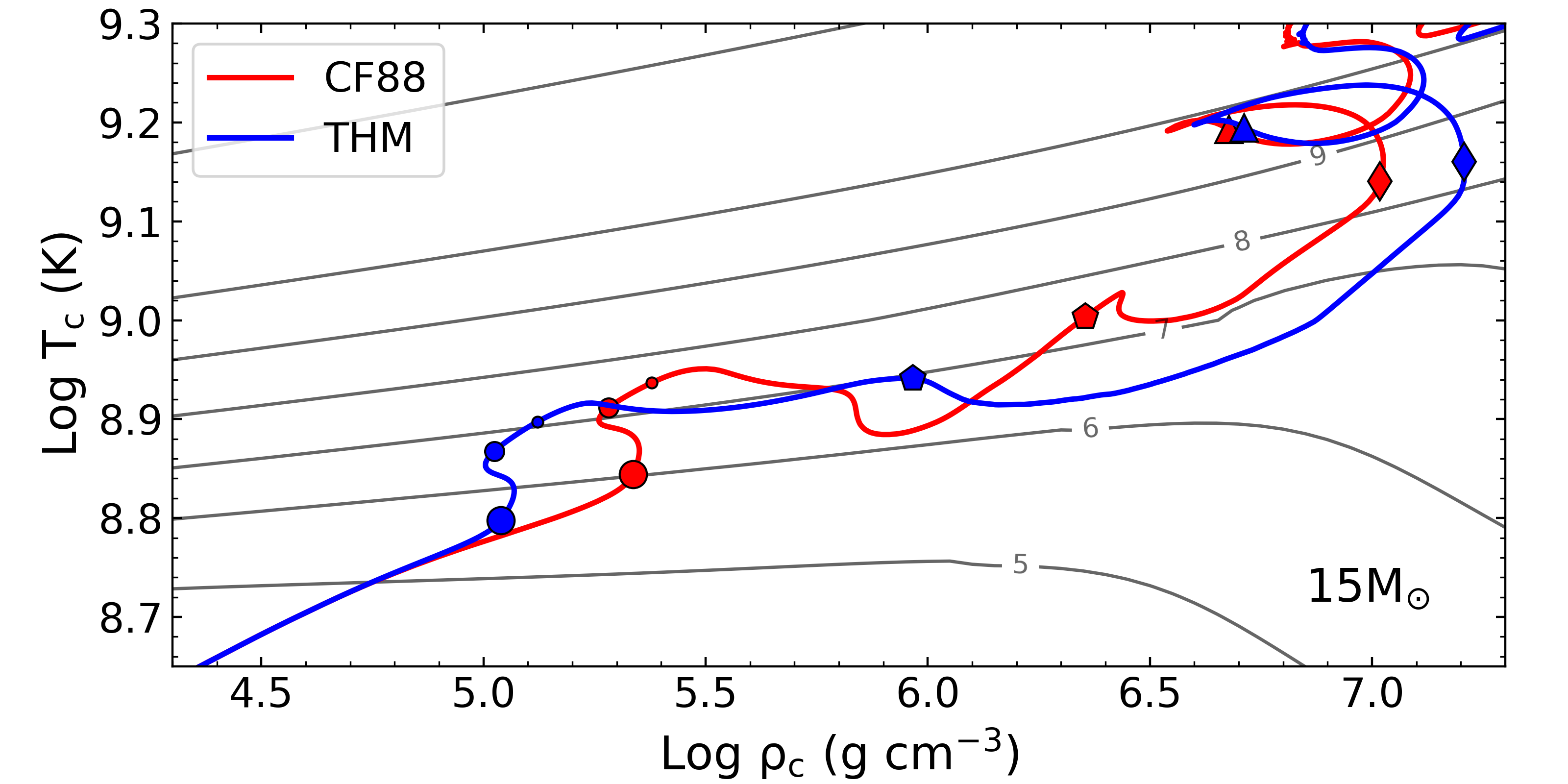}
        \includegraphics[width=0.49\linewidth]{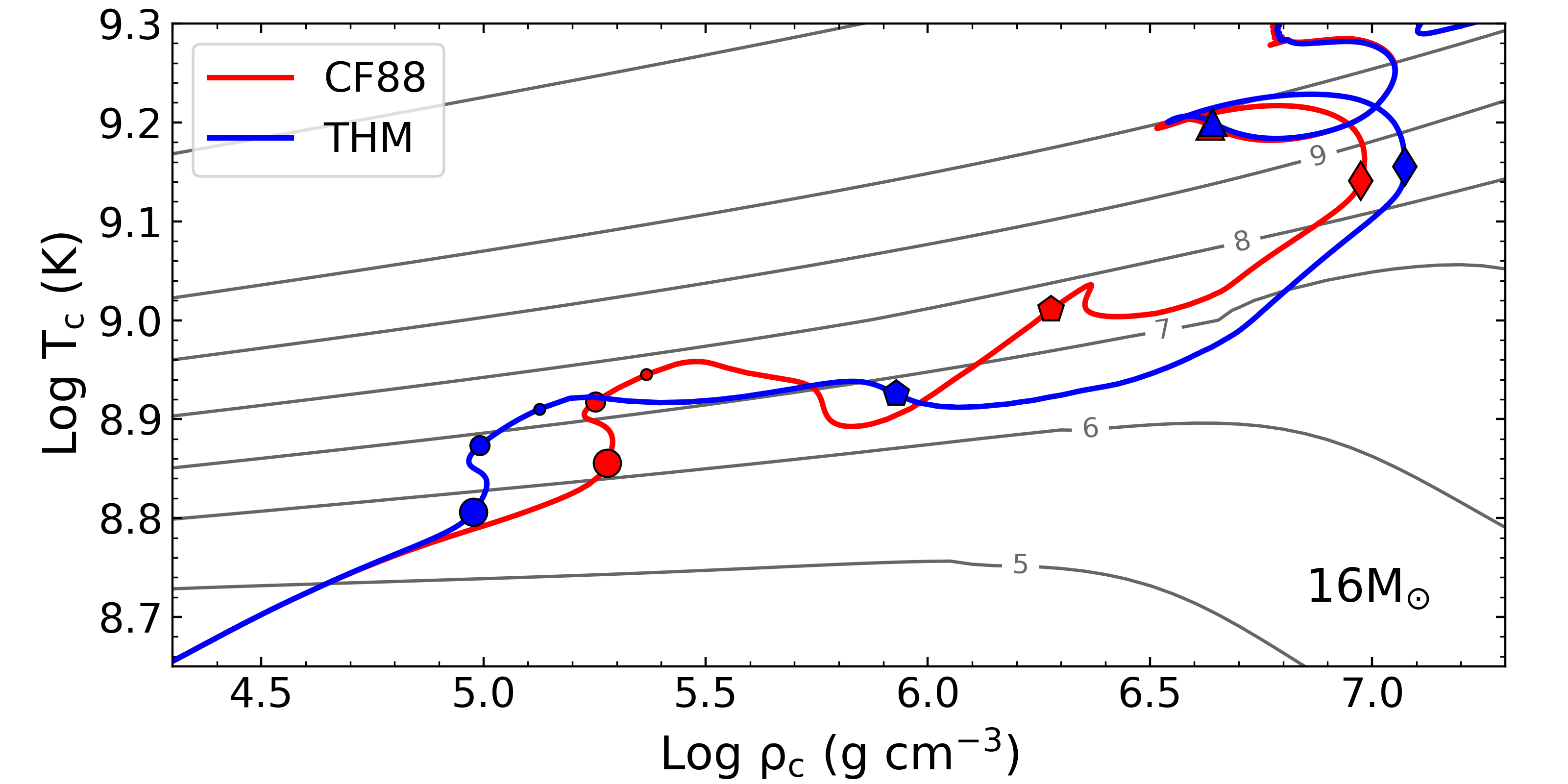}
        \includegraphics[width=0.49\linewidth]{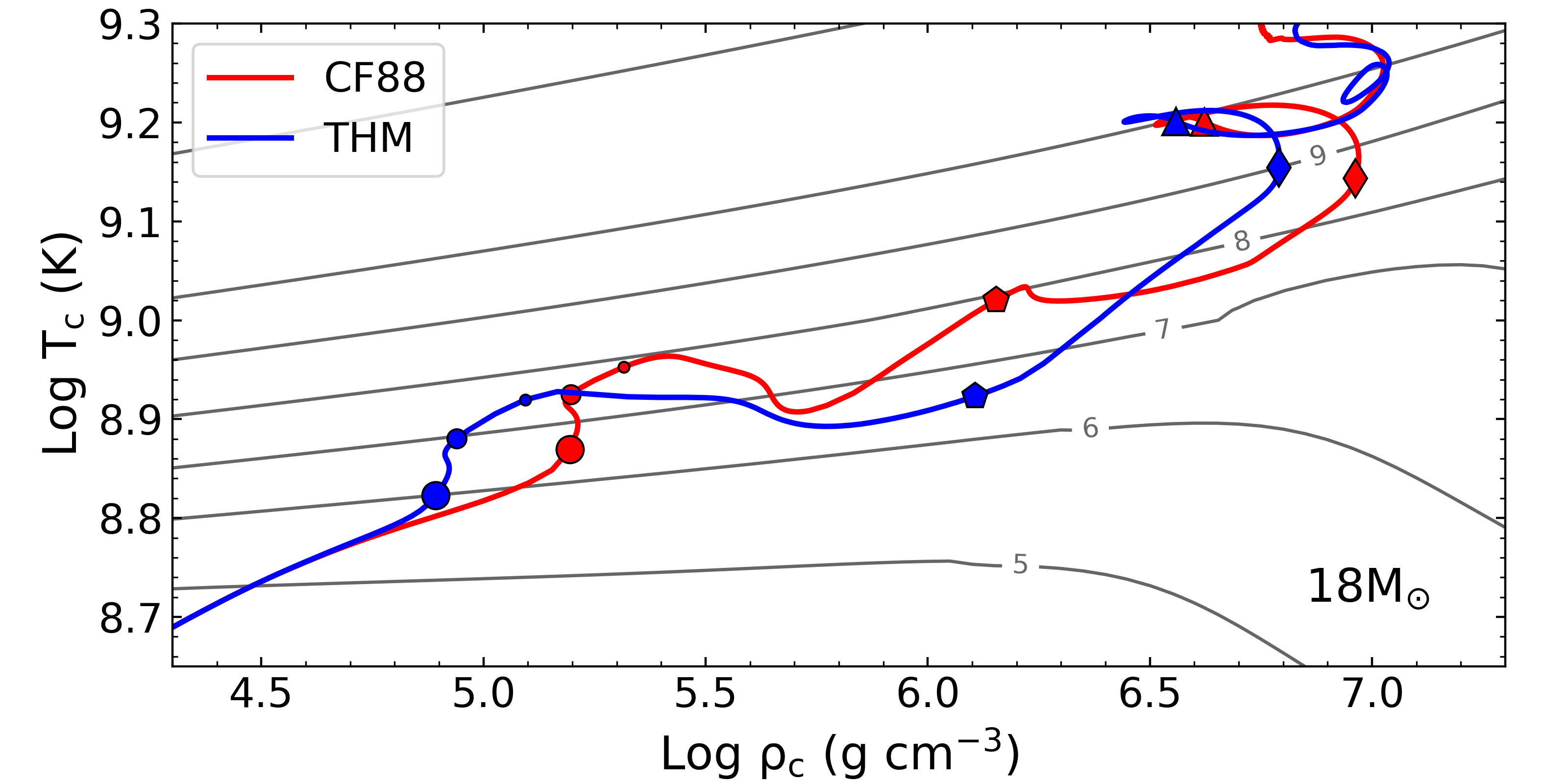}
        \includegraphics[width=0.49\linewidth]{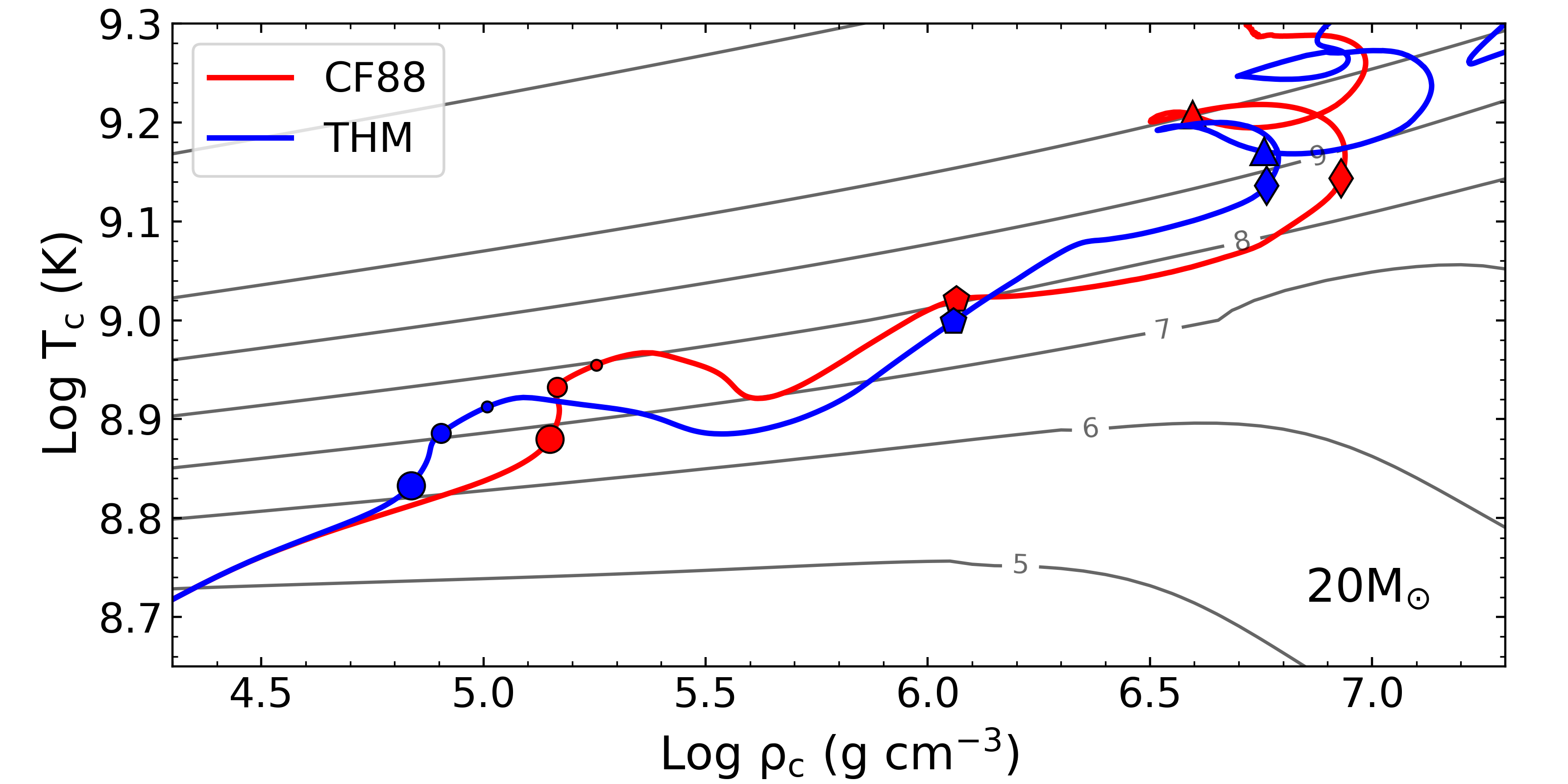}
        \includegraphics[width=0.49\linewidth]{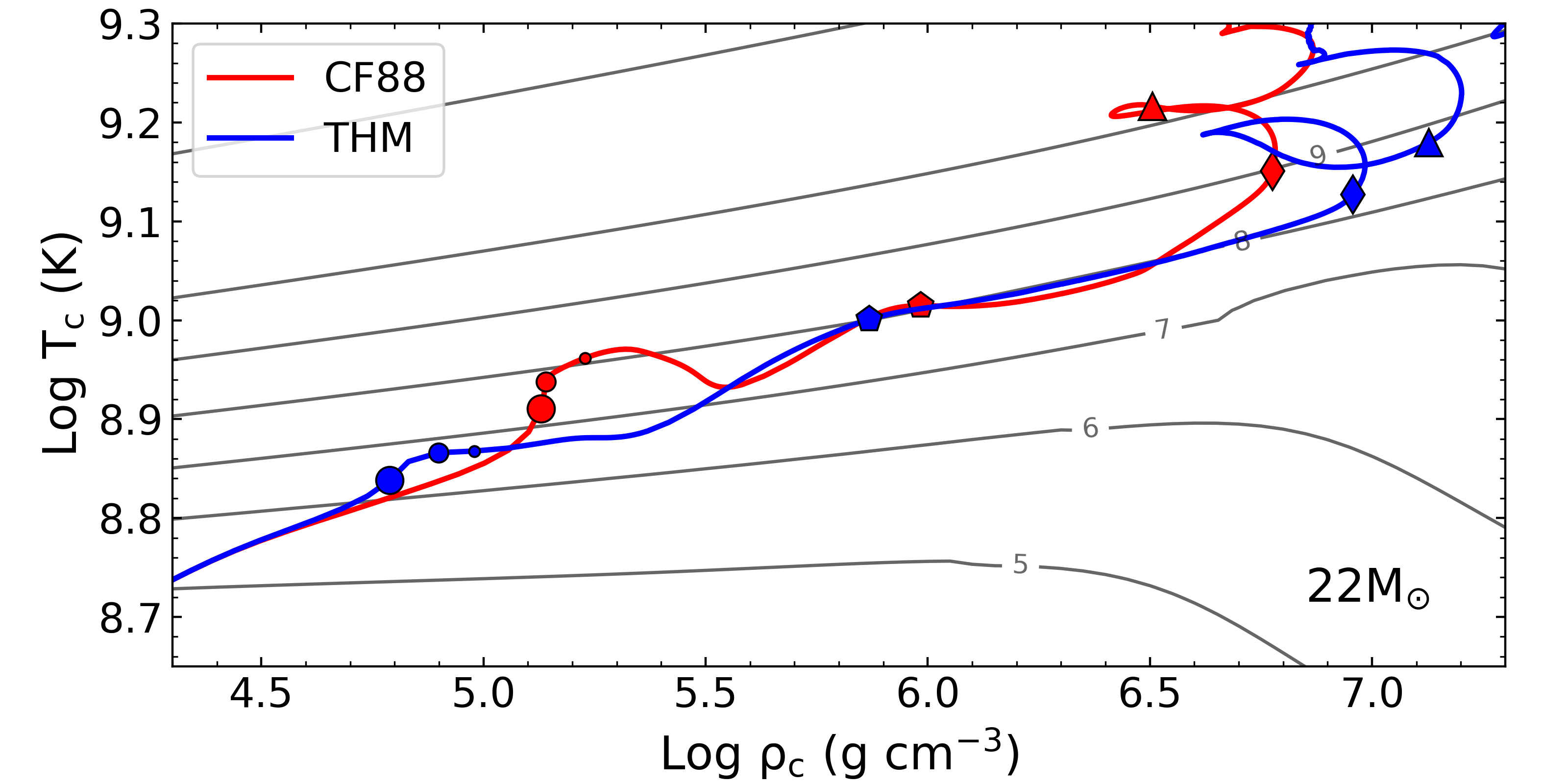}
        \includegraphics[width=0.49\linewidth]{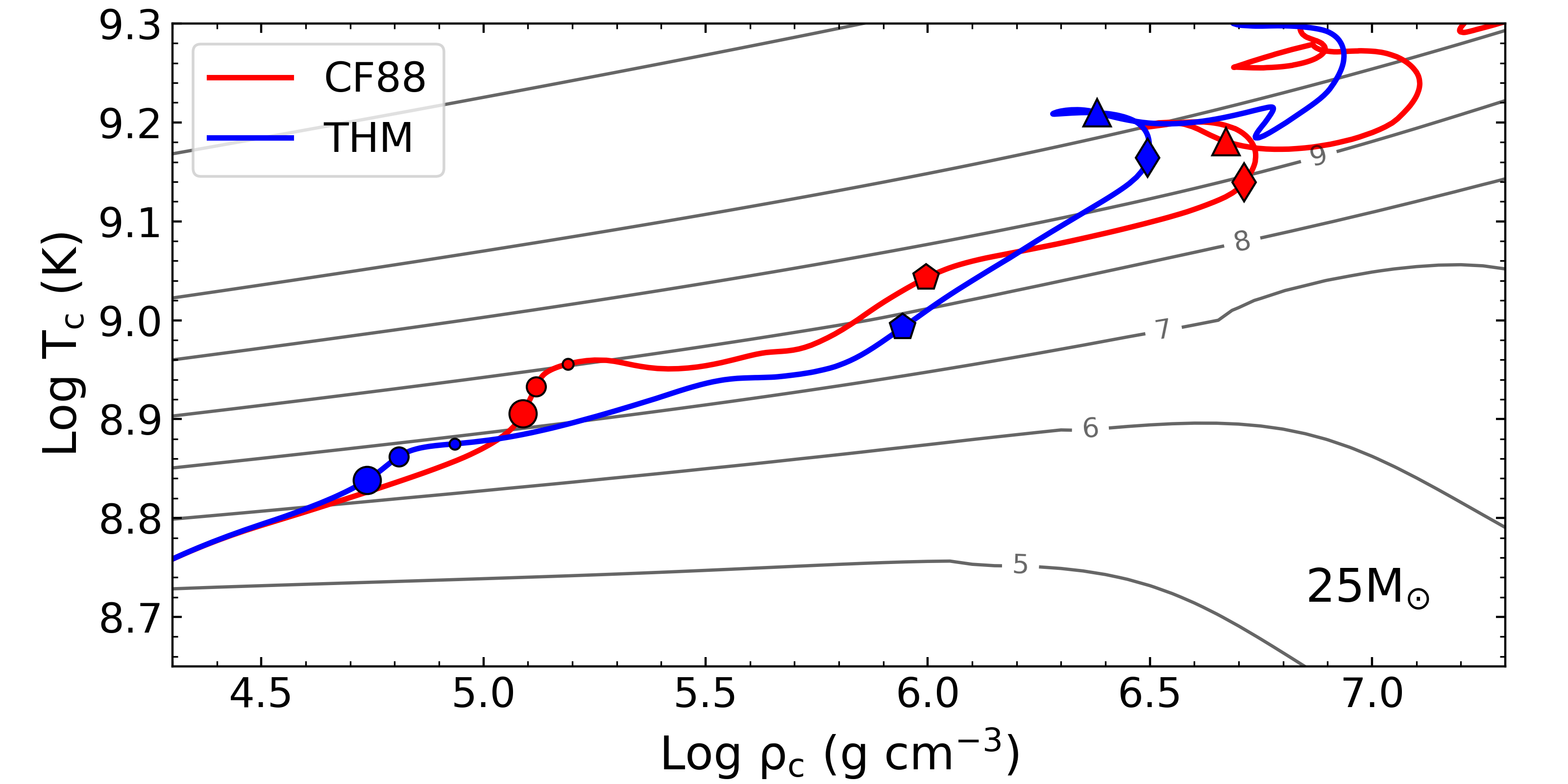}
        \includegraphics[width=0.49\linewidth]{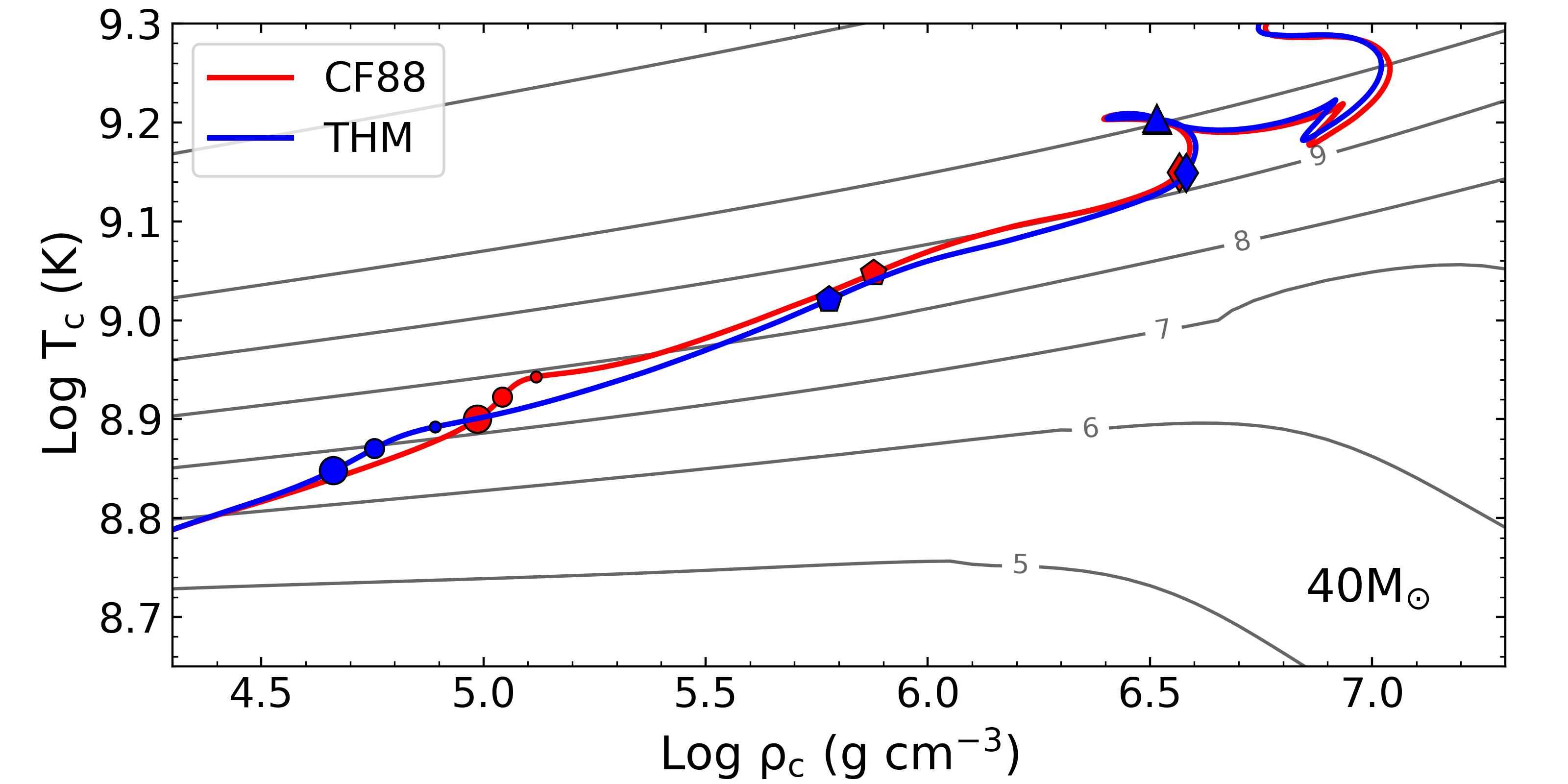}
        \caption{Central temperature versus central density diagrams for each mass. The progressively smaller dots represent: C ignition (see text), X(\isotope[12]{C})$<0.2$, and X(\isotope[12]{C})$<0.1$. The pentagon represents C exhaustion (X(\isotope[12]{C})$<10^{-3}$), the diamond symbol represents the Ne ignition (formation of a convective core), and the triangle symbol represents Ne exhaustion (X(\isotope[20]{Ne})$<10^{-3}$), for CF88 (red) and THM (blue) models. The black lines correspond to the logarithm of the absolute value of isoneutrino energy losses (erg per gram per second).}
        \label{fig:tcrhoc}
    \end{figure*}
    
    \begin{figure}[!t]
        \centering
        \includegraphics[width=\linewidth]{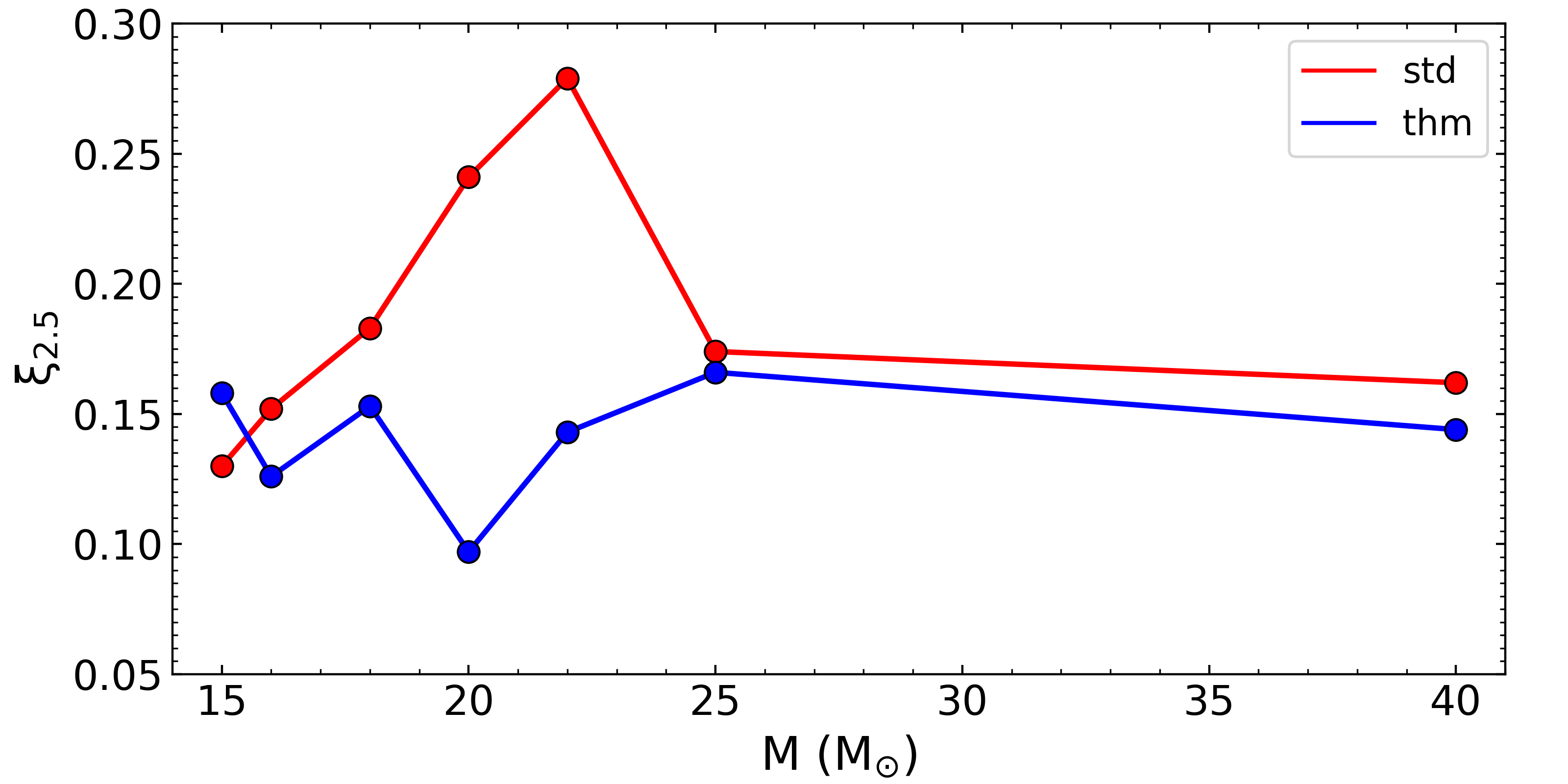}
        \caption{The compactness $\xi_{2.5}$ at the pre-supernova stage for CF88 (red) and THM (blue) models.}
        \label{fig:xi}
    \end{figure}

    \subsection{Stellar structure and evolution}
    
    The adoption of a different \cc nuclear cross section introduces significant modifications in the pre-supernova evolution and nucleosynthesis. Most of the structural changes have already been discussed in \cite{chieffi:21}; therefore, we only briefly recap the main features here. In the THM models, the higher nuclear cross section allows C burning to activate at lower temperature and density (\figurename~\ref{fig:cburn}, upper panels) with respect to the CF88 case, where the neutrino losses are less efficient. The consequence is that the convective core extends over a wider mass interval and the central C burning lifetime increases by roughly a factor of three with respect to models computed with the CF88 rate (\figurename~\ref{fig:cburn}, lower left panel). It must be noted, however, that the maximum mass that forms a C convective core reduces in the THM models: a detailed discussion of the reason why the mass interval of the models that form a C convective core scales inversely with \cc nuclear reaction rate while the mass size of the convective core scales directly with the rate may be found in \cite{chieffi:21}. In the present set of models, a convective core forms up to the 20 \msun\ in the THM models and up to the 25 \msun\ in the models employing the CF88 rate. An additional consequence of the lower temperature at which C burning occurs in the THM models is that the \isotope[16]{O}$(\alpha, \gamma)$\isotope[20]{Ne} reaction is less efficient so that the final abundance of \isotope[20]{Ne} at the central C exhaustion reduces with respect to the CF88 case (\figurename~\ref{fig:ne}, upper panel).

    Following central C exhaustion, the core contracts and the temperature increases, shifting the C burning in a shell. Shell C burning is characterized by the development of one or more convective zones. Because the nuclear energy generation depends so strongly on temperature, this burning phase almost invariably drives convection. Since a higher nuclear energy release results in a more spatially extended convective zone, the THM models develop fewer, more massive, and longer-lasting C-convective shell episodes (\figurename~\ref{fig:kip1}-\ref{fig:kip2}). Analogous properties were already noted by \cite{bennett:12, pignatari:13, chieffi:21}. Up to 22 \msun, CF88 models develop two consecutive C convective shells that vanish prior to central Ne ignition, followed later by a third one that forms concurrently with the Ne and O convective shells. In contrast, THM models up to 20 \msun\ develop a single C convective shell that disappears before central Ne ignition. As the initial mass increases, a second convective shell forms progressively earlier: alongside with the O convective shell in the 15 \msun\ model, together with the O convective core in the 16 \msun\ model and with the Ne convective core in the 18 \msun\ model, and eventually before the central Ne ignition in the 20 \msun\ model. In these latter two cases, the formation of the C convective shell slows down the contraction of the core \citep{chieffi:21}, resulting in Ne ignition at progressively lower densities. This leads to a Ne burning phase more heavily dominated by neutrino losses, as shown in \figurename~\ref{fig:tcrhoc}, and consequently to smaller convective cores (\figurename~\ref{fig:ne}, lower panel). Between 22 to 25 \msun, CF88 models maintain convective central C burning, whereas the THM models switch to a radiative regime: this leads to different C shell and central Ne ignition conditions in CF88 and THM models. By increasing further the initial mass, C ignition occurs in a radiative environment in both cases. Moreover, central and shell burning occur at progressively higher temperature, where the differences between THM and CF88 rates decrease, as shown in the lower panel of \figurename~\ref{fig:reaction}. It must also be reminded that the amount of C left by the central He burning scales inversely with the He core mass (and hence, obviously, with the initial mass) and since the nuclear reaction rate of the \cc scales with the square of the \isotope[12]{C} abundance, above a certain mass the difference between the two rates is not so critical in determining the further evolution of a star. For both these reasons the differences between models computed with the THM and the CF88 nuclear cross sections progressively reduce for stellar models having an He core mass in excess of $\sim11$ \msun\ at the core collapse, i.e., initial mass larger than roughly 40 \msun (see column "$\rm M_{He}$" in \tablename~\ref{tab:eexp}).
    
    The subsequent evolution is dictated by the relative formation timescale of the O convective core and the Ne convective shell. Oxygen burning begins shortly after central Ne exhaustion, while Ne is still present in the region left by the recession of the convective core. This allows the development of a convective Ne shell quite close to the center and to the region of maximum energy generation from O burning. In the CF88 models up to 22 \msun, as well as the 15 and 16 \msun\ THM models, the O convective core develops and exhausts the fuel before the ignition of the Ne shell. Conversely, at higher masses (between 18 and 22 \msun\ in the THM models and in the 25 \msun\ CF88 model) the Ne shell develops at the same time as the O convective core. As a consequence also due to the large active convective C shell, in the 18 \msun\ THM case, the Ne shell burns before the O core, limiting its extension in mass. In the other cases, the O burning overtakes and ingests the Ne shell. The ingested Ne then burns within a hybrid convective core,  temporarily halting O burning. Only after this second central Ne exhaustion, the O burning in the core resumes and a convective core develops again (within the same mass region as the hybrid core). At even higher masses (40 \msun\ model), the Ne shell develops and exhausts before the formation of the O convective core in both the THM and CF88 cases.

    A final key event that further amplifies the differences between the THM and CF88 models is the formation of an efficient O shell following central Si burning, immediately prior to core collapse. In the CF88 models up to 18 \msun, the O shell advances into the O- and Ne-rich zone, ingesting material enriched with the ashes of the Ne and C burning shells, whereas it does not in the higher mass models. The THM models exhibit a different behaviour. In the 15 and 18 \msun\ cases, the O shell does not ingest material from its surroundings. In the 16 \msun, however, the O shell advances in mass until it reaches the C shell, triggering a C--O shell merger. This event is likely favoured by the higher fuel depletion in the C shell (X(\isotope[12]{C})$\simeq1.3\times10^{-2}$ in the THM model, X(\isotope[12]{C})$\simeq4.4\times10^{-2}$ in the CF88 model), which lowers the entropy barrier at the O-C interface \citep{roberti:24}. In the 20 to 25 \msun\ models, the C shell ignites significantly earlier and is particularly efficient, limiting the growth in mass of the O shell and leading to the (partial) ingestion of the ONe rich zone. Finally, the 40 \msun\ THM model follows the CF88 trend, showing no ingestion or merger phenomena.

    All these differences arise from the distinct burning timescales and shell formation sequences triggered by the THM and CF88 C burning rates, which drive the stars along different evolutionary paths, ultimately resulting in significantly different chemical and thermodynamic profiles at core collapse. A key parameter summarizing these structural differences is the compactness $\xi_{2.5}$ \citep{oconnor:11}, which reflects how deep the gravitational potential well is or, equivalently, how large is the average density of the inner 2.5 \msun, and correlates with the Chandrasekhar mass and the Fe core mass \citep{timmes:96,boccioli:24,laplace:25}. Although compactness is often used to determine whether a massive star explodes as a supernova or collapses directly into a black hole \citep[e.g.,][]{oconnor:11}, recent work showed that this simplified picture is not replicated in more sophisticated simulations. More recent explodability criteria \citep{boccioli:23} showed that the density contrast at the Si/Si-O interface is a better predictor of the explodability. Moreover, high-compactness progenitors lead to strong and successful explosions \citep{boccioli:25b}, which is the exact opposite of what was previously assumed, although in the most extreme cases of stars with the highest compactness, it has been shown that successful CCSN explosions can eject a significant amount of matter while still forming BHs above 20 \msun\ \citep{chan:18,burrows:23,sykes:25,andersen:25}. Nonetheless, the compactness $\xi_{2.5}$ is still an informative quantity that summarizes the main properties of the pre-supernova progenitor.
    
    \figurename~\ref{fig:xi} shows the compactness of the THM and CF88 models at the pre-supernova stage. It is quite evident that the main, global effect of a more efficient \cc nuclear reaction rate is that of slowing down the contraction of the star so that the THM models in most cases reach the core collapse with a structure less compact than their CF88 counterparts. The most significant discrepancies appear between models that develop an efficient C shell before central Ne ignition and those that do not, specifically 20 and 22 \msun\ models. Moreover, in the latter two cases, the ONe (or, alternatively, the C-free) core mass at the core collapse is considerably smaller than in their CF88 counterparts, while having roughly the same CO core mass (see "$\rm M_{ONe}$" and "$\rm M_{CO}$" columns in \tablename~\ref{tab:eexp}), indicating a more extended C convective shell. Such differences in the compactness and, in general, in the pre-supernova structure can significantly affect the strength of the supernova explosion and the propagation of the shock wave, as discussed in Sect. \ref{sec:exp}.

    \subsection{Hydrostatic nucleosynthesis}

        The nucleosynthetic differences at the end of the hydrostatic evolution primarily reflect the differences in the structure as described above and, in particular, in the widths of the Si, O, Ne, and C burning shells. Contrary to the findings of \cite{pignatari:13}, we do not observe any nucleosynthetic products from central C burning being subsequently transported outward into the CO core during the shell-burning phases. 
        This discrepancy with the results of \cite{pignatari:13} arises from their parametric inclusion of a strong \cc cluster resonance at 1.5 MeV, which boosted the CF88 rate by up to four orders of magnitude at $T_9 \sim 0.8$. Interestingly, while the THM data also reveal a resonance structure in the same energy range ($E_{cm} \le 2$~MeV), its measured strength leads to a much more moderate increase in the reaction rate compared to the extreme scenario assumed in \cite{pignatari:13}. Consequently, in our models, all C shells develop and evolve well outside the exhausted carbon core, and the material within the newly formed ONe core is entirely processed by the subsequent central and shell Si, O, and Ne burning stages.

        \begin{figure*}[!t]
            \centering
            \includegraphics[width=.49\linewidth]{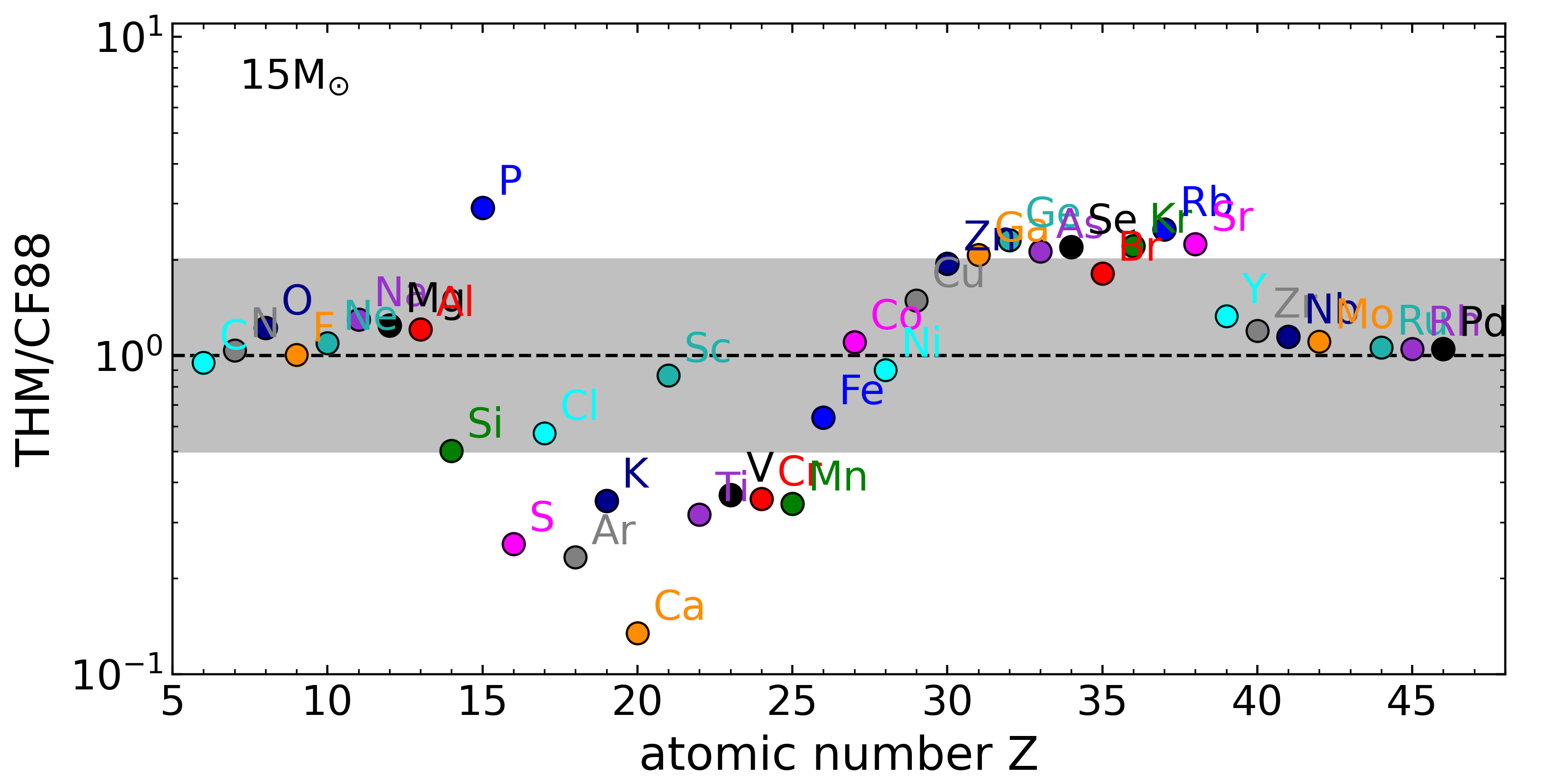}
            \includegraphics[width=.49\linewidth]{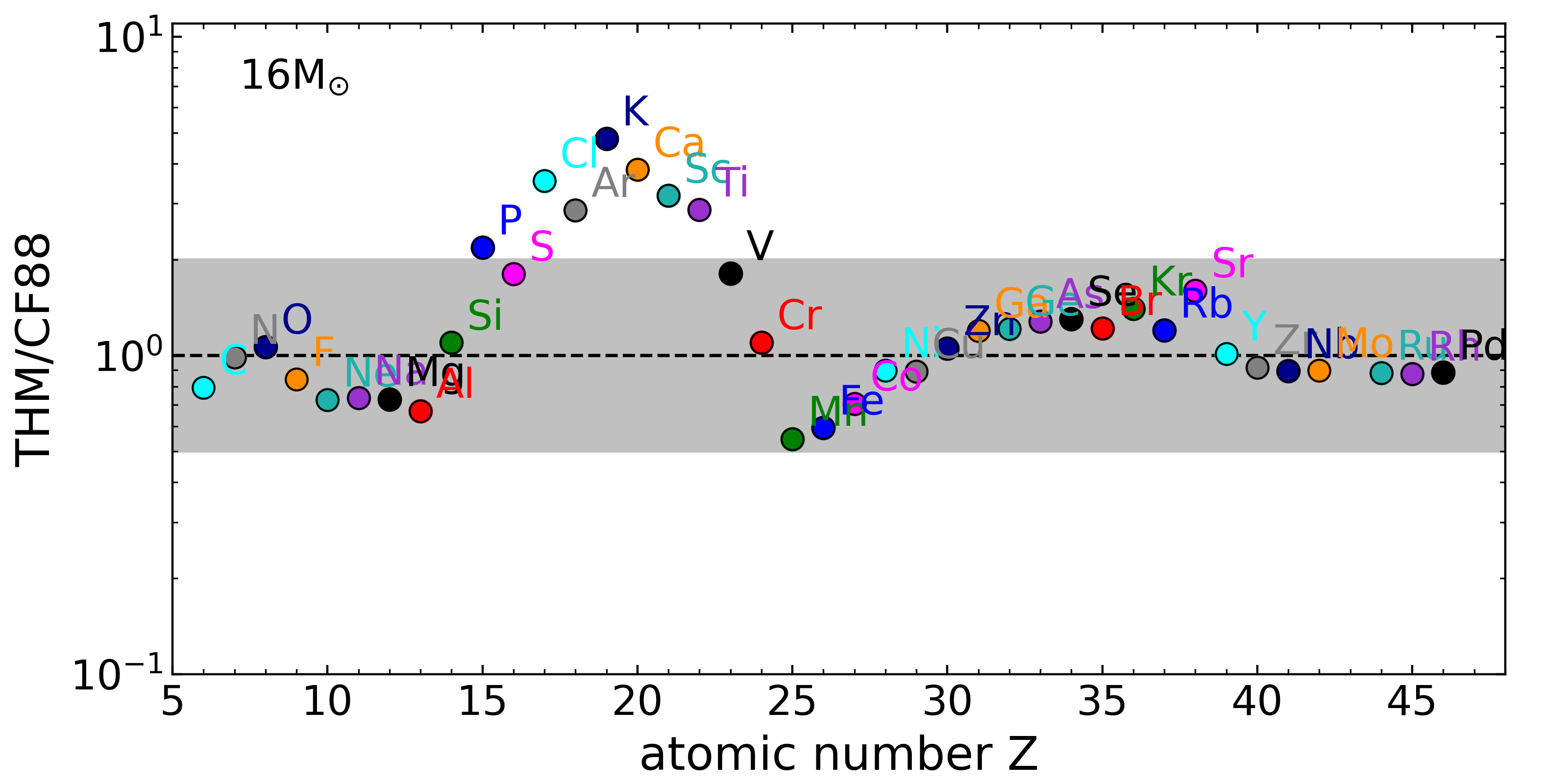}
            \includegraphics[width=.49\linewidth]{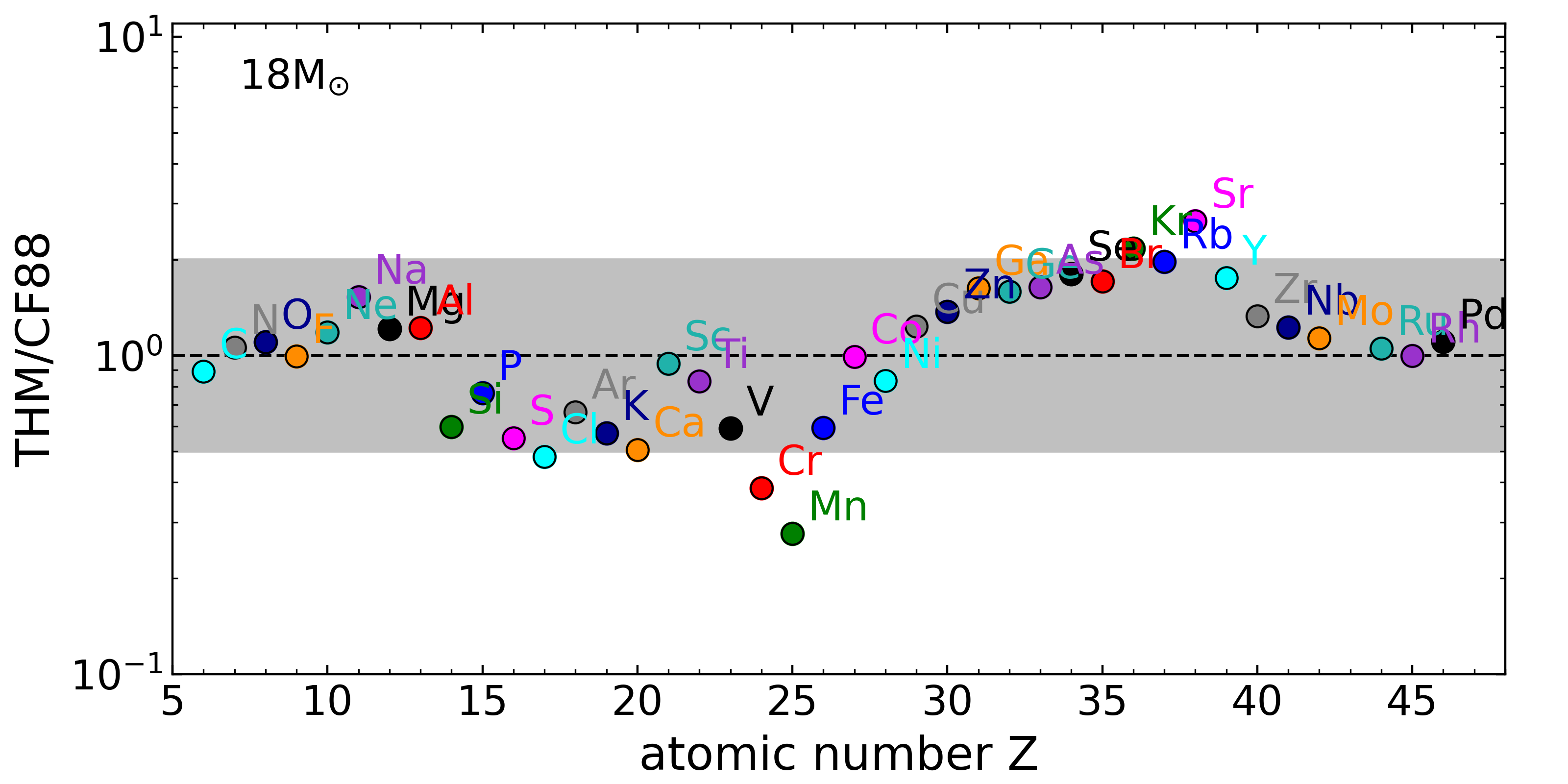}
            \includegraphics[width=.49\linewidth]{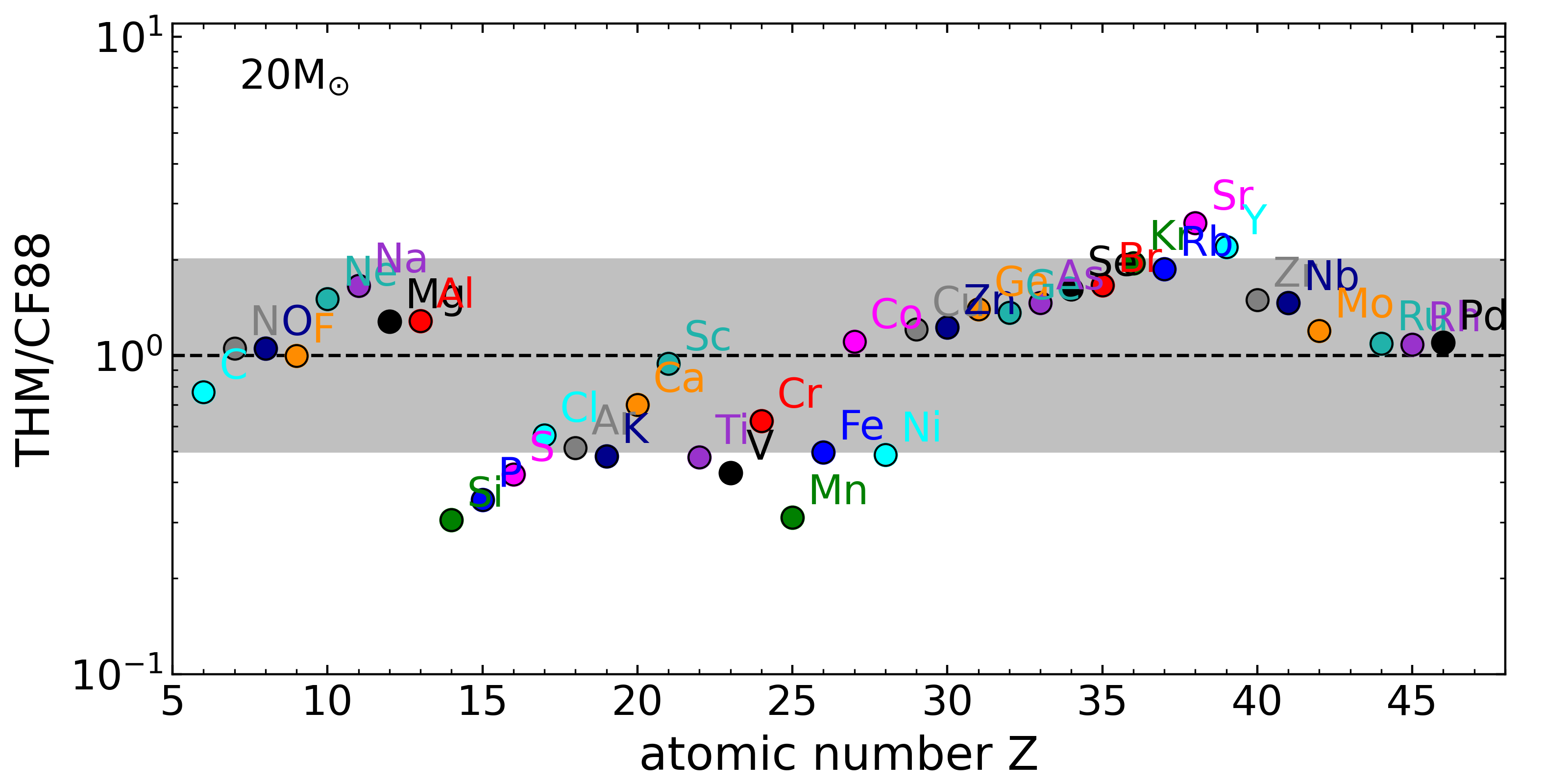}
            \includegraphics[width=.49\linewidth]{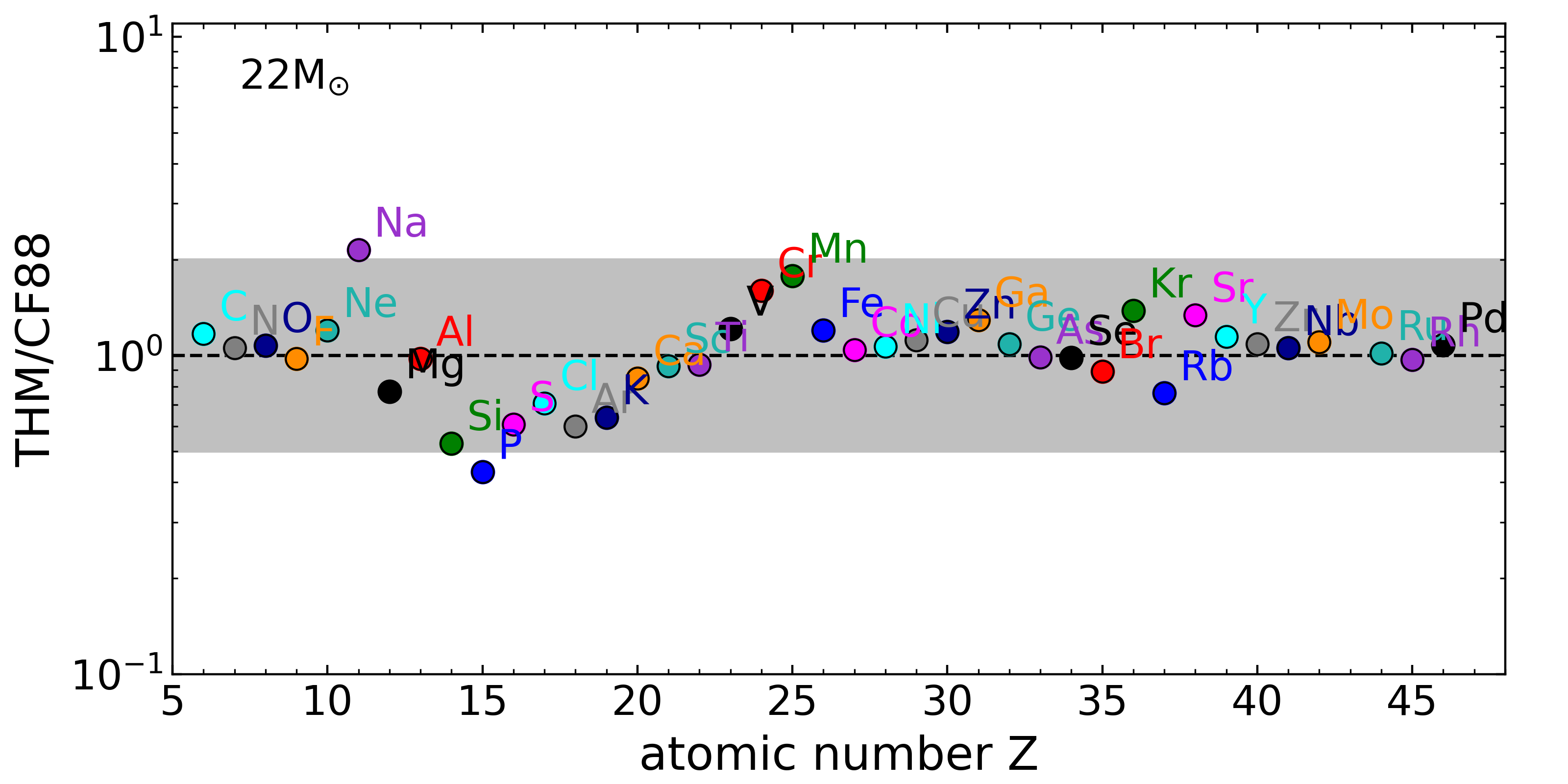}
            \includegraphics[width=.49\linewidth]{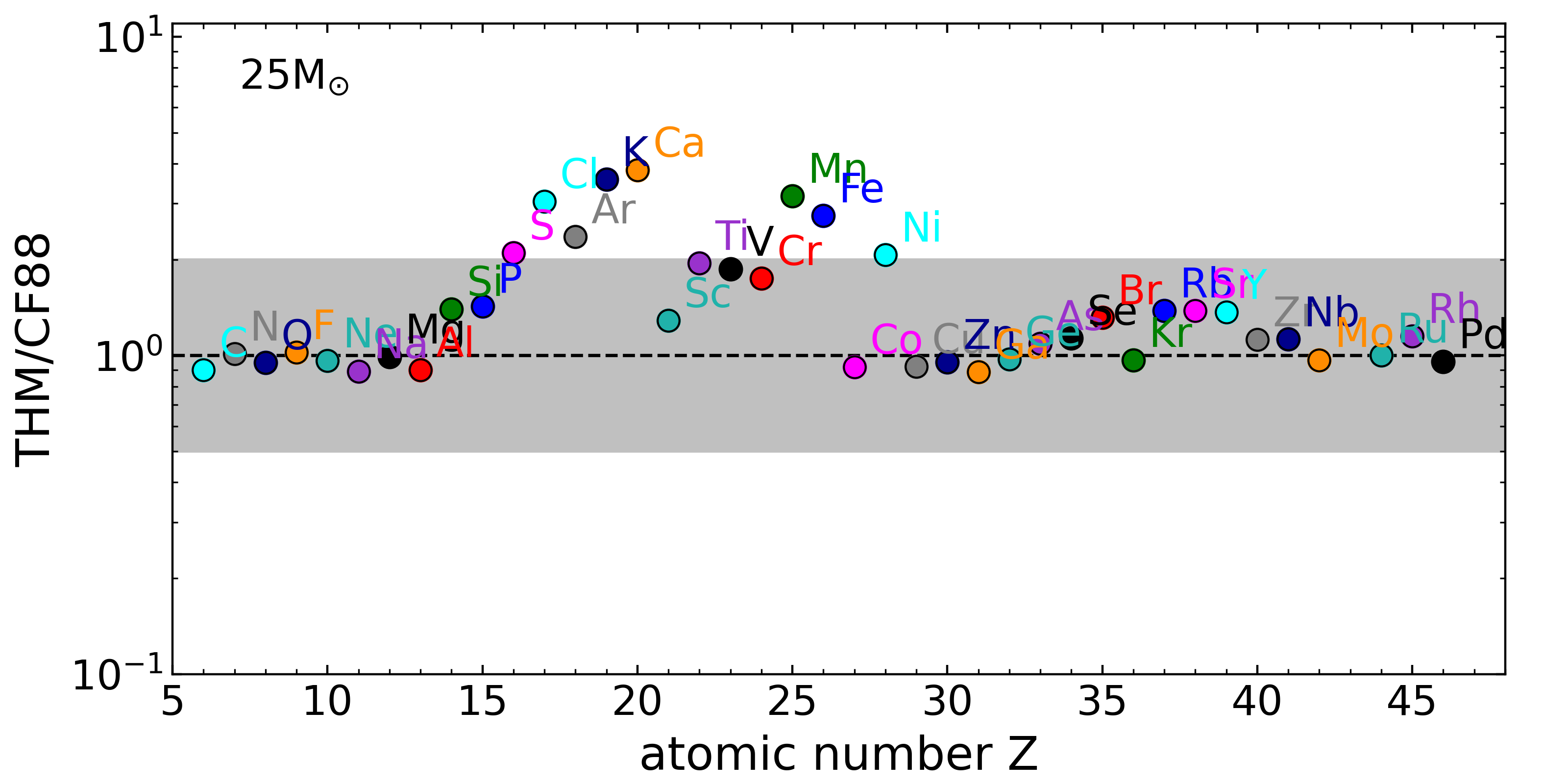}
            \includegraphics[width=.49\linewidth]{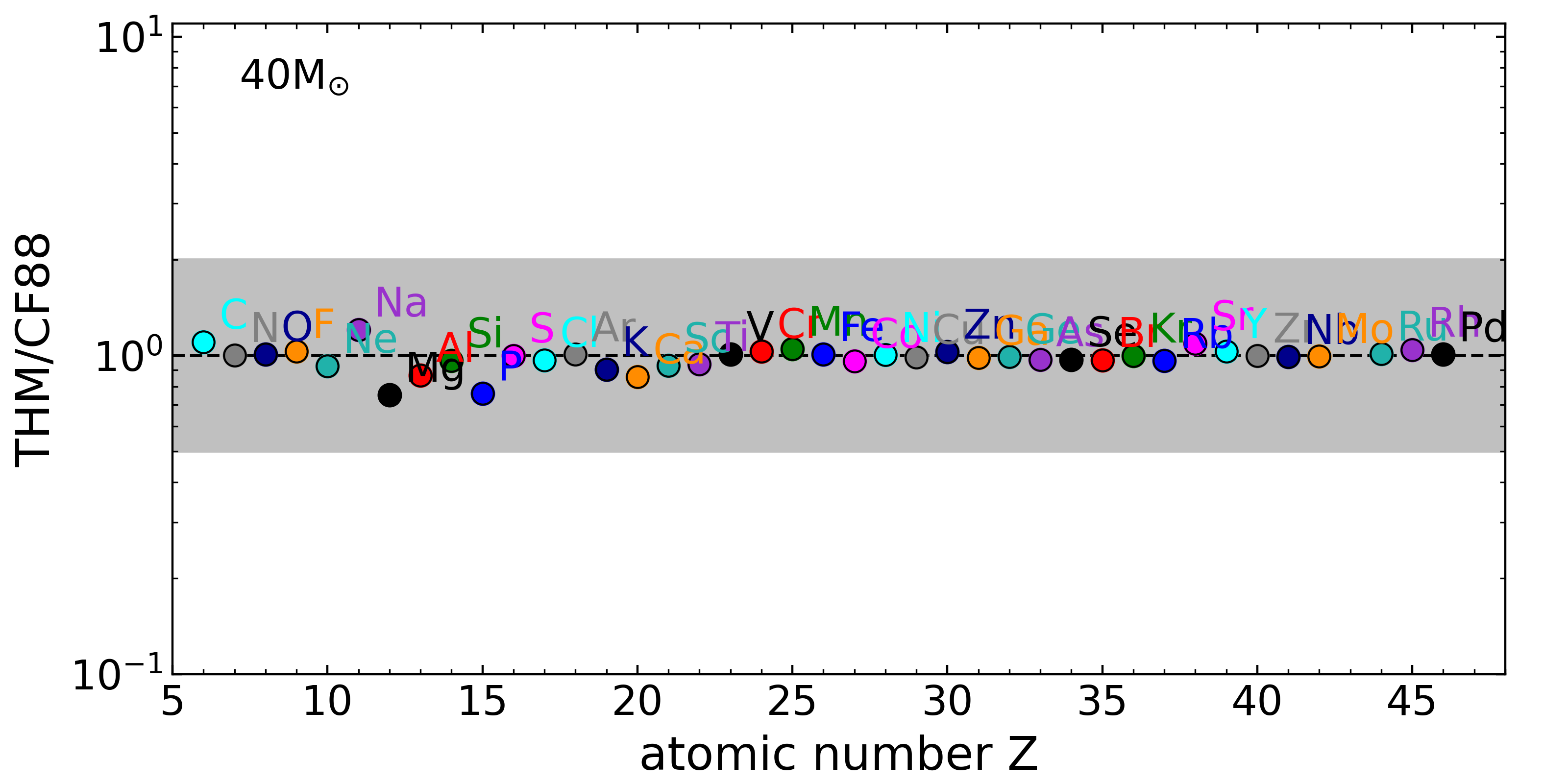}
            \caption{Ratio between THM and CF88 yields for each model at the pre-supernova stage (see text). The gray band identifies a factor of 2 variation.}
            \label{fig:presn}
        \end{figure*}

        To evaluate the different chemical composition at the pre-supernova stage, we calculated the integrated yields from the bottom of the O shell up to the surface of the star, including the contribution from the wind. We neglect the contribution from the Si rich zone, as it will be fully reprocessed by the explosion. \figurename~\ref{fig:presn} shows the elemental THM/CF88 ratios for each model. In the THM models, at the pre-supernova stage the C shells are generally more extended in mass compared to their CF88 counterparts, while it is generally the opposite for the O and Ne shells (see Sect. \ref{sec:strut}). This leads, on average, to an overproduction of elements between Ne and Al and an underproduction of elements between Si and Ca in THM with respect to CF88 models. There are two exceptions: the 16 \msun\ and the 25 \msun\ THM models. Specifically, the C--O merger that occurs in the 16 \msun\ model drastically increases the O-burning products and the odd-Z elements, and the larger O shell in the 25 \msun\ model produces a similar, though smaller, effect. In the 40 \msun\ models, the structures are very similar; therefore, the differences are quite negligible. All these features are largely preserved during the explosion. 

        \begin{figure*}[!t]
            \centering
            \includegraphics[width=0.32\linewidth]{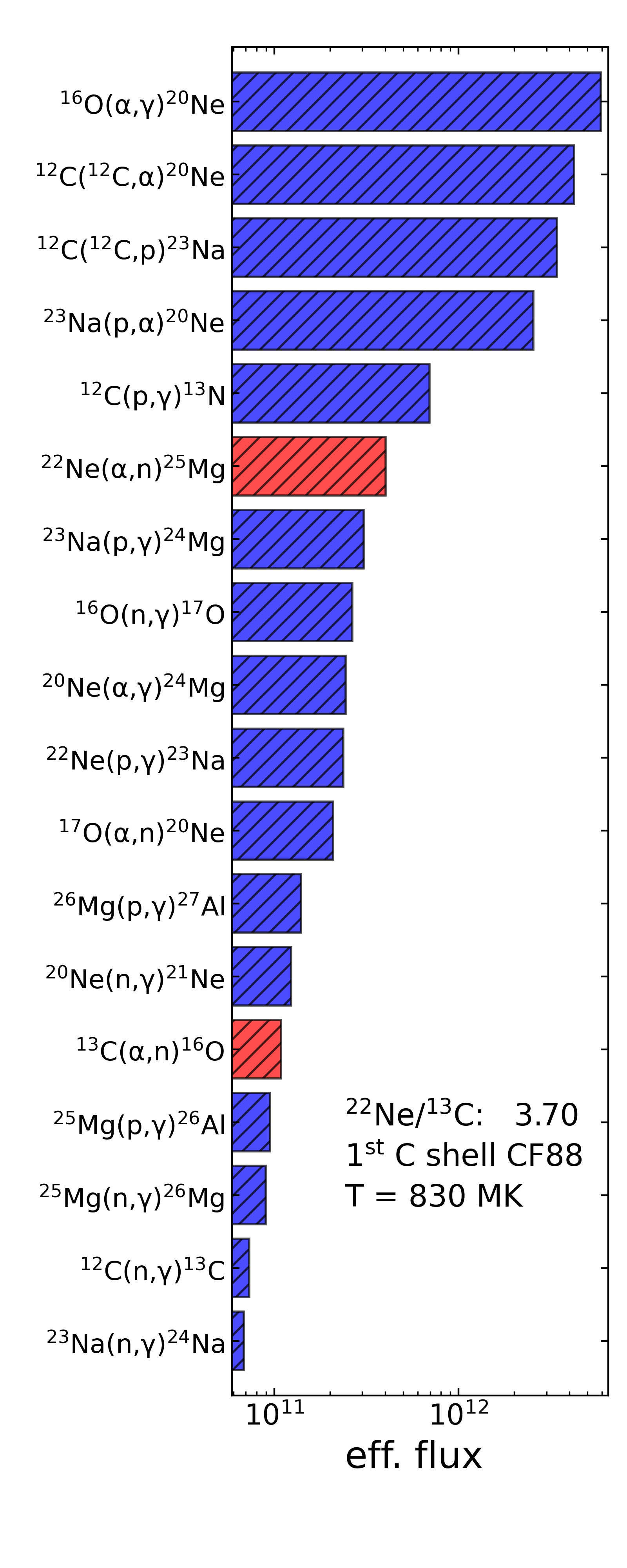}
            \includegraphics[width=0.32\linewidth]{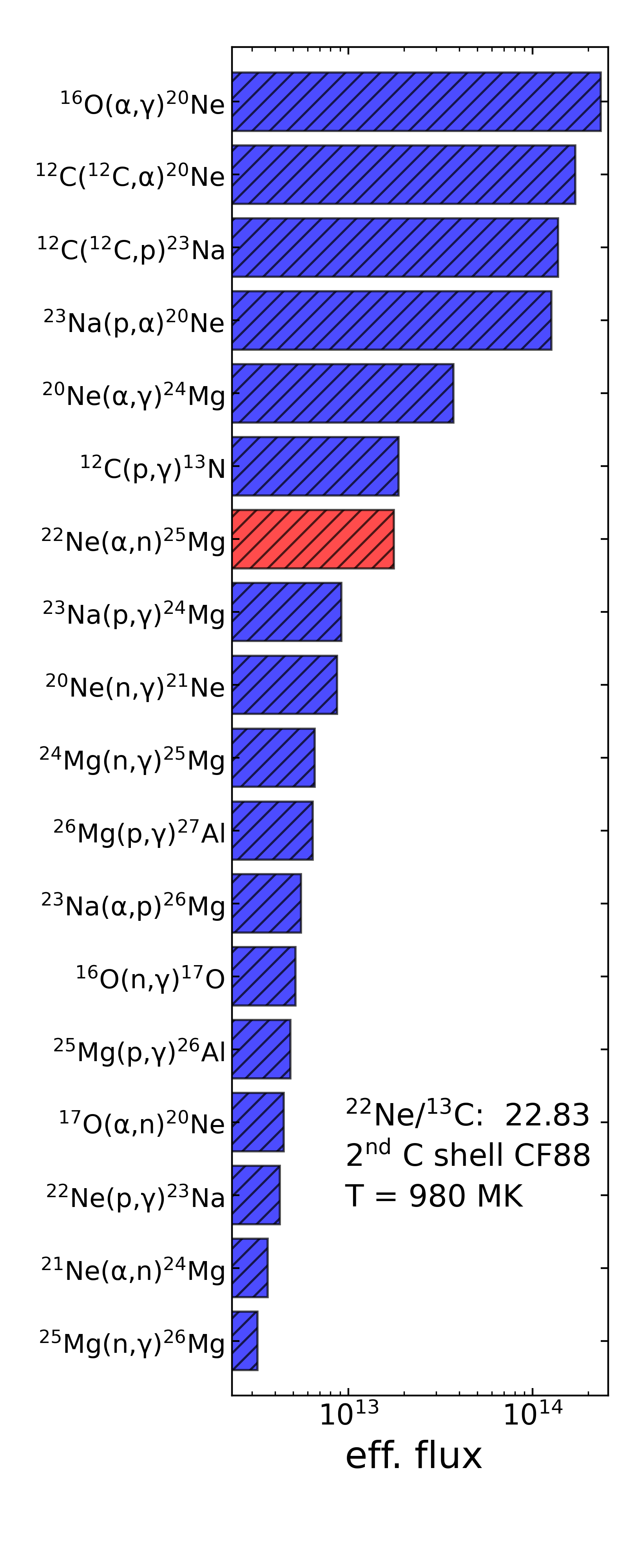}
            \includegraphics[width=0.32\linewidth]{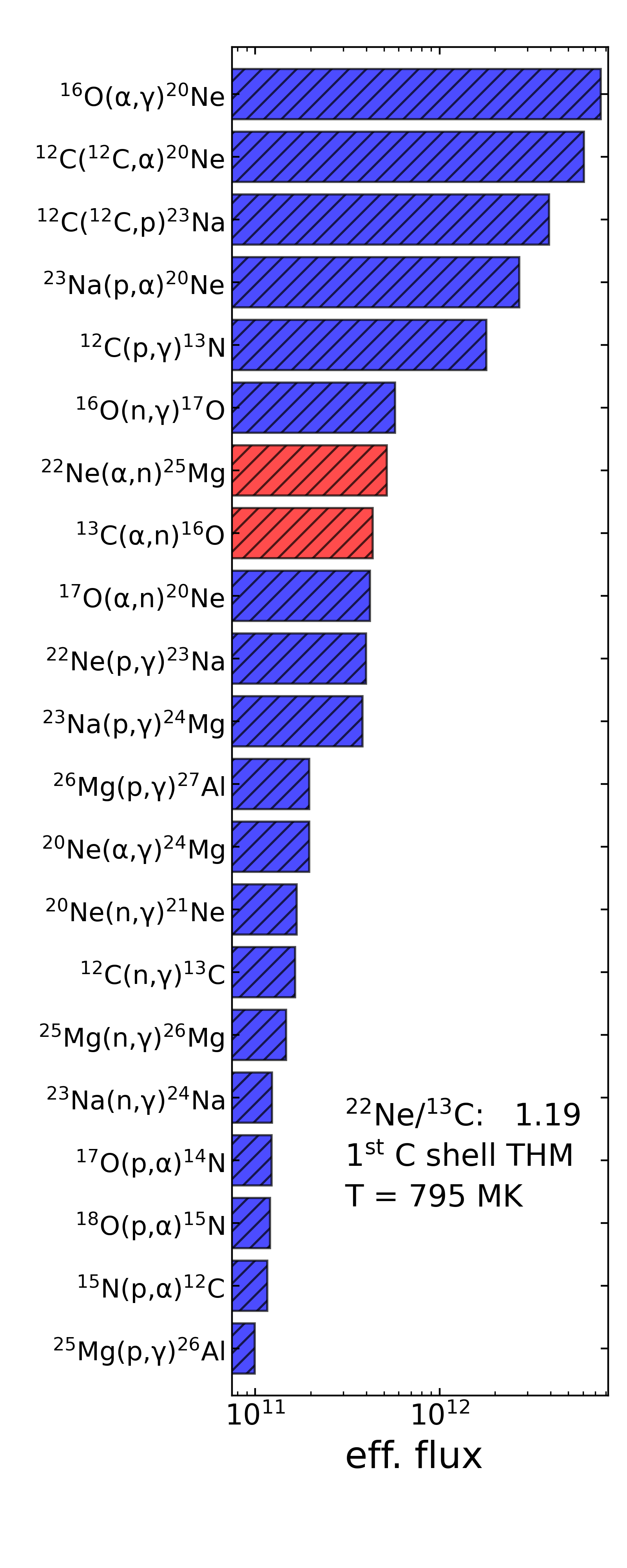}
            \caption{Effective fluxes (see text) at the bottom of the first and second C shell in the CF88 case (left and central panels) and in the THM case (right panel) in the 15 \msun\ models. Red bars identify the \nen\ and \cn\ neutron sources.}
            \label{fig:flux}
        \end{figure*}

        \subsubsection{The s-process nucleosynthesis} \label{subsec:sproc}
        
            An important feature regards the elements heavier than Fe produced via the $slow$ neutron captures. While the main component of the \s process is synthesized in low-mass AGB stars through the \cn\ reaction \citep[e.g.,][]{pal21}, the weak component, spanning from Ni to Sr, is mostly produced during the central He-burning phase of massive stars via the \nen\ reaction \citep[e.g.,][]{pignatari:10}. The subsequent C burning shell, which develops within the CO core containing these weak \s process products, can further enhance this nucleosynthesis where the $\alpha$ particles released by the \cca\ channel interact with the \isotope[22]{Ne} left over from He-burning. \cite{pignatari:13} demonstrated that a more efficient C burning reaction rate may significantly favour the \s process in C burning region, by the activation of the \cn\ neutron source. In fact, the earlier activation of the C burning reduces the efficiency of the \isotope[13]{N}$(\gamma,p)$\isotope[12]{C} reaction that competes with the \isotope[13]{N}$(\beta^{+})$\isotope[13]{C} decay for the destruction of \isotope[13]{N}, resulting in a larger production of \isotope[13]{C}. Let us show as an example the 15 \msun\ case. In the CF88 case, the \s process elements are mostly produced during the first and second convective shells, igniting at 0.65 \msun, $\rm T=830\ MK$, $\rm \rho=1.53\times10^{5}\ g\ cm^{-3}$ and at 1.20 \msun, $\rm T=980\ MK$, $\rm \rho=1.66\times10^{5}\ g\ cm^{-3}$, respectively. In the THM case, instead, they are mostly produced at lower temperature and density by the first C shell, igniting at 1.10 \msun, $\rm T=795\ MK$, $\rm \rho=0.95\times10^{5}\ g\ cm^{-3}$. We extracted the nuclear fluxes\footnote{$\rm f_{ab} = Y_aY_b \rho\ N_A\left \langle \sigma v\right \rangle_{ab}$ (and $\rm f_{a\gamma} = Y_a N_A\lambda_{a}$ in the case of a photodisintegration or a decay), with $N_A\left \langle \sigma v\right \rangle_{ab}$ the reaction rate of the reaction a+b, $\rm \lambda_{a}=ln\ 2/\tau^a_{1/2}$ is related to the inverse of the half-life $\tau^a_{1/2}$ in case of a decay or a photodisintegration for the nucleus $a$, and $\rm Y_i$ is the abundance by number, with $\rm Y_i=X_i/A_i$ where $\rm X_i$ and $\rm A_i$ are the mass fraction and the atomic weight of the species $i$, respectively.} for each reaction at the bottom of each shell in the moment of maximum production of the \s process elements. The effective fluxes, namely the difference between the fluxes of the forward and reverse reactions, are shown in \figurename~\ref{fig:flux} and the neutron sources are highlighted in red. Note that in this case the contribution of the \isotope[13]{N}$(\gamma,p)$\isotope[12]{C} reaction is contained in the effective flux of the forward reaction \isotope[12]{C}$(p,\gamma)$\isotope[13]{N}, which is much higher in the THM case.  \figurename~\ref{fig:flux} shows that in the two shells of the CF88 model, the \nen\ neutron source is 3 and 22 times more efficient than the \cn\ neutron source, while in the shell of the THM model the two neutron sources equally contribute to the neutron production. Note also that the \isotope[17]{O}$(\alpha,n)$\isotope[20]{Ne}\footnote{In this work, the adopted \isotope[17]{O}$(\alpha,\gamma)$\isotope[21]{Ne} and \isotope[17]{O}$(\alpha,n)$\isotope[20]{Ne} reaction rates are from \cite{Best11,Best13}.} is also more efficient in the THM model and may play a role in the neutron flux: however, it acts more as a neutron modulator (being \isotope[17]{O} mostly produced by the neutron capture on the neutron poison \isotope[16]{O}, which is also more efficient in the THM model) rather than a pure neutron source.
            This effect results in an overproduction with respect to CF88 models of the \s elements up to Sr in models with $M\leq 20$ \msun. For higher masses, the C shell ignition occurs at higher temperature, therefore the \isotope[13]{N}$(\gamma,p)$\isotope[12]{C} reaction starts to be efficient despite the higher C burning reaction rate. It is worth mentioning that the increase of the yields of nuclei with $\rm A \lesssim 88$ from massive stars could lead to a better agreement with those from low mass stars when combined in studies of galactic chemical evolution. Indeed, the most recent models of AGB nucleosynthesis show enhanced yields of nuclei belonging to the main component of the \s process but, at the same time, an important reduction of the tail contribution to the abundances of elements lighter than Sr \citep[see e.g.][]{bus22}. 


\section{The explosion and explosive yields} \label{sec:exp}
        
    \begin{figure*}[!t]
        \centering
        \includegraphics[width=.49\linewidth]{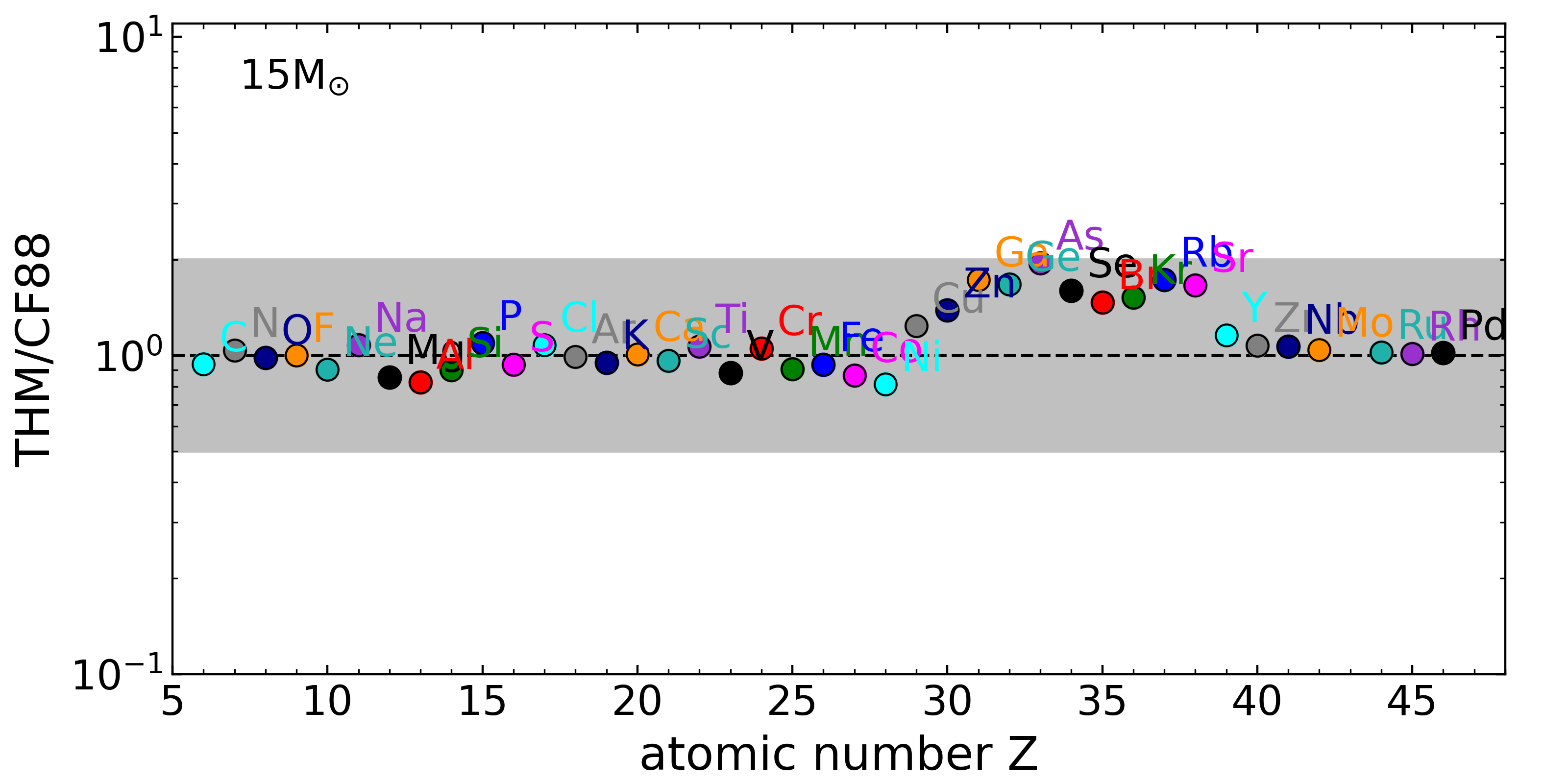}
        \includegraphics[width=.49\linewidth]{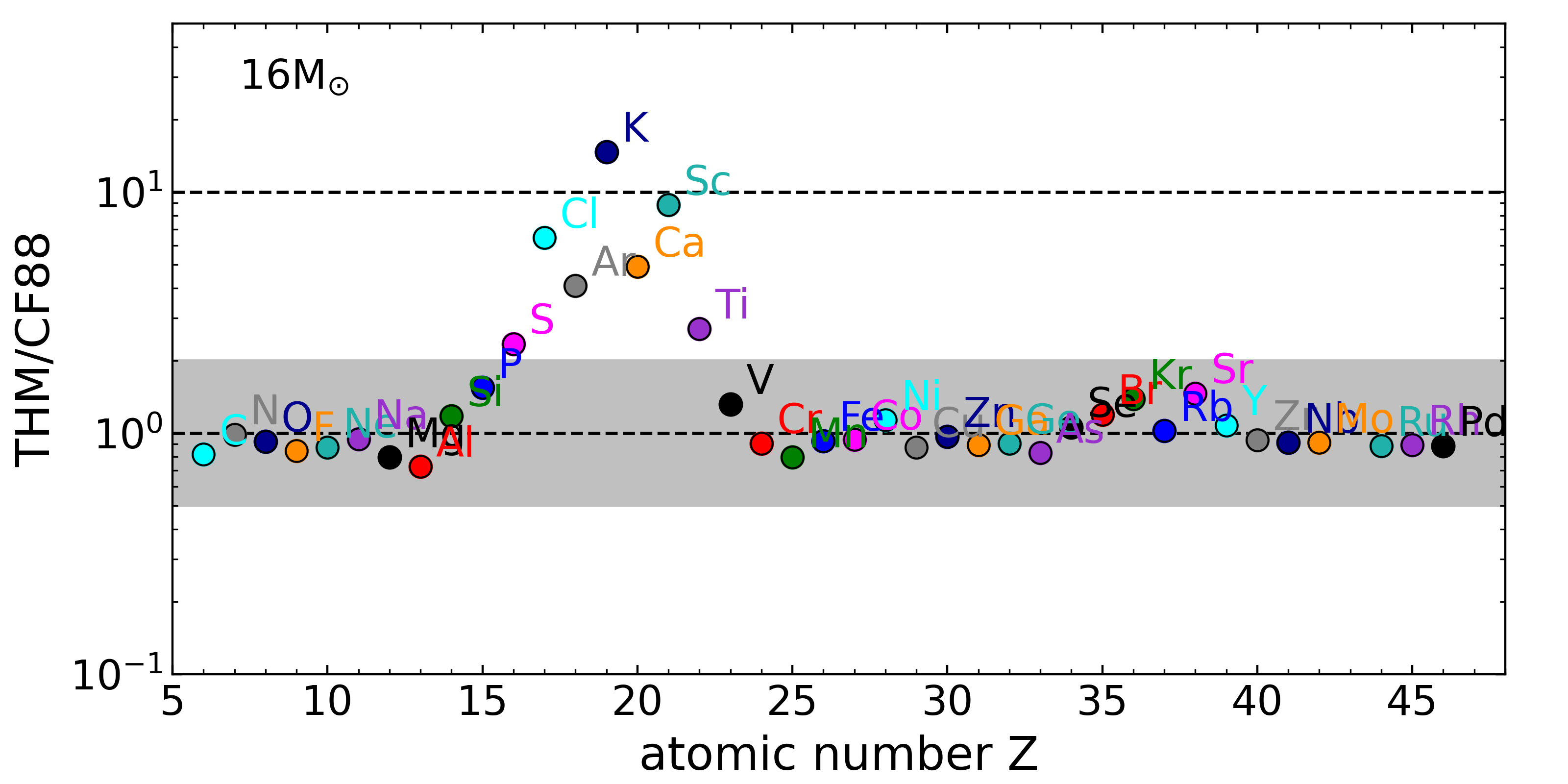}
        \includegraphics[width=.49\linewidth]{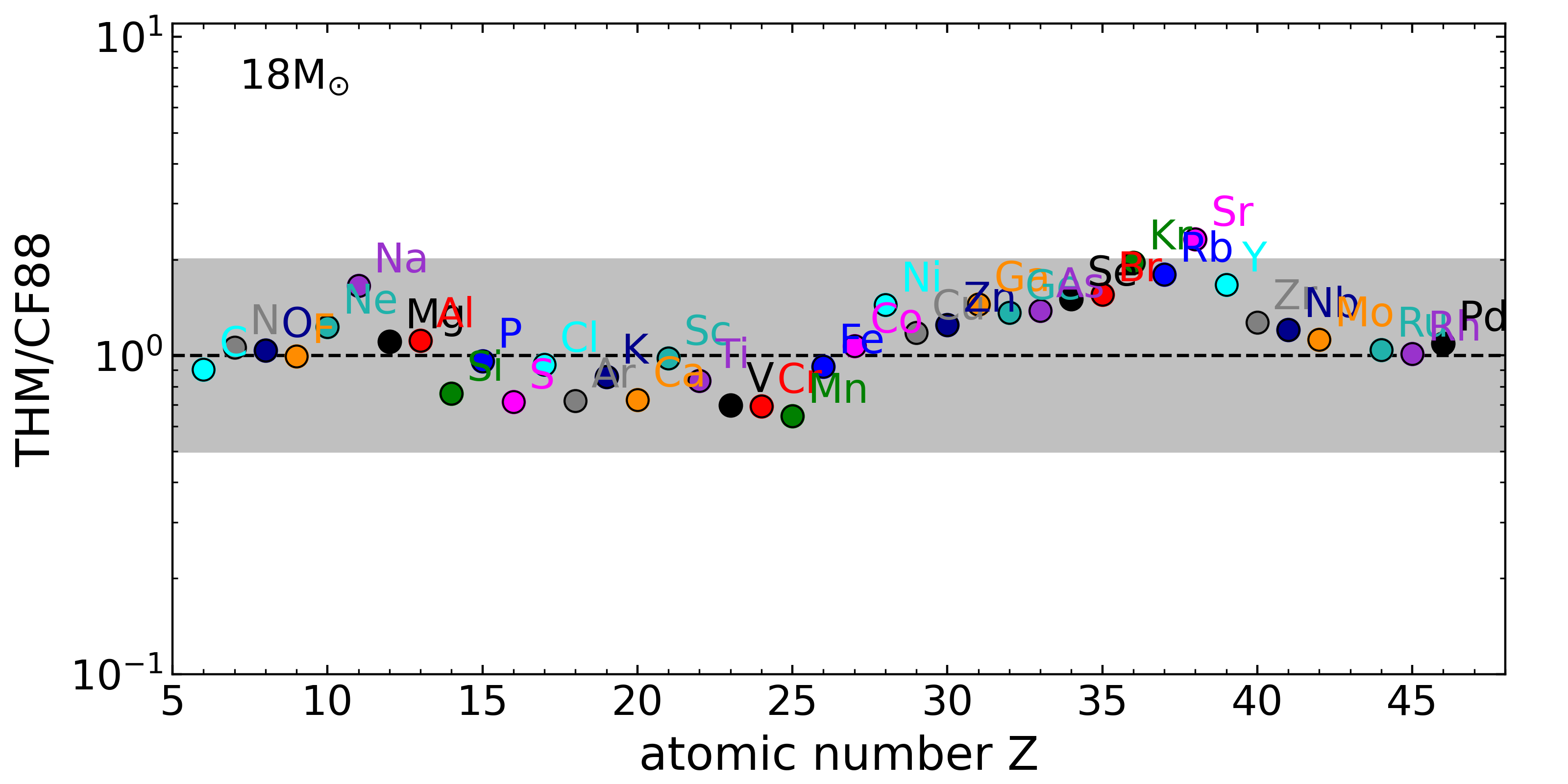}
        \includegraphics[width=.49\linewidth]{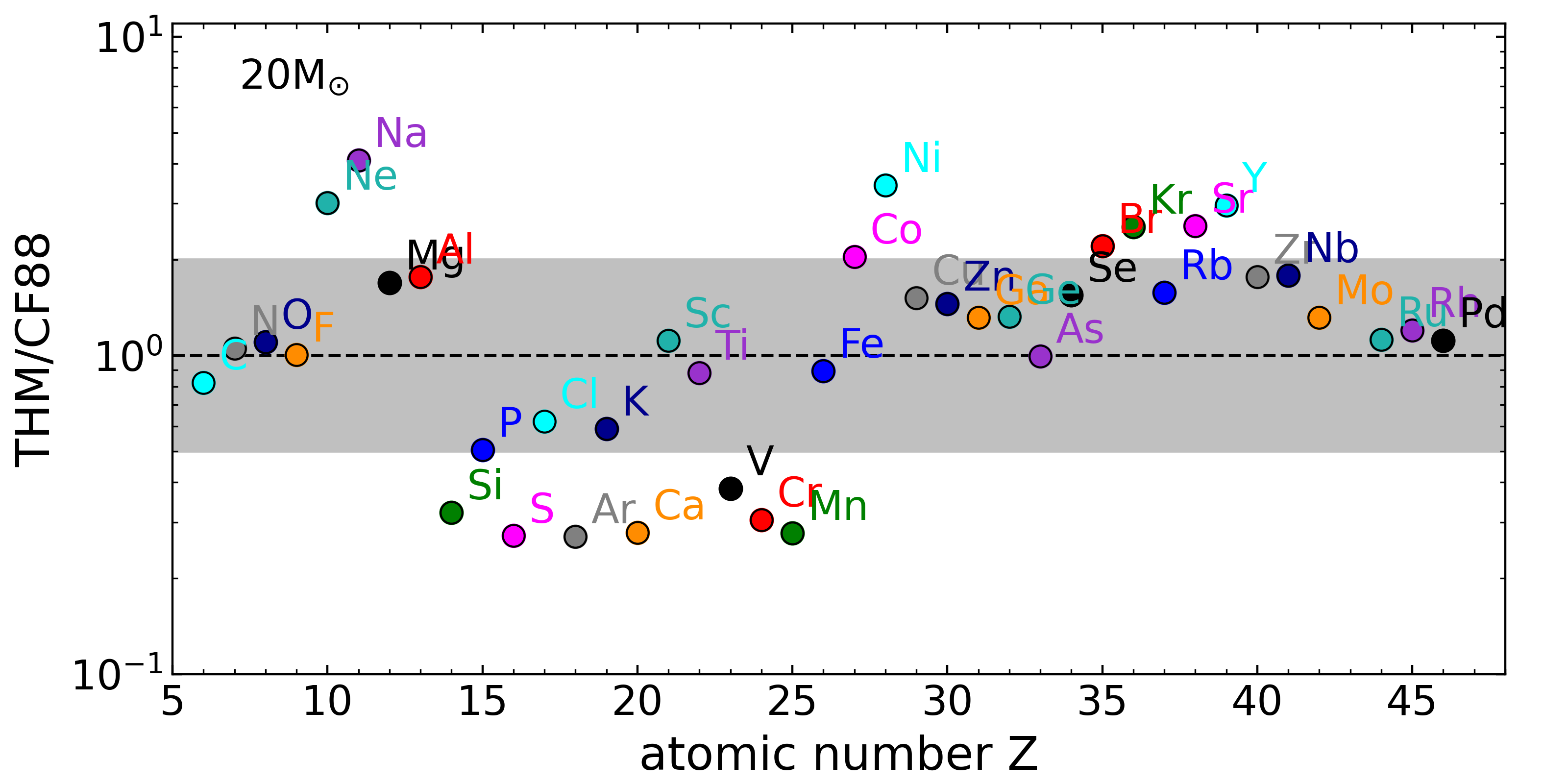}
        \includegraphics[width=.49\linewidth]{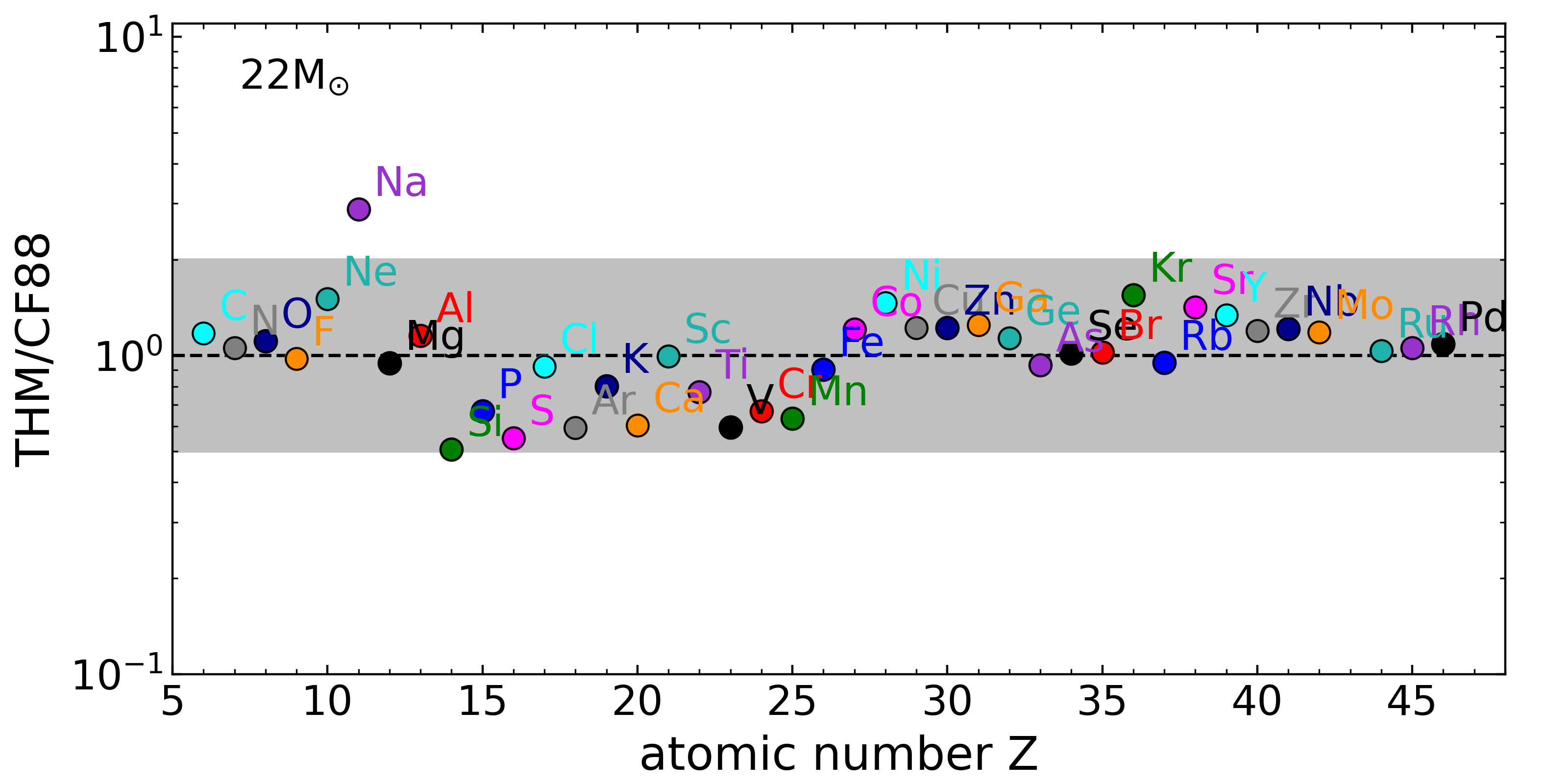}
        \includegraphics[width=.49\linewidth]{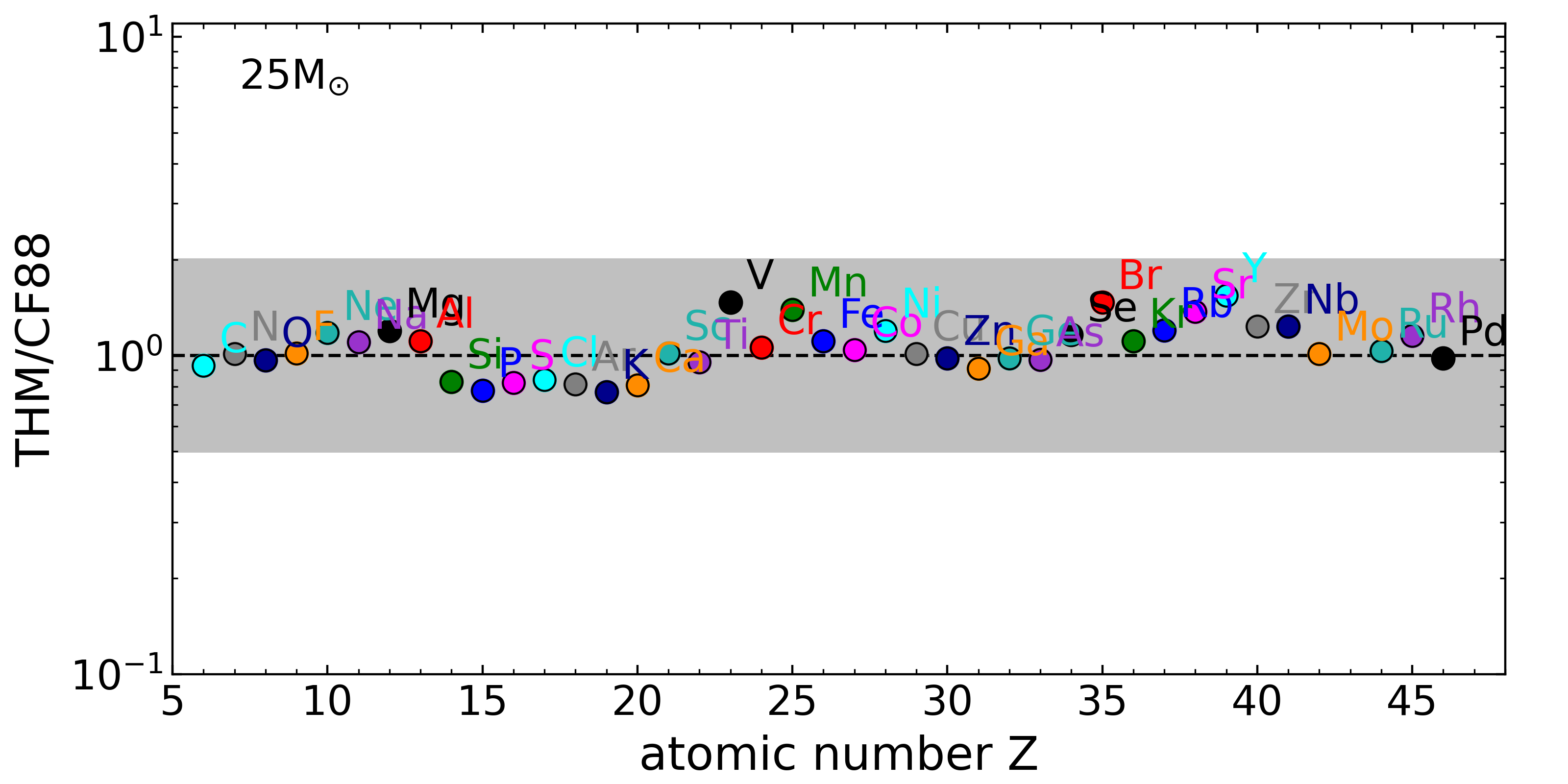}
        \includegraphics[width=.49\linewidth]{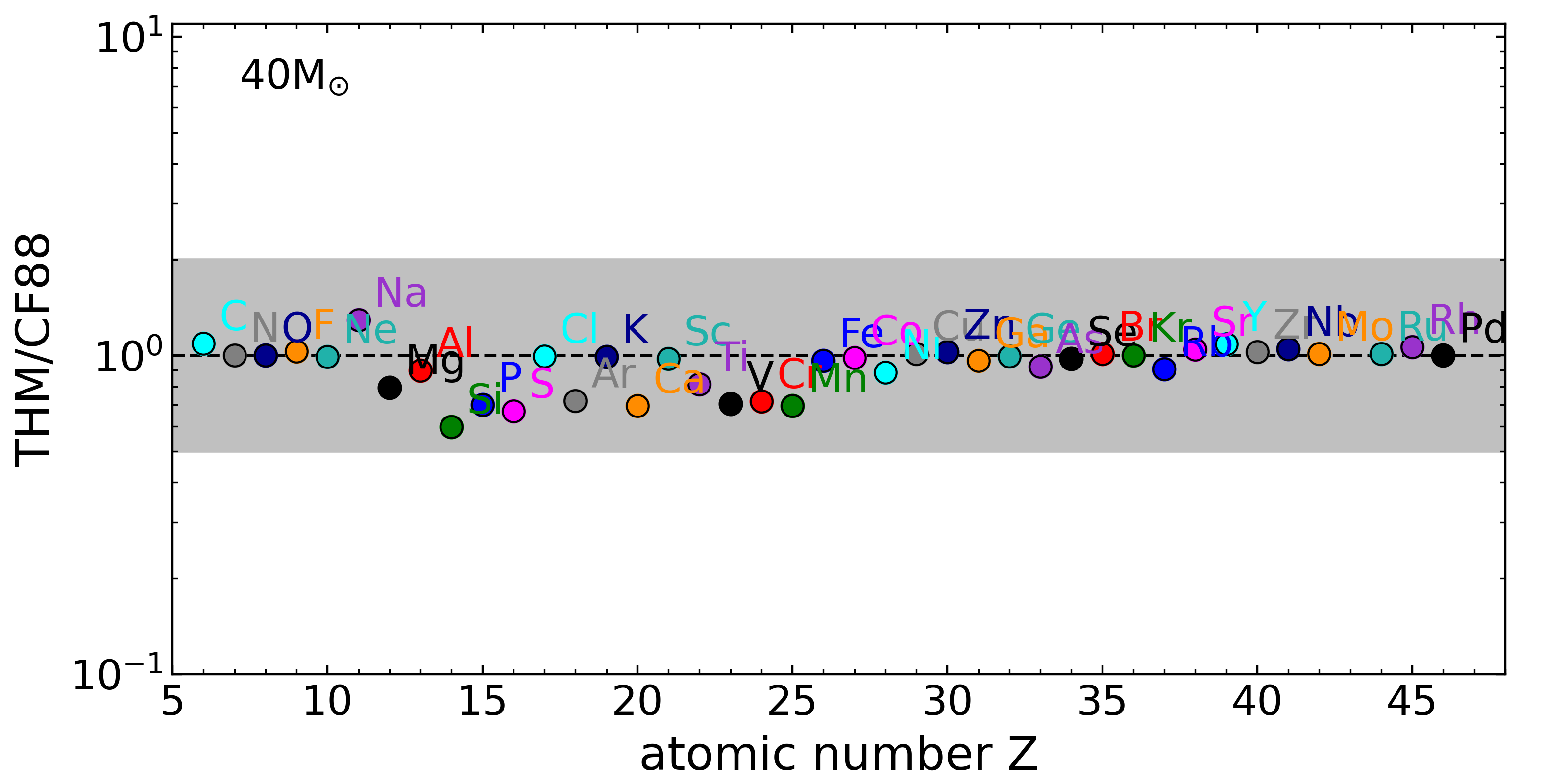}
        \includegraphics[width=.49\linewidth]{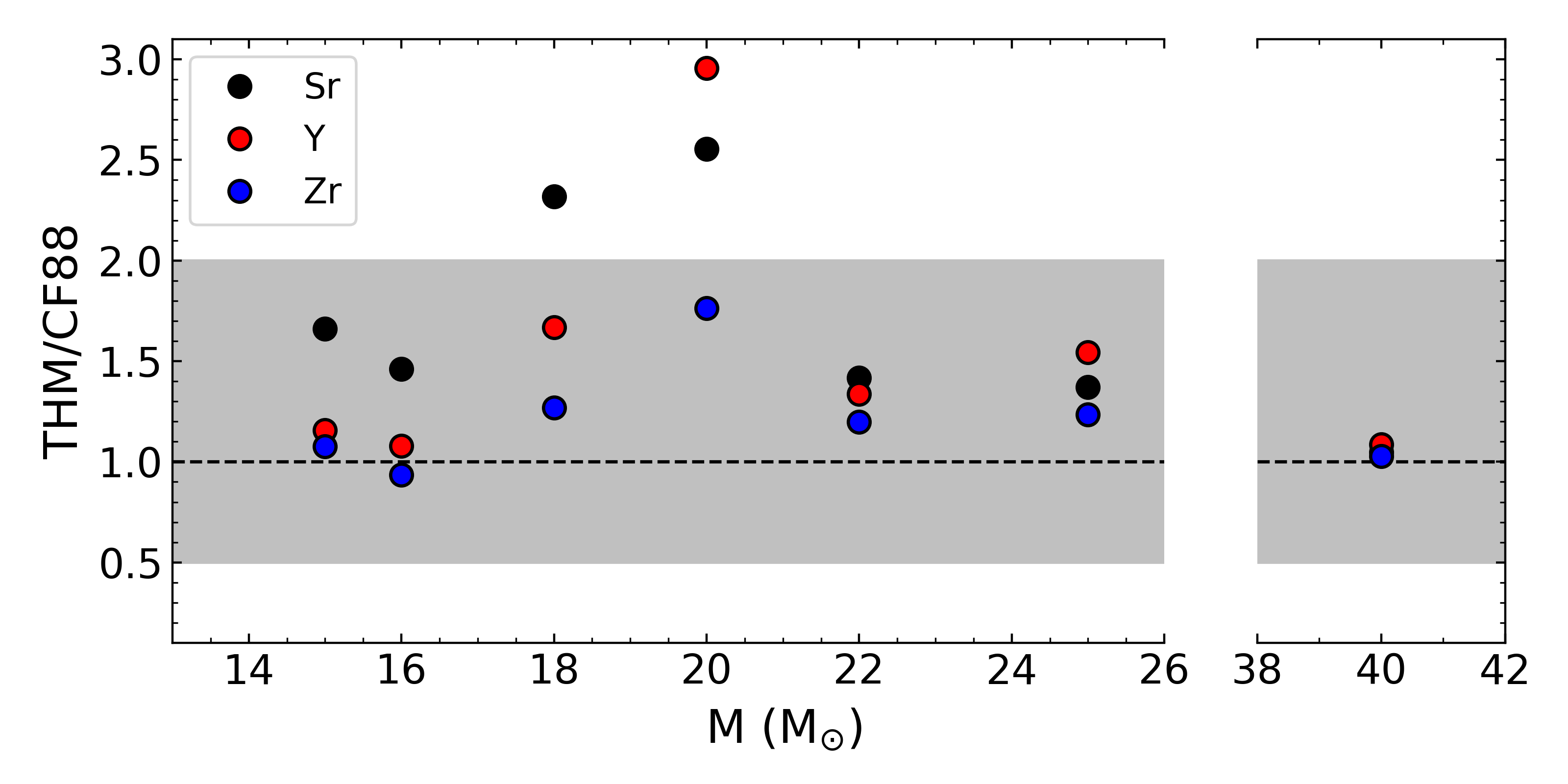}
        \caption{Ratio between THM and CF88 yields for each model, in the case of Set A (see text). The gray band identifies a factor of 2 variation.}
        \label{fig:yields}
    \end{figure*}

    \begin{figure*}[!t]
        \centering
        \includegraphics[width=.49\linewidth]{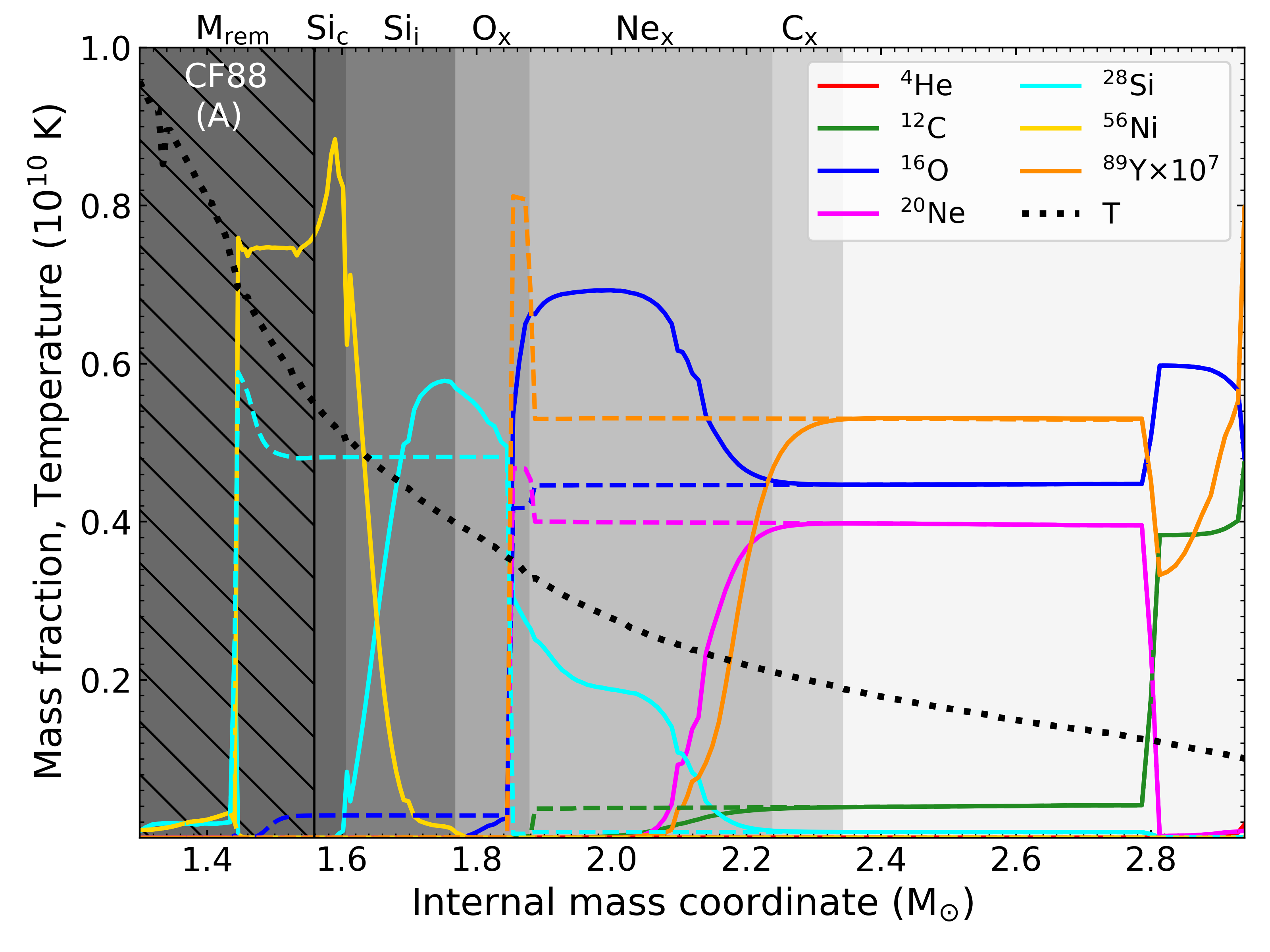}
        \includegraphics[width=.49\linewidth]{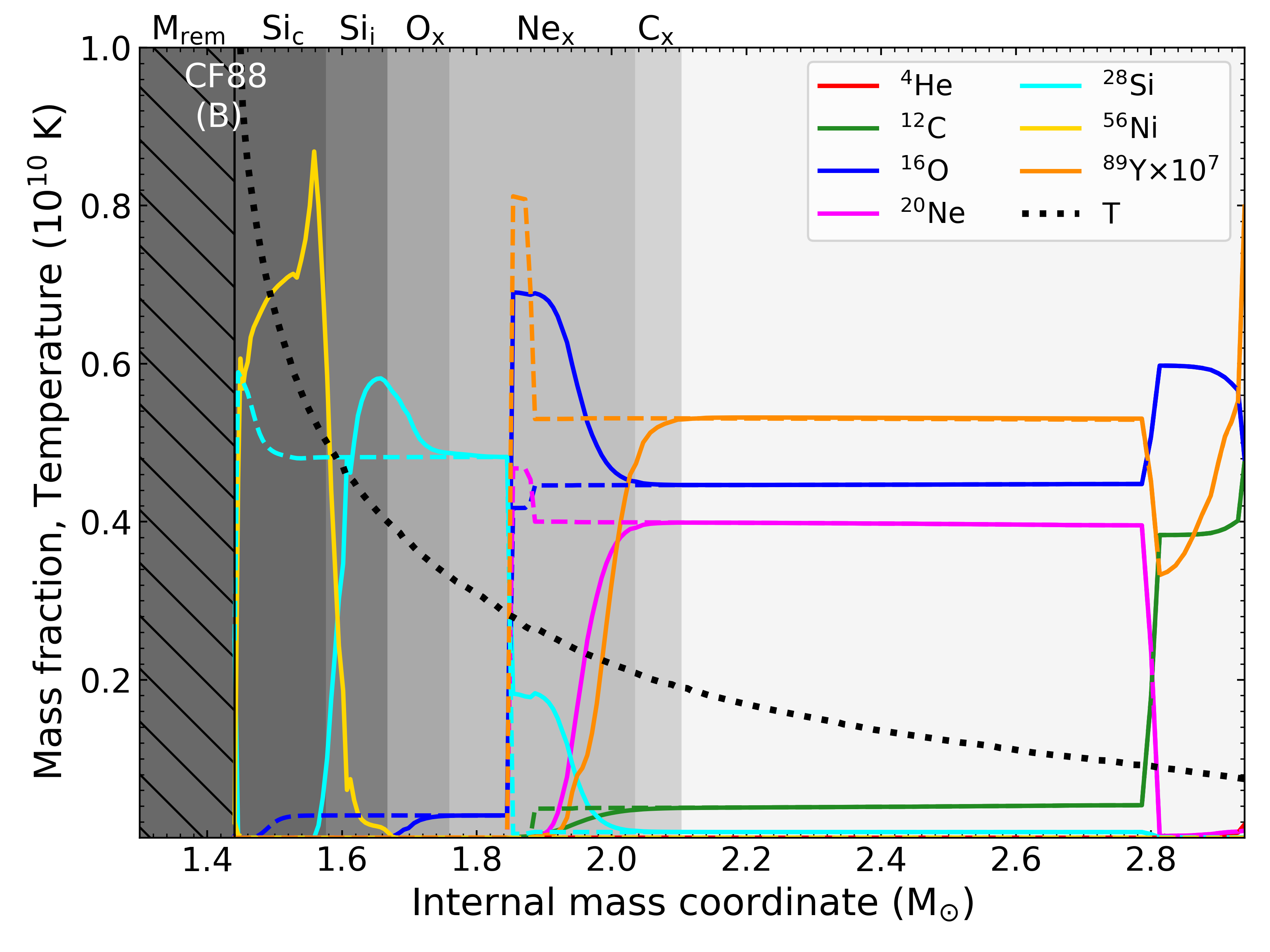}
        \includegraphics[width=.49\linewidth]{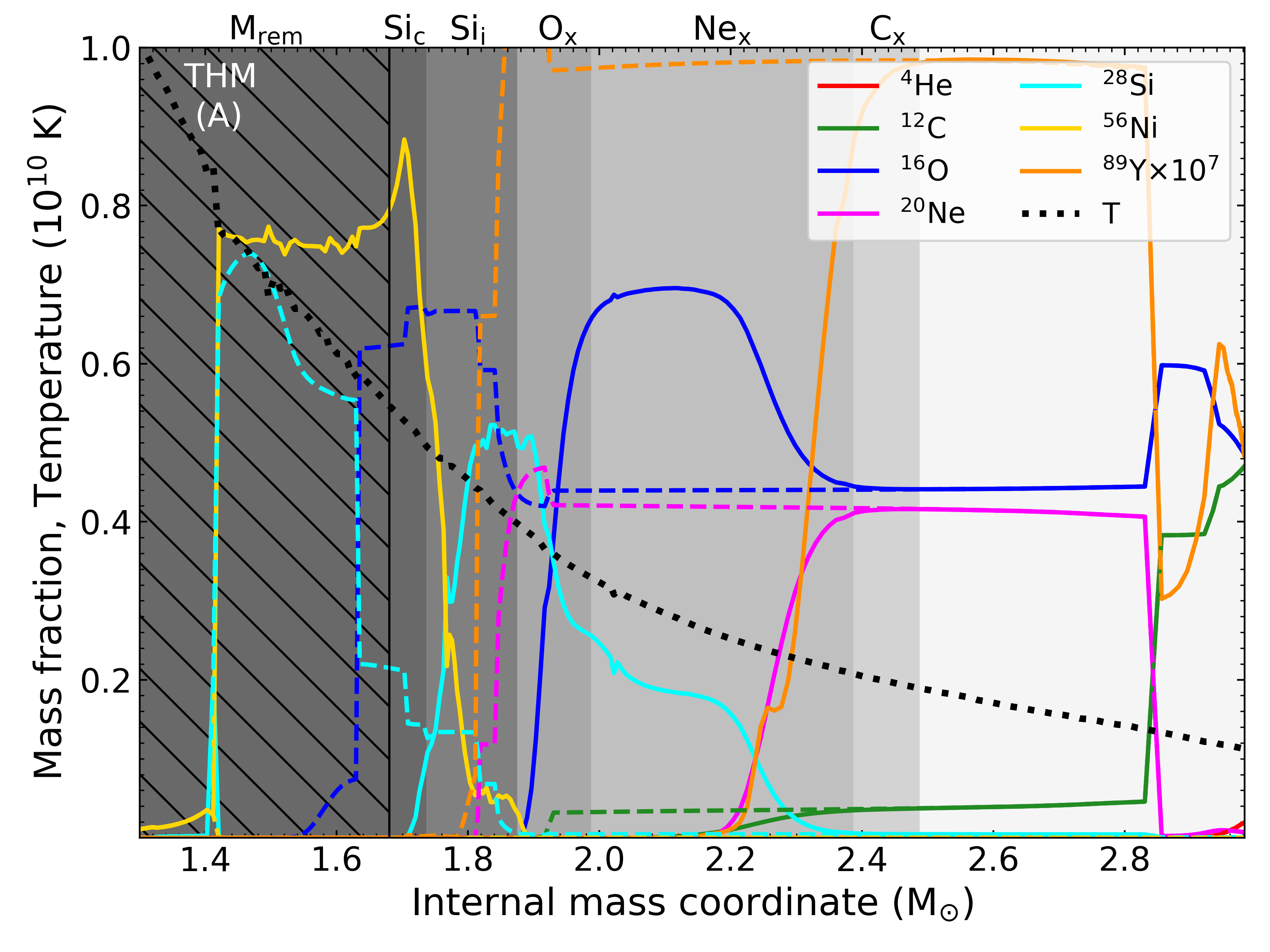}
        \includegraphics[width=.49\linewidth]{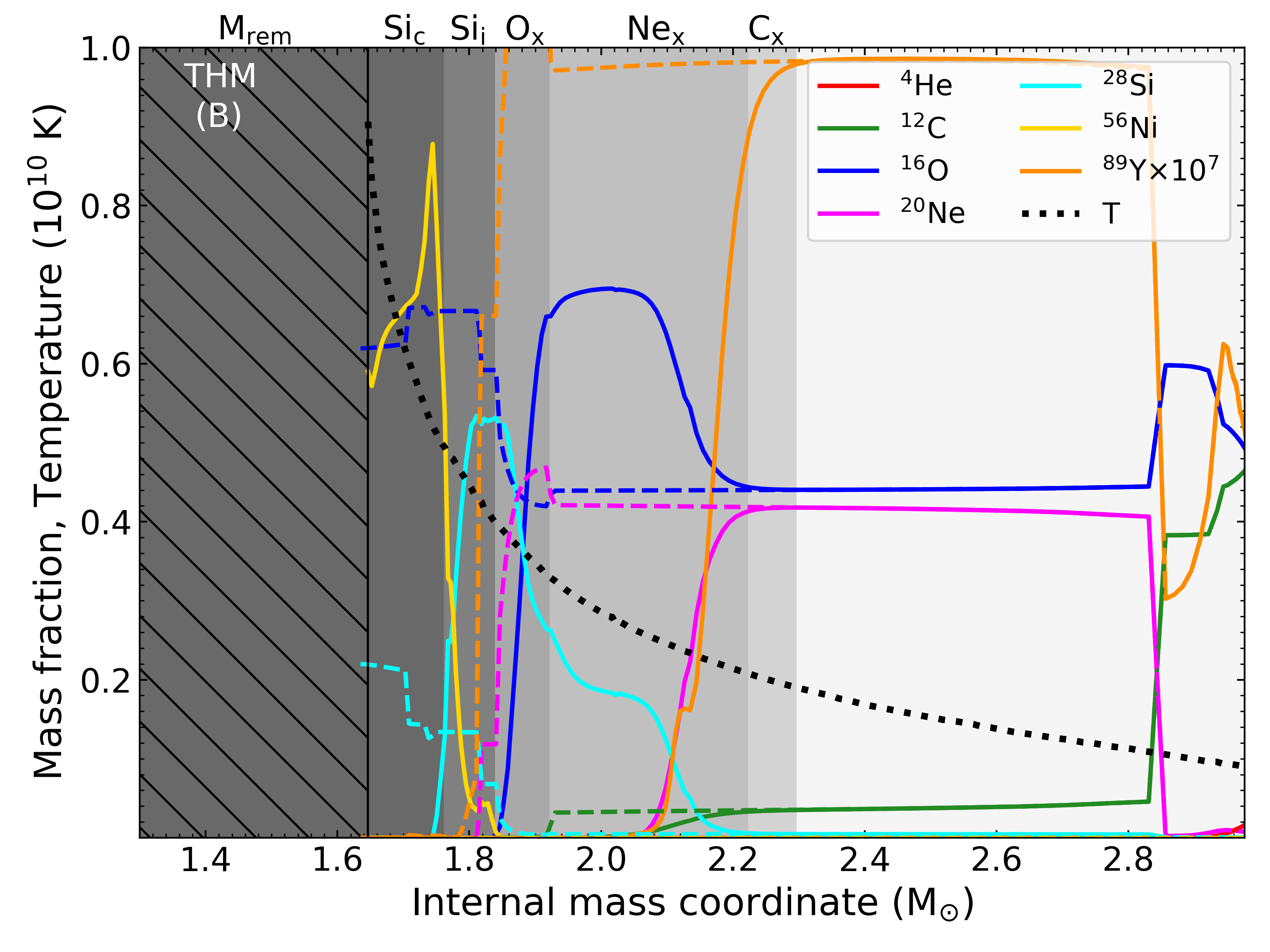}
        \caption{Comparison between the abundances before (dashed lines) and after (solid lines) the explosion in the case of the 15 \msun\ CF88 (upper panels) and THM (lower panels) model for the Set A (left panels) and Set B (right panels) explosions. The vertical solid black line represent the mass-cut that divides the supernova ejecta from the remnant mass. The gray bands in the plots mark each explosive burning stage in the corresponding mass coordinate (i.e., complete and incomplete Si burning, explosive O, Ne, and C burning). The dotted line represents the peak temperature of the shock in units of $\rm 10^{10}\ K$ throughout the stellar structure}
        \label{fig:expstrut}
    \end{figure*}
    
    At the time of explosion, the THM and CF88 models exhibit different structures, as discussed in the previous Sect. \ref{sec:strut}. Consequently, the propagation of the shock wave also differs. The chemical composition of the ejecta reflects these differences, both in the material reprocessed by the shock within $\rm \sim 10^4\ km$ from the Fe core (explosive nucleosynthesis) and in the material above, exposed to temperatures too low to be modified and thus ejected preserving the signature of the previous evolution of the star (hydrostatic nucleosynthesis). We also account for the contribution from stellar winds during the lifetime of the star, which is nearly identical in both THM and CF88 cases since mass loss after He exhaustion is negligible.

    The ratio between the THM and CF88 yields (in solar masses) for Set A is shown in \figurename~\ref{fig:yields}. In general, we note that in models with $\rm M \le 20\ M_\odot$, the THM yields of the weak \s process component (specifically Kr, Rb, and Sr) are higher by about a factor of 2. Since the shock tends to destroy most \s process nuclei, this result primarily reflects the hydrostatic nucleosynthesis, as discussed above (see the case of \isotope[89]{Y} in \figurename~\ref{fig:expstrut}). For $\rm M \ge 18\ M_\odot$, there is a slight overproduction of Ne, Na, and Mg (C burning products) and a slight underproduction of $\alpha$ and odd-Z elements between Si and Ca (Ne and O burning products). This effect is particularly evident in the 20 and 22 \msun\ cases, where a large C shell limits the growth of the Ne and O shells and reduces the overall size of the ONe core (see also Sect. \ref{sec:strut}). The 25 and 40 \msun\ models show the smallest differences, as the higher C shell ignition temperature reduces the impact of the different rates. Finally, the most significant discrepancies occur in the 16 \msun\ model, where elements between P and Ti are overproduced by a factor of 10--20. This is the consequence of a C-O shell merger in the THM model, which does not occur in the CF88 case, boosting the production of odd-Z elements and $\alpha$-nuclei from O burning \citep[][and references therein]{roberti:25,roberti:25b,boccioli:26}. The yields of short-lived radioactive (SLR) nuclei do not exhibit differences between CF88 and THM cases, except for \isotope[107]{Pd}, that follows the same trend with the mass as the \s process elements shown at the bottom of \figurename~\ref{fig:yields}. The only exceptions are in the 16 THM model, where the radioactive nuclei \isotope[36]{Cl}, \isotope[40]{K}, \isotope[41]{Ca}, and \isotope[44]{Ti} are largely overproduced compared to the CF88 model due to the C--O shell merger nucleosynthesis.
        
    \subsection{Impact of the explosion prescription} \label{subsec:exp}

       Set A yields were calculated assuming that all the stars would eject the same amount of \isotope[56]{Ni}, which corresponds to that estimated for SN 1987A \citep{arnett:89}. However, collection of observations \citep[see, e.g., Fig. 2 from][]{nomoto:13} and light curve inference studies showed that the \isotope[56]{Ni} that powers the SN light-curve varies significantly among different progenitors \citep{kasen:09,pejcha:15,goldberg:19}. In particular, the amount of \isotope[56]{Ni} ejected positively correlates with the explosion energy and compactness of the progenitor \citep{bruenn:16,ertl:16,curtis:19,burrows:24,janka:25,BR25}. Therefore, we performed another series of explosions (Set B), imposing the final kinetic energy of the ejecta dependence on the pre-supernova compactness following Eq.~\eqref{eq:eexp}. As discussed in Sect. \ref{sec:strut}, the compactness correlates with the Fe core mass, therefore we expect this equation to be independent from the choice of the \cc\ reaction rate. More specifically, we adopted the thermal bomb technique with the \verb|HYPERION| code, instantaneously injecting an amount of energy in a narrow region (0.1 \msun) above the Si/Si-O interface, such that the energy at infinity satisfies Eq. \ref{eq:eexp} (see Sect. \ref{sec:methods}). We do not impose any constraints on the mass-cut and therefore we adopt the result of the hydrodynamic simulation for the remnant mass, which, in most of the cases, is very close to the mass of the Si/O interface (see columns "Si/Si-O" and "$\rm M_{rem}^B$" from \tablename~\ref{tab:eexp}). In only two cases (22 and 25 \msun\ THM models) the explosion energy is low enough to allow a significant fallback within the first $\sim10^5$ seconds ($\sim$ two days), resulting in a remnant mass of 4.57 and 5.02 \msun, respectively. 

        \begin{deluxetable*}{llrrllllllll}
        \tablecaption{Properties of the exploding models for set A and B (see text).\label{tab:eexp}}
        \tablehead{
            \colhead{Mass} & \colhead{Set} & \colhead{$\rm M_{TOT}$} & \colhead{$\rm M_{He}$} & \colhead{$\rm M_{CO}$} & \colhead{$\rm M_{ONe}$} & \colhead{Si/Si-O} & \colhead{$\xi_{2.5}$} & \colhead{$\rm E_{\rm exp}^B$} & \colhead{$\rm ^{56}Ni^B$} & \colhead{$\rm M_{rem}^B$} & \colhead{$\rm M_{rem}^A$} \\
            \colhead{(\msun)} & \colhead{} & \colhead{(\msun)} & \colhead{(\msun)} & \colhead{(\msun)} & \colhead{(\msun)} & \colhead{(\msun)} & \colhead{} & \colhead{(foe)} & \colhead{(\msun)} & \colhead{(\msun)} & \colhead{(\msun)}}
            \startdata
            15   & CF88 & 12.89 &  5.01 & 2.94 & 1.86 & 1.44 & 0.130 & 0.532   & 1.043E-01 & 1.45  & 1.56  \\
                 & THM  & 12.73 &  5.01 & 2.96 & 1.91 & 1.56 & 0.158 & 0.649   & 8.994E-02 & 1.65  & 1.68  \\
            16   & CF88 & 13.58 &  5.49 & 3.26 & 1.95 & 1.47 & 0.152 & 0.626   & 3.986E-02 & 1.49  & 1.58  \\
                 & THM  & 13.44 &  5.49 & 3.28 & 1.87 & 1.45 & 0.126 & 0.514   & 5.388E-02 & 1.45  & 1.48  \\
            18   & CF88 &  6.49 &  6.42 & 3.94 & 2.10 & 1.50 & 0.183 & 0.760   & 3.944E-02 & 1.51  & 1.68  \\
                 & THM  &  6.47 &  6.42 & 3.96 & 1.91 & 1.50 & 0.153 & 0.629   & 4.289E-02 & 1.50  & 1.66  \\
            20   & CF88 &  7.38 &  7.31 & 4.64 & 2.34 & 1.61 & 0.241 & 1.010   & 1.410E-01 & 1.65  & 1.86  \\
                 & THM  &  7.37 &  7.31 & 4.65 & 1.67 & 1.58 & 0.097 & 0.389   & 1.076E-02 & 1.60  & 1.59  \\
            22   & CF88 &  8.19 &  8.19 & 5.36 & 2.53 & 1.67 & 0.279 & 1.170   & 1.968E-01 & 1.67  & 2.01  \\
                 & THM  &  8.19 &  8.19 & 5.34 & 1.89 & 1.45 & 0.143 & 0.585   & -         & 4.57  & 1.66  \\
            25   & CF88 &  8.88 &  8.88 & 6.24 & 2.01 & 1.55 & 0.174 & 0.721   & 2.613E-02 & 1.55  & 1.85  \\
                 & THM  &  8.88 &  8.88 & 6.29 & 2.19 & 1.59 & 0.166 & 0.686   & -         & 5.02  & 1.88  \\
            40   & CF88 & 11.34 & 11.34 & 8.57 & 1.99 & 1.53 & 0.162 & 0.669   & 2.753E-02 & 1.57  & 1.83  \\
                 & THM  & 11.34 & 11.34 & 8.57 & 1.92 & 1.55 & 0.144 & 0.590   & 8.242E-03 & 1.69  & 1.89  \\   
            \enddata
            \tablecomments{All the masses are defined at the pre-supernova stage. He core mass (or H-free mass) $\rm M_{He}$ defined where $\rm X(H)<0.01$; CO core mass (or He-free mass) $\rm M_{CO}$ defined where $\rm X(^4He)<0.01$; ONe core mass (or C-free mass) $\rm M_{ONe}$ defined where $\rm X(^{12}C)<0.001$.}
        \end{deluxetable*}

        Figure~\ref{fig:expstrut} compares the shock wave propagation for the 15~$M_\odot$ models (Set A and Set B) using both CF88 and THM rates. The primary difference lies in a steeper peak temperature profile ($T_{\text{peak}}$), which defines the extent of the various explosive burning regimes: complete Si burning ($\rm T_9 \gtrsim 5$, "$\rm Si_c$"); incomplete Si burning ($5\gtrsim\rm T_9\gtrsim4$, "$\rm Si_i$"); explosive O burning ($4\gtrsim\rm T_9\gtrsim3.3$, "$\rm O_x$"); explosive Ne burning ($3.3\gtrsim\rm T_9\gtrsim2.1$, "$\rm Ne_x$"); explosive C burning ($2.1\gtrsim\rm T_9\gtrsim1.9$, "$\rm C_x$"); where $\rm T_9 = T/10^9 K$. Specifically, Set B explosions tend to have more narrow $\rm Si_i$, $\rm O_x$, $\rm Ne_x$, and $\rm C_x$ burning regions, but slightly larger $\rm Si_c$ regions. As reported in \tablename~\ref{tab:eexp}, this is due to, on average, smaller remnant masses while ejecting, in most of the cases, less \isotope[56]{Ni} than in Set A.

        \begin{figure*}[!t]
            \centering
            \includegraphics[width=.49\linewidth]{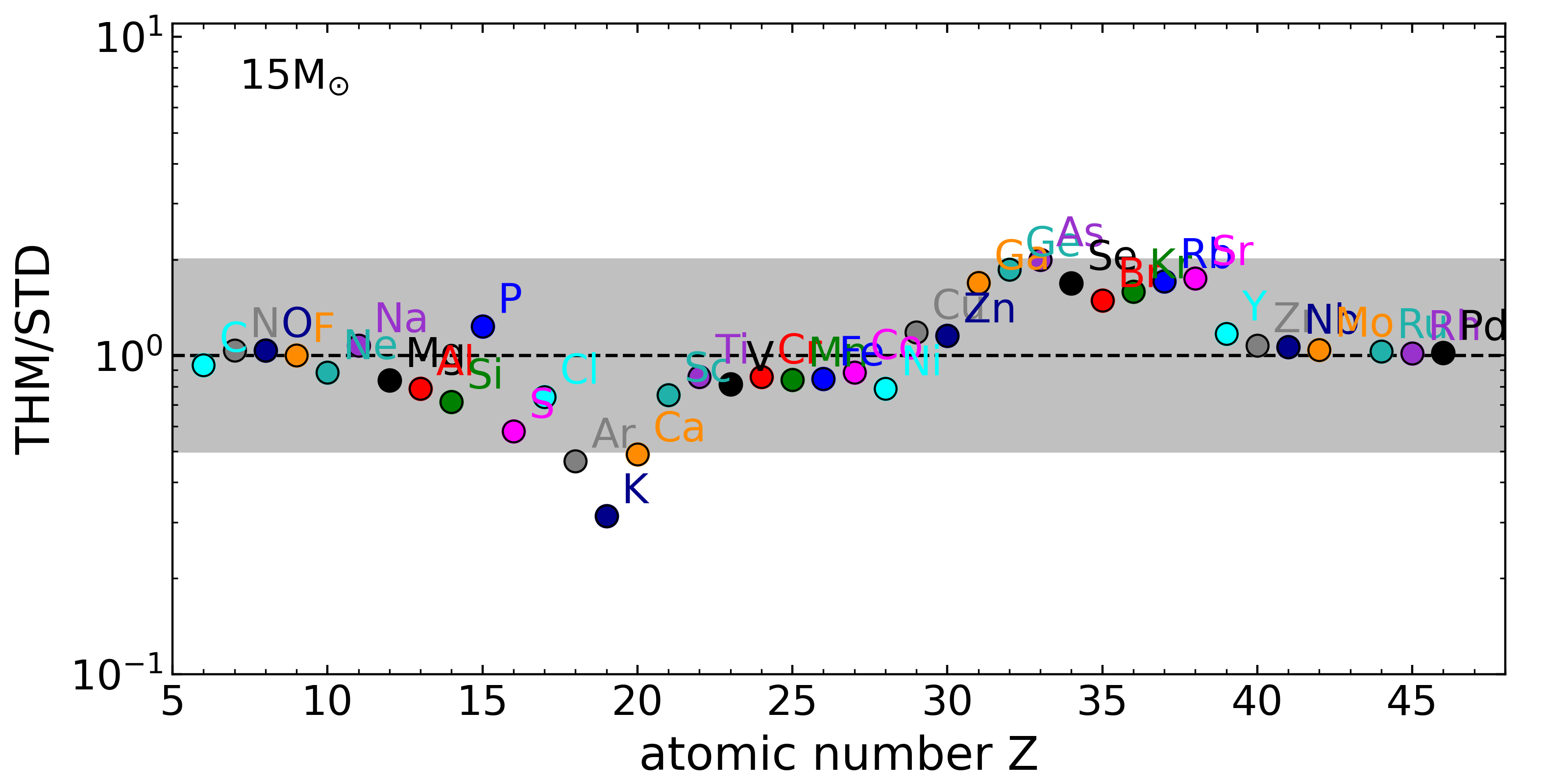}
            \includegraphics[width=.49\linewidth]{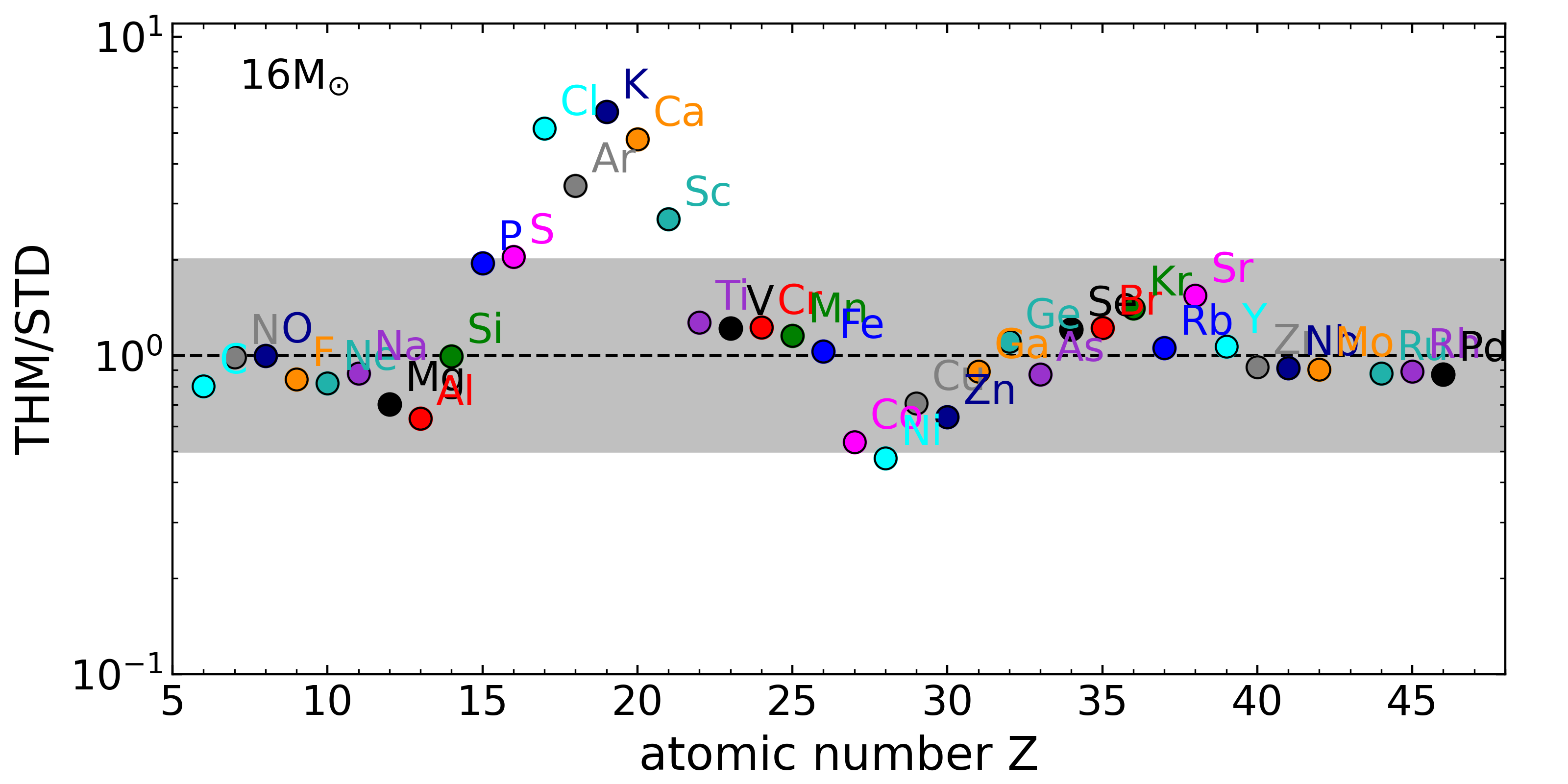}
            \includegraphics[width=.49\linewidth]{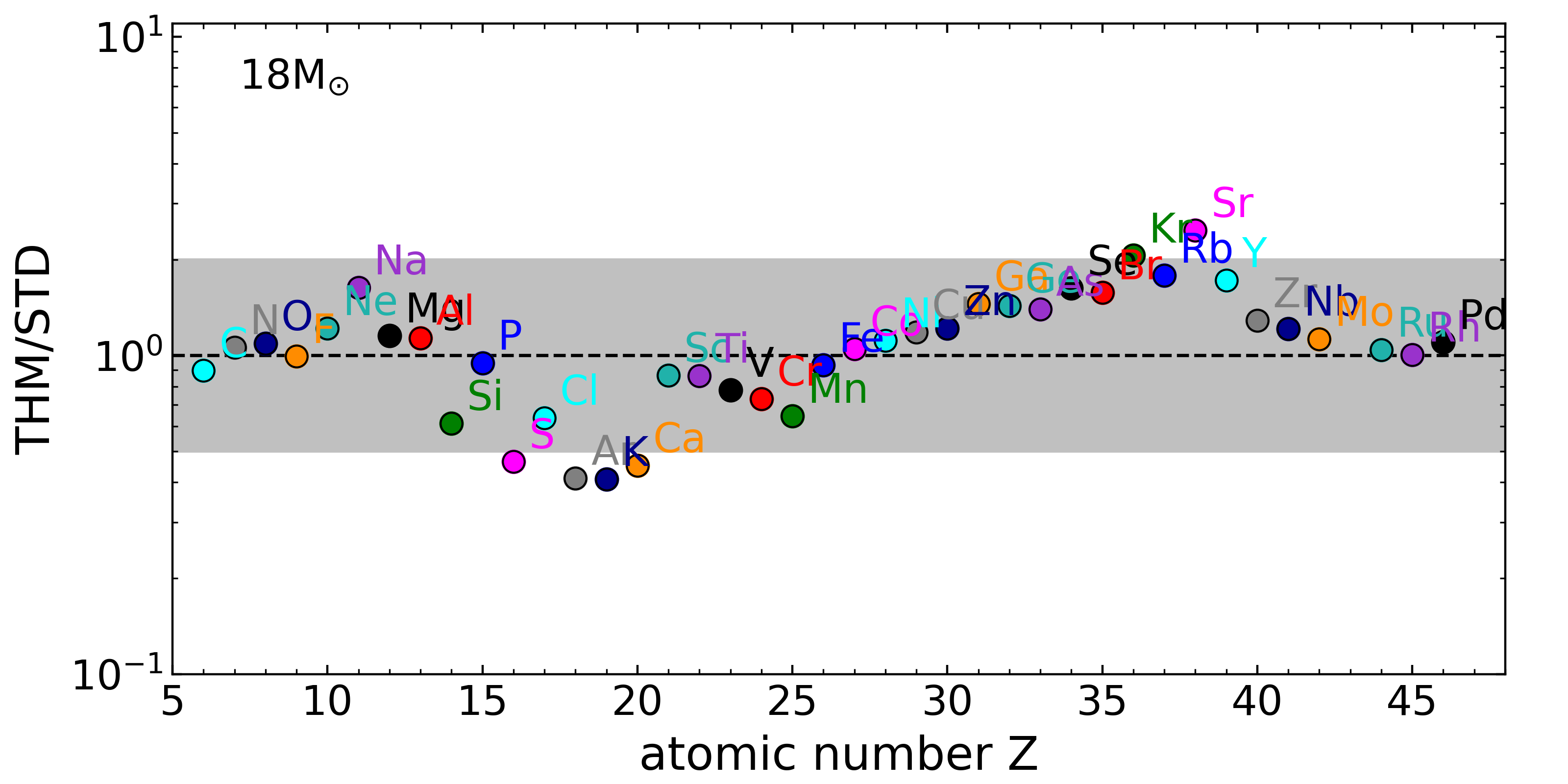}
            \includegraphics[width=.49\linewidth]{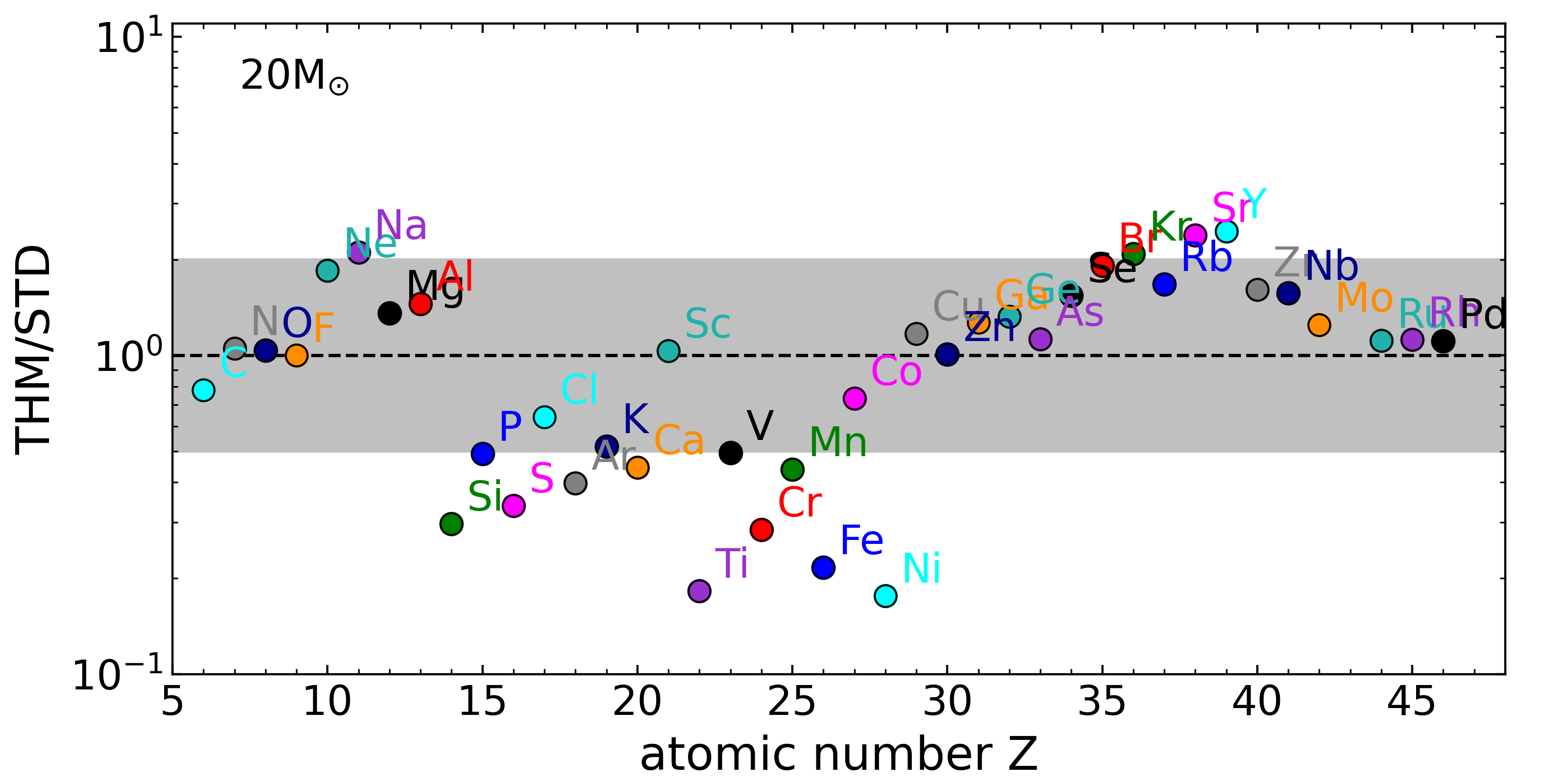}
            \includegraphics[width=.49\linewidth]{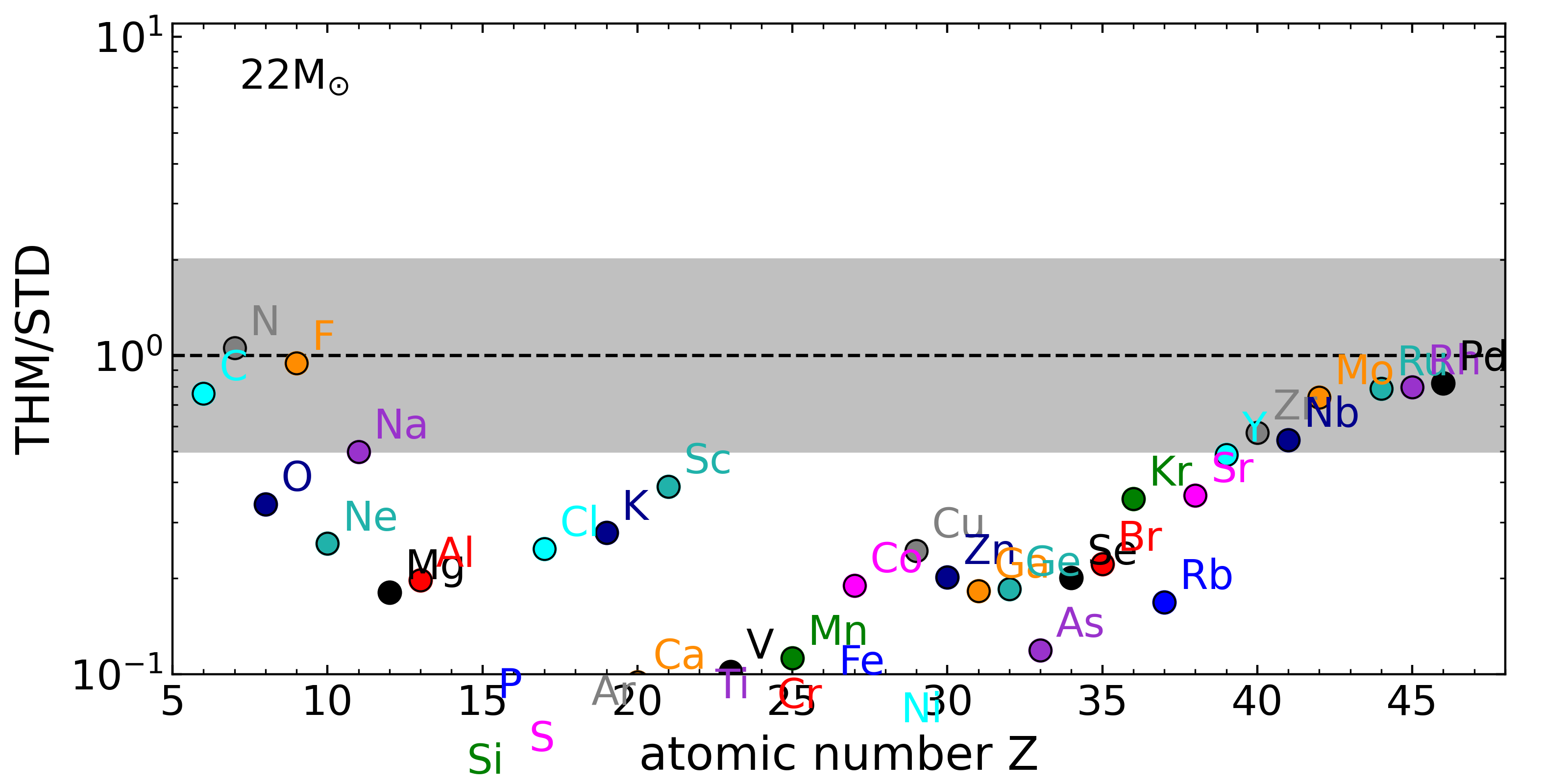}
            \includegraphics[width=.49\linewidth]{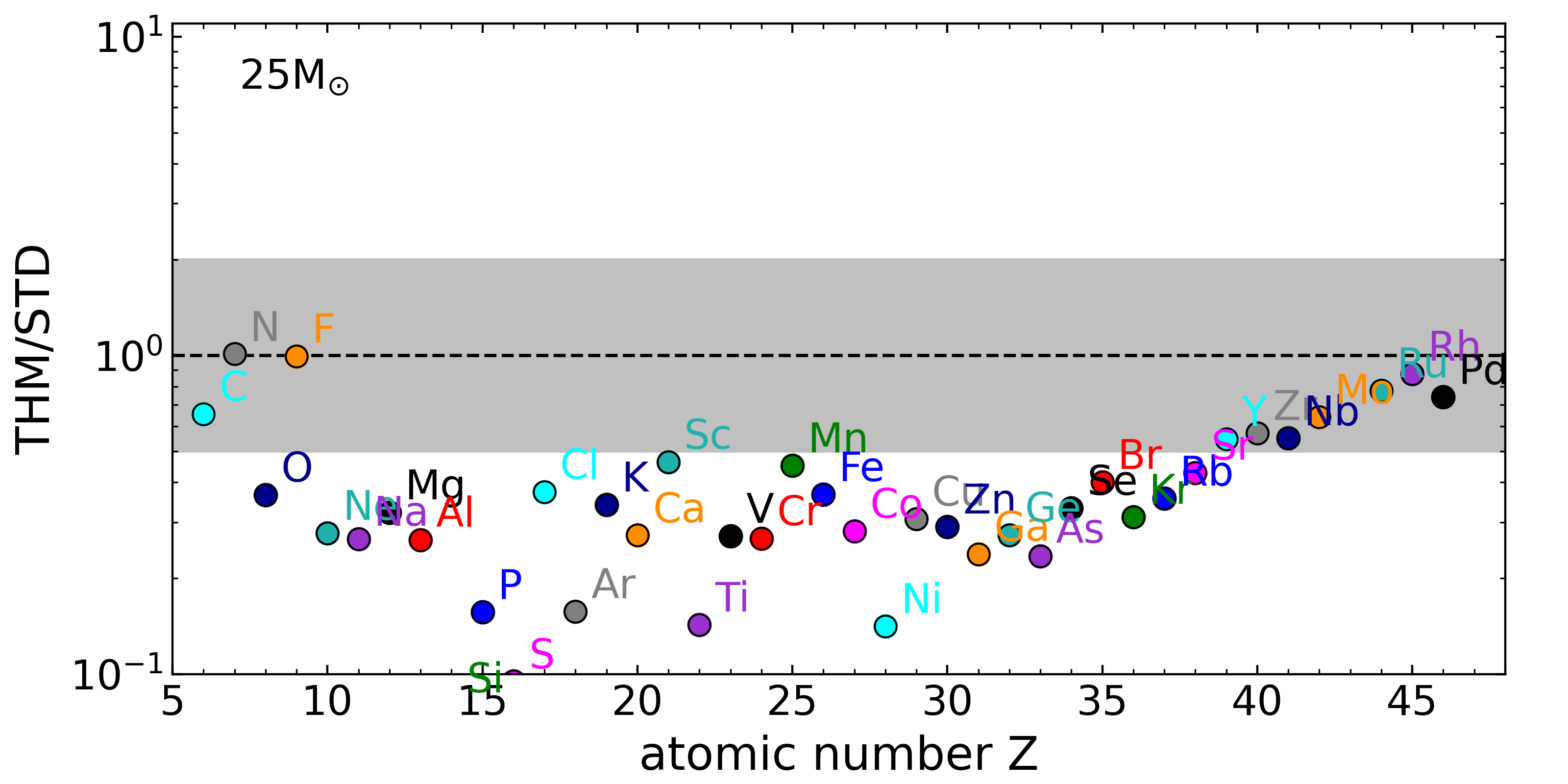}
            \includegraphics[width=.49\linewidth]{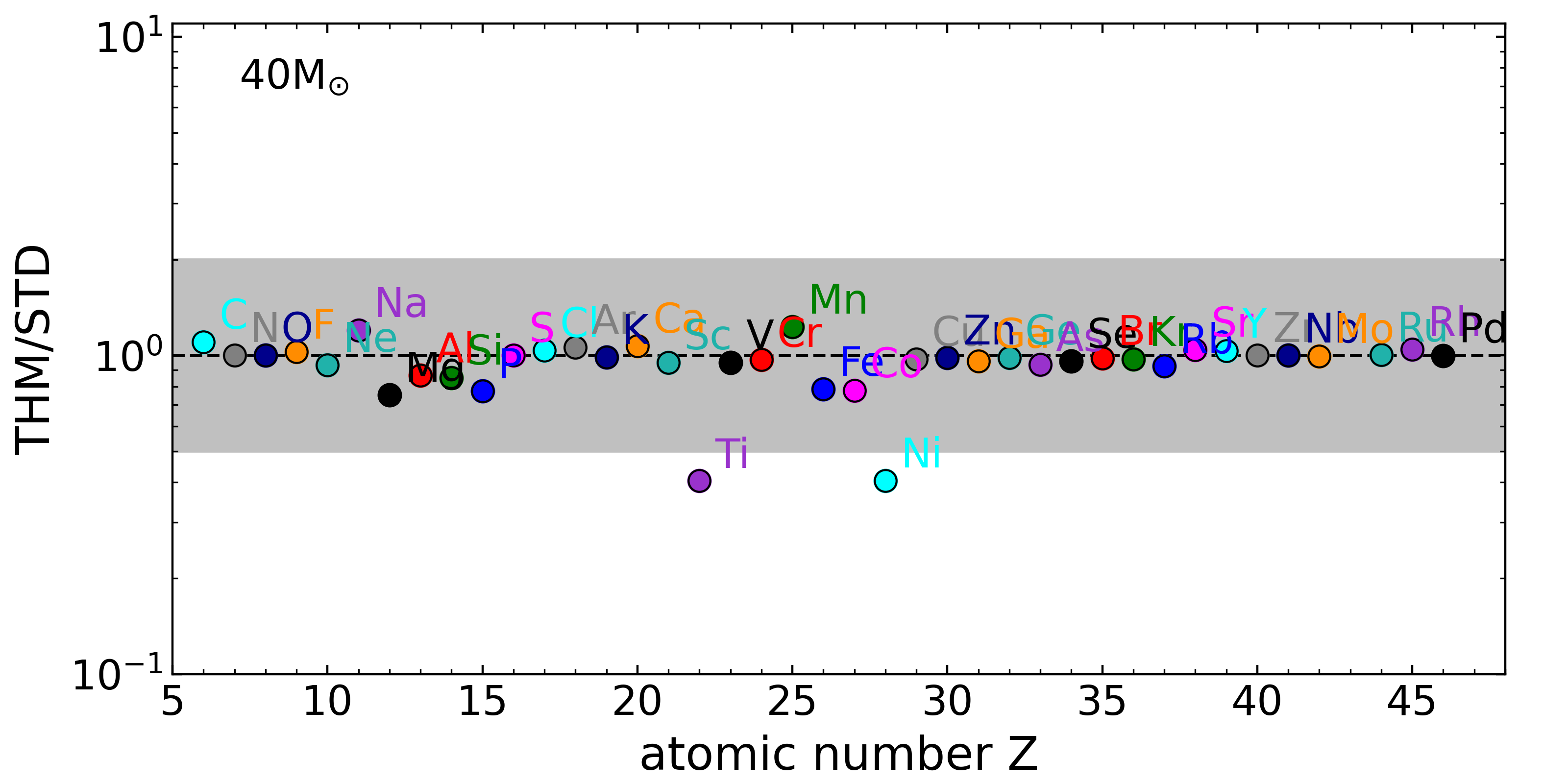}
            \includegraphics[width=.49\linewidth]{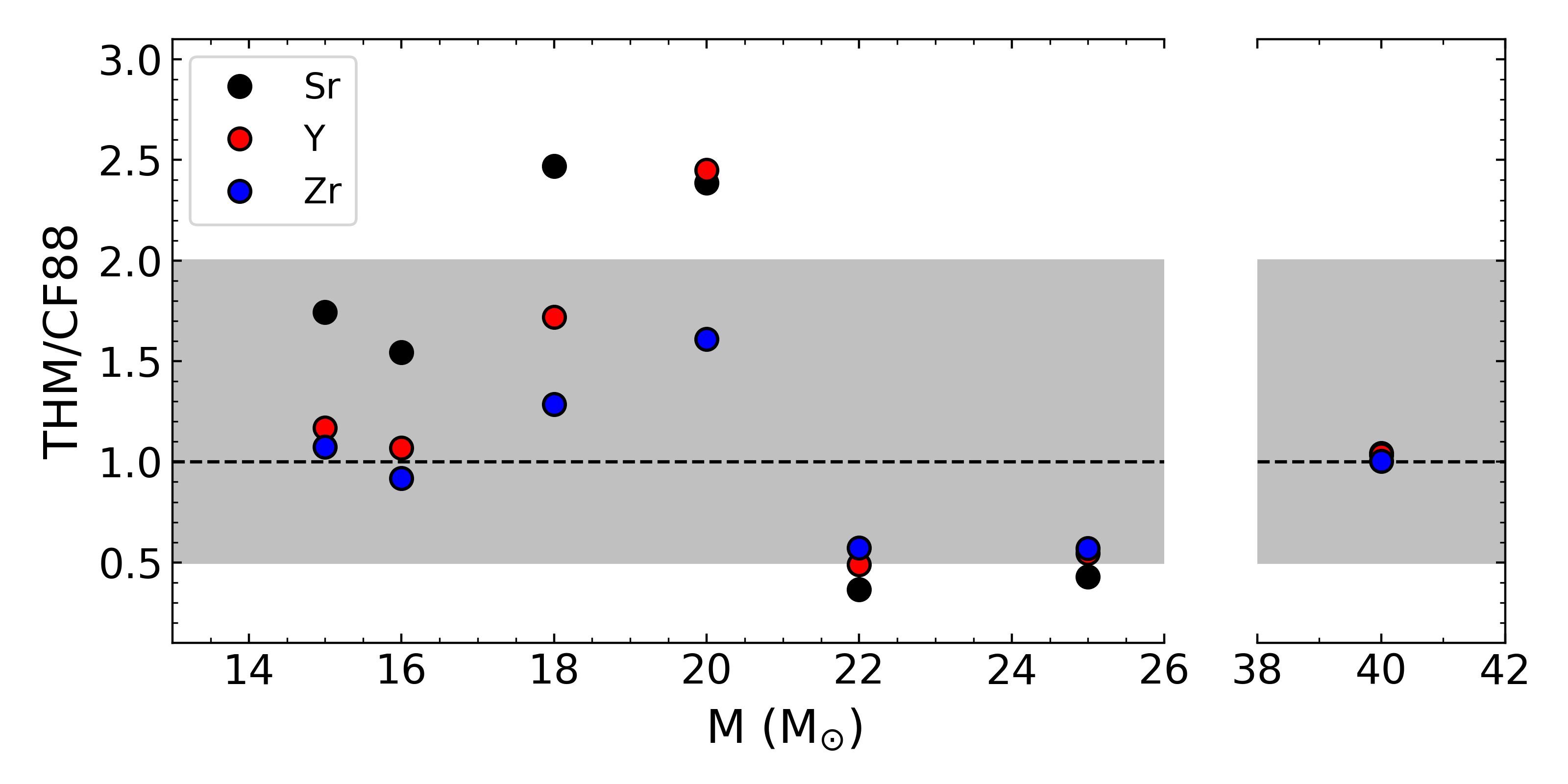}
            \caption{Ratio between THM and CF88 yields for each model, in the case of Set B (see text). The gray band identifies a factor of 2 variation.}
            \label{fig:yieldsB}
        \end{figure*}
    
        Despite its mild effect on the total yields \citep{limongi:03a,roberti:24b}, the explosion parametrization may significantly impact the elemental ratios. In particular, by construction, Set A explosions are more energetic than those of Set B. Therefore, many of the structural differences between CF88 and THM models that are clearly visible in the yields of Figure~\ref{fig:presn} are attenuated by the explosion, and appear less prominent in \figurename~\ref{fig:yields}. Conversely, in Set B, some of these features persist because the explosive burning zones are less extended, thus leading to a more limited reprocessing of the pre-supernova structure. \figurename~\ref{fig:yieldsB} shows the same ratios of \figurename~\ref{fig:yields} but in the case of Set B. In particular, the underproduction with respect to CF88 models of Si to Ca elements is more evident in the 15, 18, and 20 \msun\ models. 
        In the 16 \msun\ model, the impact of the explosion is marginal for Set B, as the differences remain largely dominated by the pre-supernova C--O shell merger. Nevertheless, elements around the K peak appear slightly underproduced compared to Set A: this occurs because the more energetic Set A explosions synthesize a non negigible explosive component for these species, which is otherwise missing in Set B.
        The two THM models 22 and 25 \msun\ in the Set B case instead have a large fallback, that allows only the ejection of the outer region of the C shell, locking into the remnant the explosive nucleosynthesis products and the ashes of the more internal zones of the star. Finally, in the case of the 40 \msun\ model, the only differences are Ni and Ti, that are produced by complete Si burning in the most internal parts of the ejecta and that are not ejected in the THM case.


\section{Discussion and conclusions} \label{sec:disc}

    We studied the impact of the \cc reaction rate by \citet[][THM]{tumino:18} on the evolution, explosion, and nucleosynthesis of non-rotating models of massive stars at solar metallicity. We find significant structural differences compared to models calculated with the reaction rate by \citet[][CF88]{caughlan:88}, which is the main reference for C burning in stellar models still today. The differences in the structure between CF88 and THM models also impact the core-collapse supernova explosion and hence the chemical composition of the ejecta. Our findings can be summarized as follows: 
    
    \begin{itemize}
        \item The duration of the central C burning computed with the THM reaction rate is about three times longer than that in CF88 models;
        \item the convective core in central C burning of THM models is more massive than that of CF88 ones, while the transition to radiative C burning occurs earlier as the initial mass increases for the THM models;
        \item the formation of C, Ne, and O shells is influenced by the ignition condition of the central C and Ne burning phases in both CF88 and THM models, leading to different pre-supernova structures (e.g., mass extent of the convective shells, ONe core masses, compactness);
        \item in most cases, the THM models have a lower compactness $\xi_{2.5}$ at the onset of core collapse compared to the CF88 models, reflecting a systematically less massive ONe core;
        \item the differences between THM and CF88 models become negligible for He core mass at core collapse above $\sim11$ \msun\ (corresponding to an initial mass equal to 40 \msun), because the \cc efficiency scales with the available \isotope[12]{C} abundance, which in turn inversely scales with the He core mass;
        \item the products of C burning nucleosynthesis are more abundant than those of Ne and O burning in THM models compared to CF88 ones by a factor of 1.5--3, as the THM models have on average wider C burning shells;        
        \item THM models produce about a factor of 2 more elements heavier than Fe via the \s process nucleosynthesis due to the more efficient activation of the \cn\ reaction in the early C burning shell;
        \item different explosion prescriptions can alter some elemental ratios. However, the overall elemental and isotopic distribution is influenced for the most part by the pre-supernova structure.
    \end{itemize}
    
    Advances in experimental techniques have enabled new and updated measurements of the \cc reaction. Currently, two competing scenarios are under debate: the THM rate (the subject of this work), which is higher than the classical CF88 rate, and the hindrance scenario, where the rate is lower than the CF88 value \citep{monpribat:22}. However, it is worth emphasizing that the hindrance effect may influence the overall behaviour of the reaction rate at low energies, but it does not call into question the existence of resonant structures at $\rm{E_{cm} \le 2~ MeV}$. We have shown that these resonances can lead to significant changes in the structure, evolution, and nucleosynthesis of stellar models compared to those calculated with a \cc\ rate such as the CF88 one, which does not take them into account. For a comprehensive review of the \cc reaction rate status and its impact on stellar models, we refer the reader to \cite{chieffi:25} and references therein.

    As discussed in Sect. \ref{sec:intro} and \ref{sec:methods}, the aim of this work is to study the impact of the more efficient \cc reaction on the nucleosynthesis and in particular on the \s process. While massive stars primarily contribute to the galactic chemical evolution with the weak component of the \s process \citep[corresponding to nuclei from Fe to the neutron shell closure N=50, e.g., $\rm ^{88}Sr$,$\rm ^{89}Y$, $\rm ^{90}Zr$,][]{prantzos:18,prantzos:20}, they can also moderately contribute to the main and strong components at low metallicity \citep{frischknecht:16,choplin:20,roberti:24,rizzuti:21,lombardo:25}. This is due to the increased neutron-to-seed ratios in fast rotators, which are more frequent at low metallicities, whereas massive stars closer to solar metallicity tend to be slower rotators and are characterized by a lower neutron-to-seed ratio that favors the weak component. Most of the \s process production occurs during the He burning phase, with C burning playing a more marginal role, as discussed in Sect. \ref{subsec:sproc}. Moreover, the C burning phase is not expected to be directly affected by rotation, since the timescales of rotational instabilities are significantly longer than the nuclear timescales of the advanced evolutionary stages. Instead, rotation plays an indirect role by altering the conditions at the end of He burning, specifically the CO core mass and the central \isotope[12]{C} mass fraction. However, the entanglement between H and He burning regions in rotating stars allows for an enhanced synthesis of \isotope[22]{Ne} from \isotope[14]{N} \citep{roberti:24}. This increased \isotope[22]{Ne} abundance partially survives the He burning into the advanced phases, where it may affect the balance between the \nen\ and \cn\ neutron sources. Therefore, we expect the direct impact of the C burning rate to be more prominent on the \s process nucleosynthesis than on the overall stellar structure in rotating massive star models.  An increase in the \s process production is expected also to enhance the synthesis of $p$-nuclei through the $\gamma-$process nucleosynthesis during the supernova explosion \citep{arnould:03,rauscher:13,pignatari:16,roberti:23}. However, it should be noted that a comprehensive $\gamma-$process nucleosynthesis study would require a significantly larger nuclear network than the one currently employed in our models. Due to this numerical limitation, we do not observe any significant variations in the $p$-nuclei abundances between the two different sets of explosion models. A detailed study of these effects will be presented in a forthcoming paper.\\
\\
    L.R. and S.P. acknowledge the support from the PRIN2022 project entitled “$\beta-$DE\-cays and NEutrons captures for astrophysical Branchings (DE\-NEB)" (2022THRKMK). L.R. acknowledges support of INAF grant GO-GTO2024 ”Pre-supernova
    outbursts in Galactic RSGs, the support from the PRIN URKA Grant Number 2022rjlwhn, the Lend\"ulet Program LP2023-10 of the Hungarian Academy of Sciences, the Hungarian NKFIH via K-project 138031 and NKKP Advanced grant 153697. L.B. is supported by the U.S. Department of Energy under Grant No. DE-SC0004658 and SciDAC grant, DESC0024388. L.B. would like to thank the N3AS center for its hospitality and support. M.L. aknowledges the support from the INAF Theory Grant "Massive stars as cosmic clocks: shaping the evolution of infant galaxies", Ob.Fu. 1.05.24.05.07.

\vspace{5mm}

\textit{Data availability} 
Data is available at the Online Repository for the Franec Evolutionary Output (ORFEO)\footnote{\url{https://orfeo.oa-roma.inaf.it}} and upon reasonable request to the corresponding author.

\appendix

\section{Explosion comparison}

        \begin{figure}[!t]
            \centering
            \includegraphics[width=.49\linewidth]{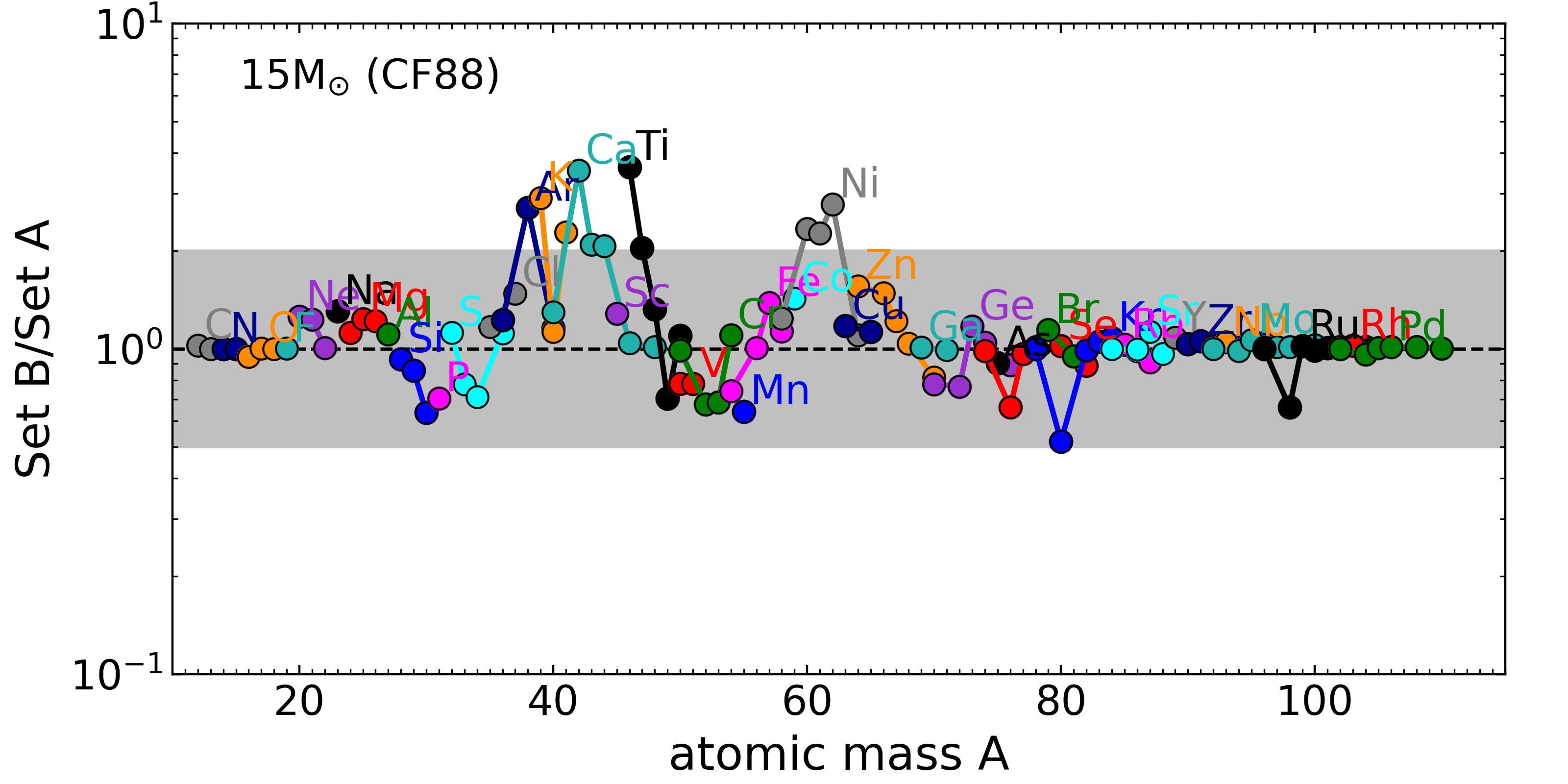}
            \includegraphics[width=.49\linewidth]{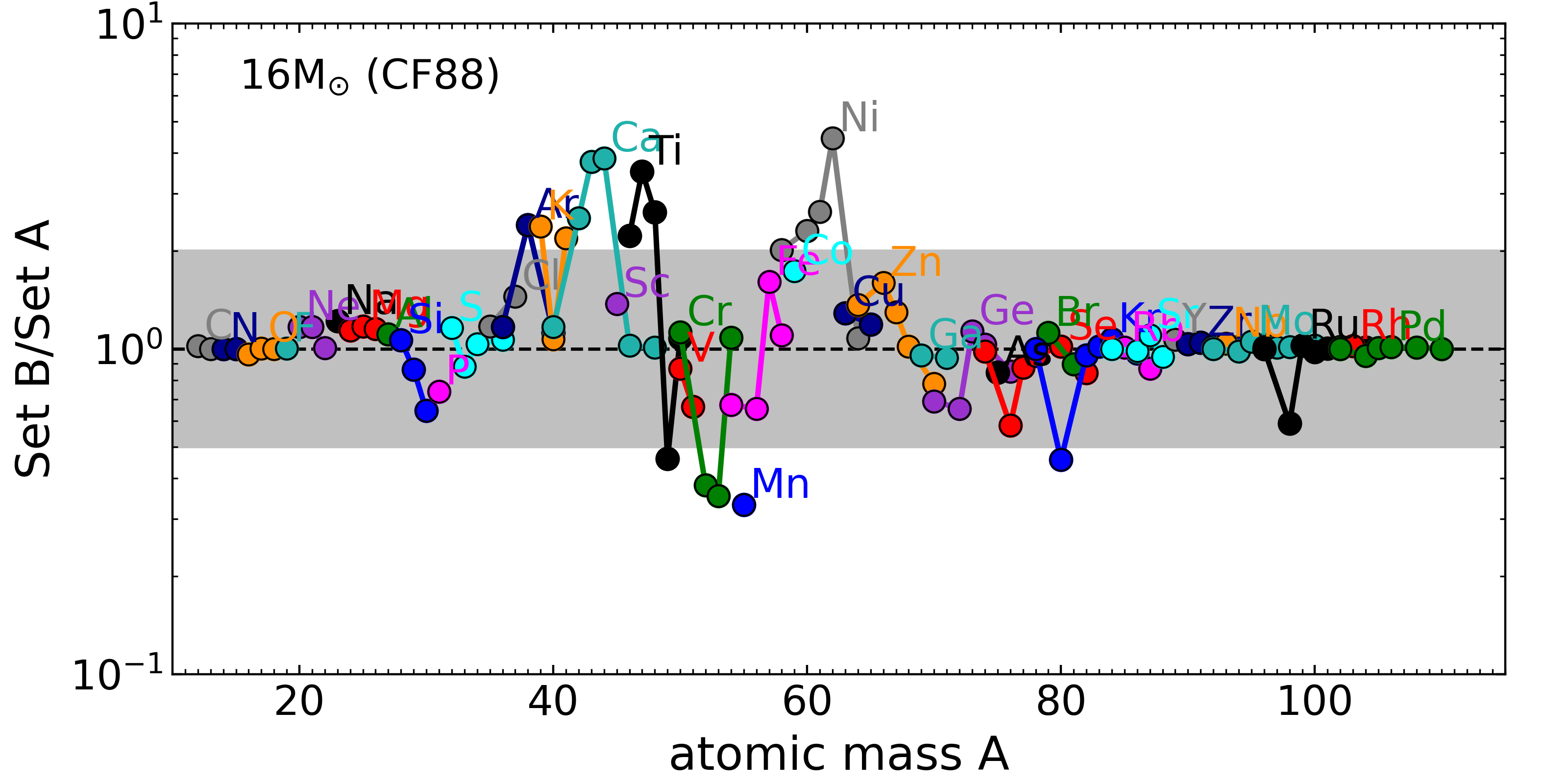}
            \includegraphics[width=.49\linewidth]{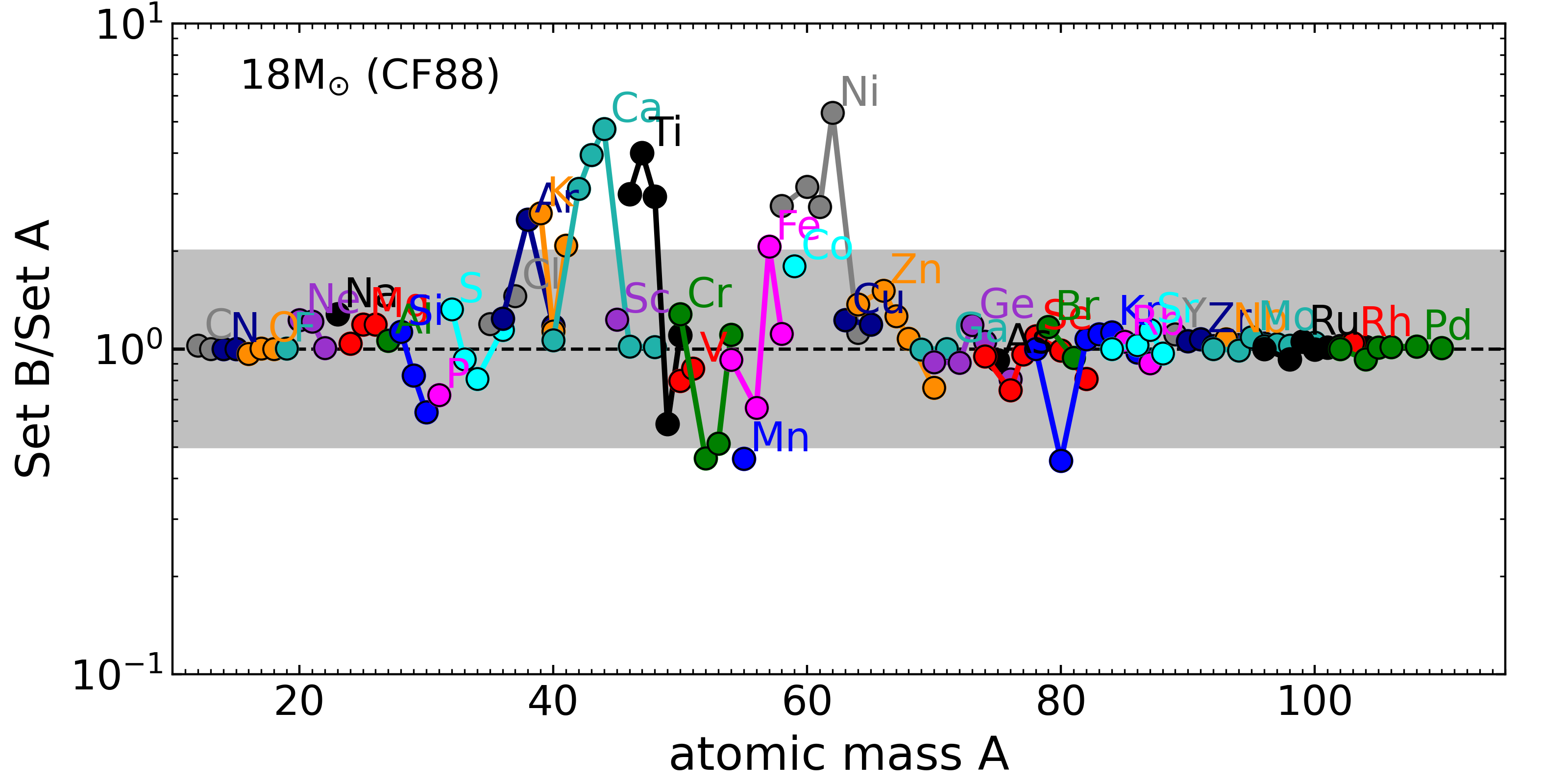}
            \includegraphics[width=.49\linewidth]{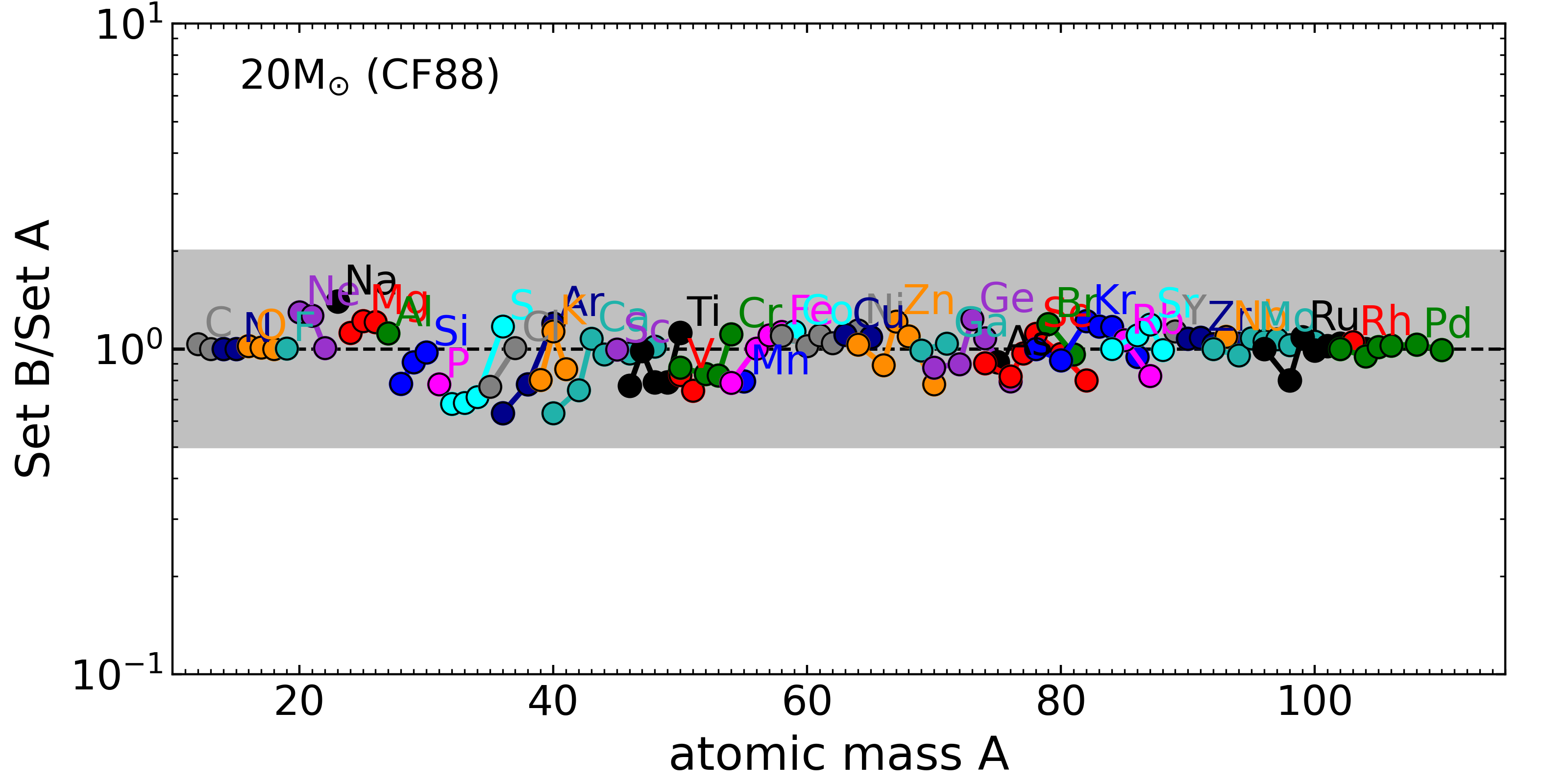}
            \includegraphics[width=.49\linewidth]{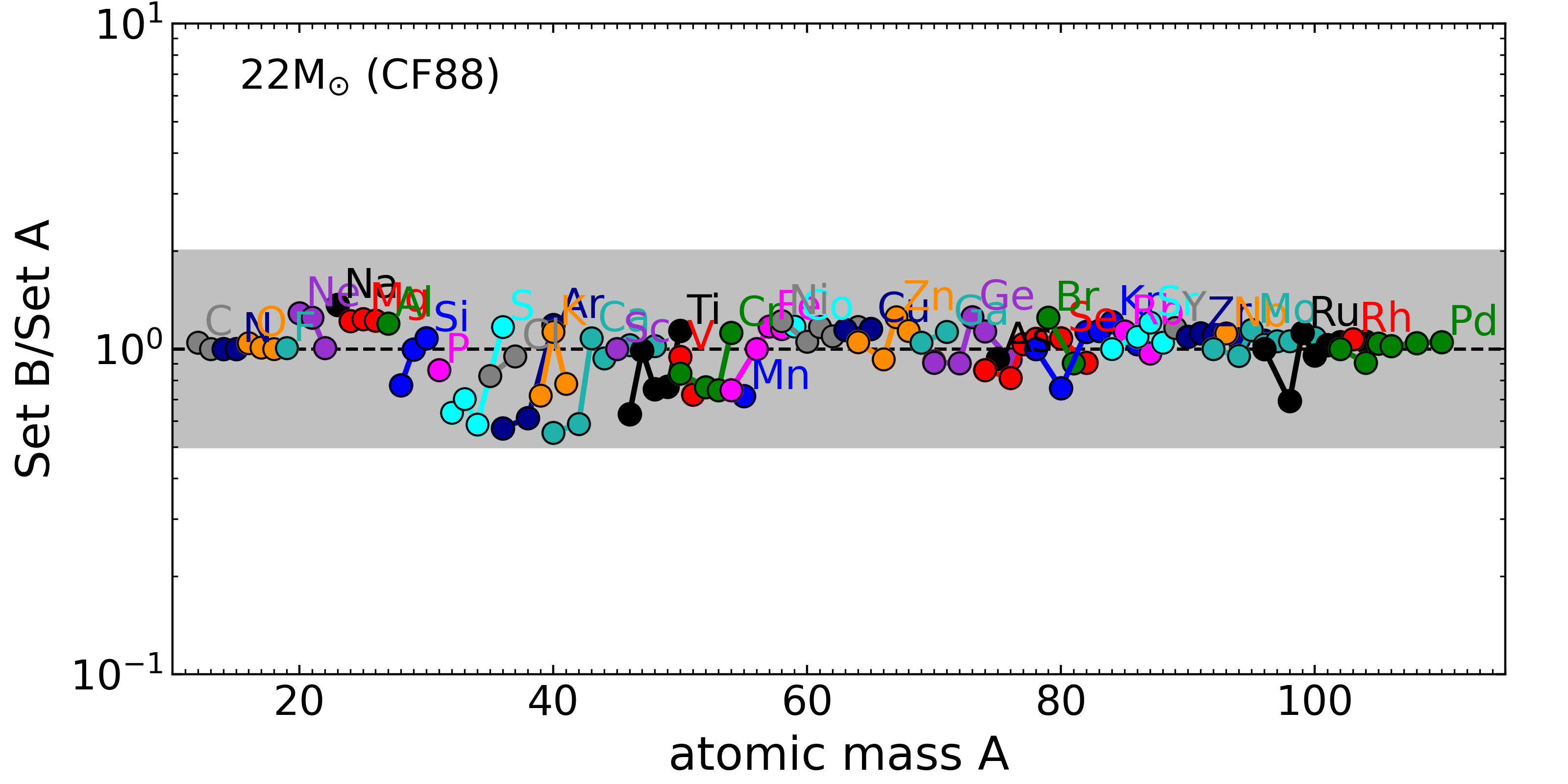}
            \includegraphics[width=.49\linewidth]{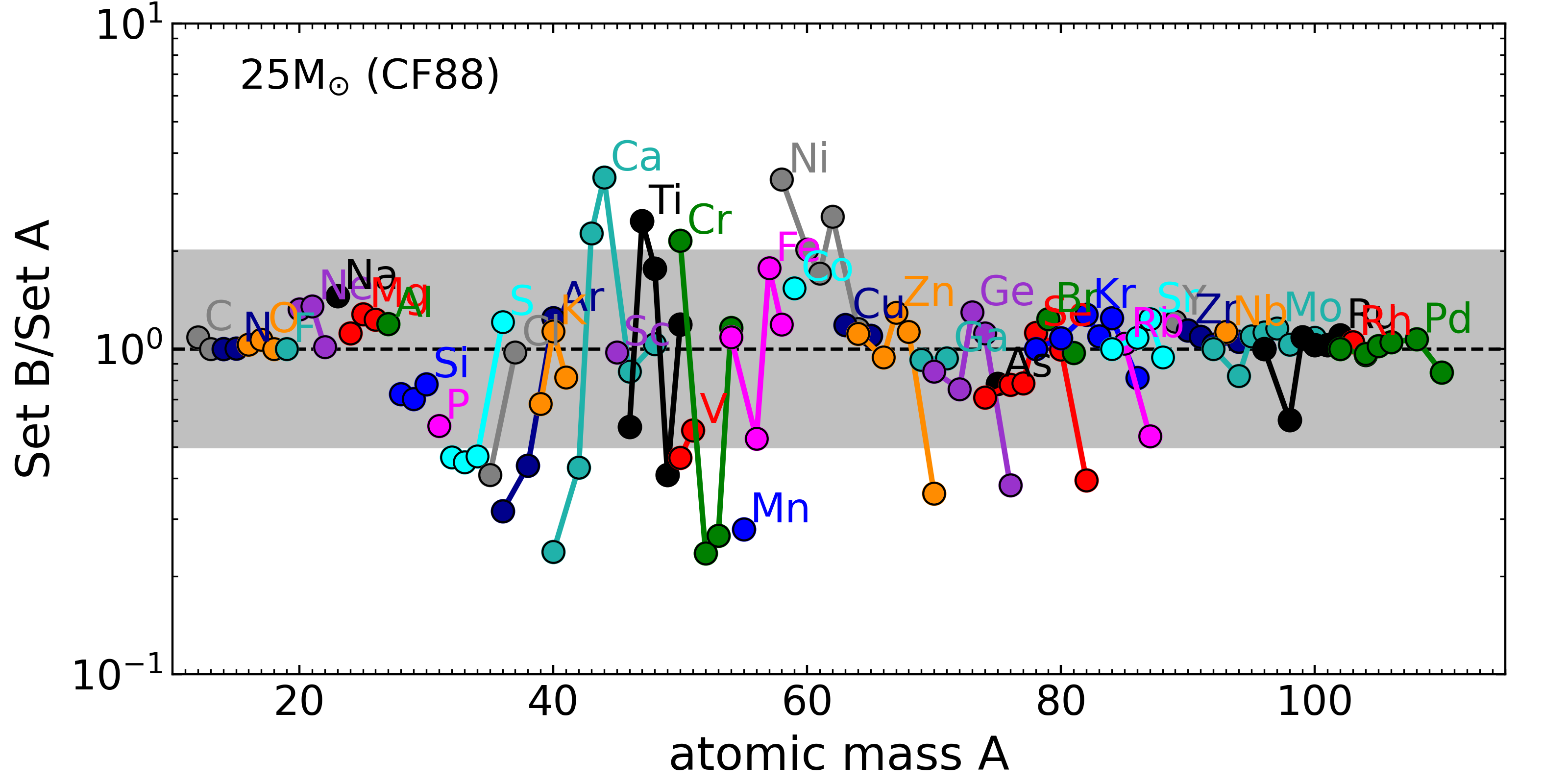}
            \includegraphics[width=.49\linewidth]{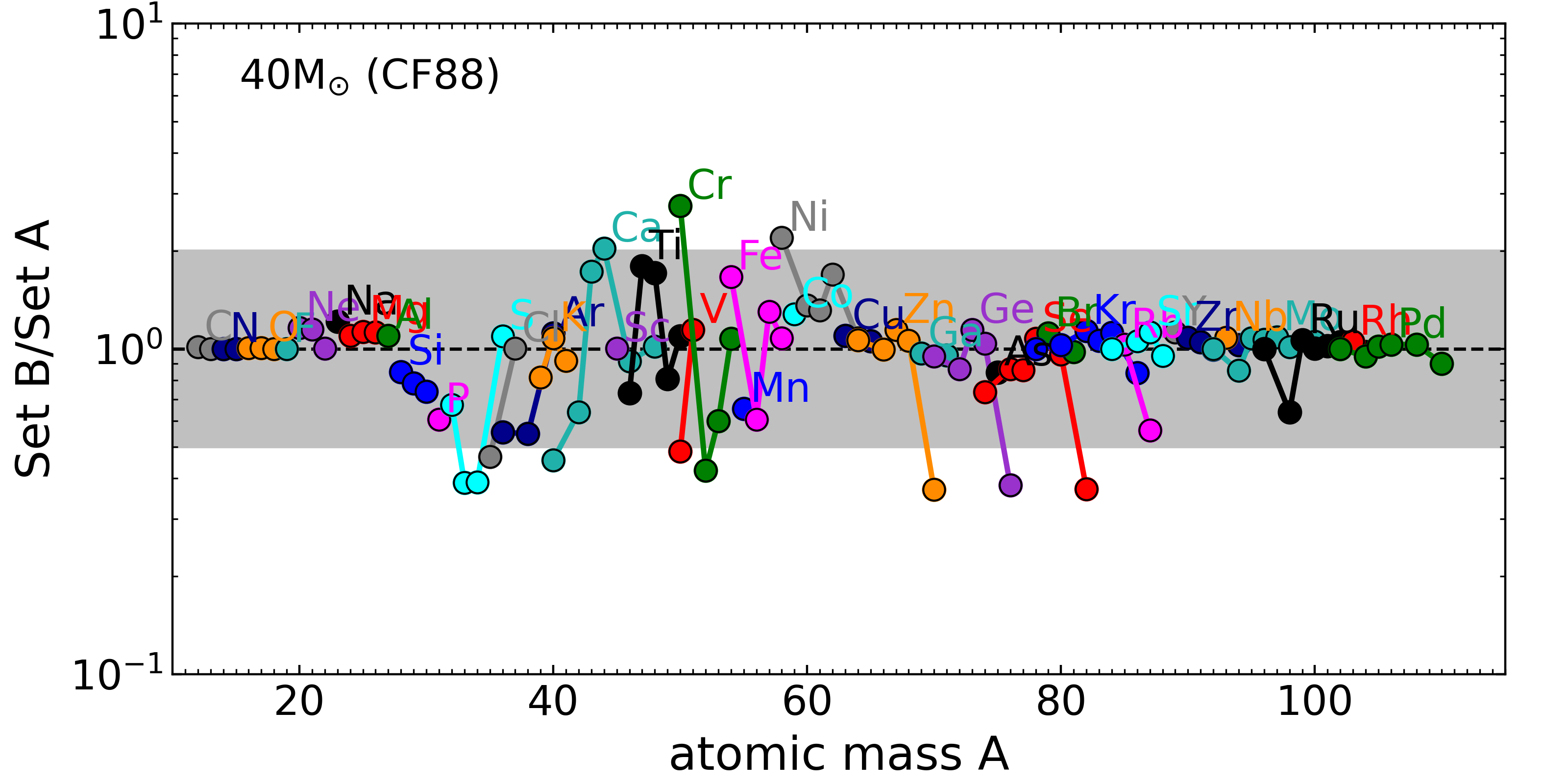}
            \caption{Stable isotopic yield ratios between Set A and Set B for each CF88 model. The gray band identifies a factor of 2 variation.}
            \label{fig:setabcf88}
        \end{figure}

        \begin{figure}[!t]
            \centering
            \includegraphics[width=.49\linewidth]{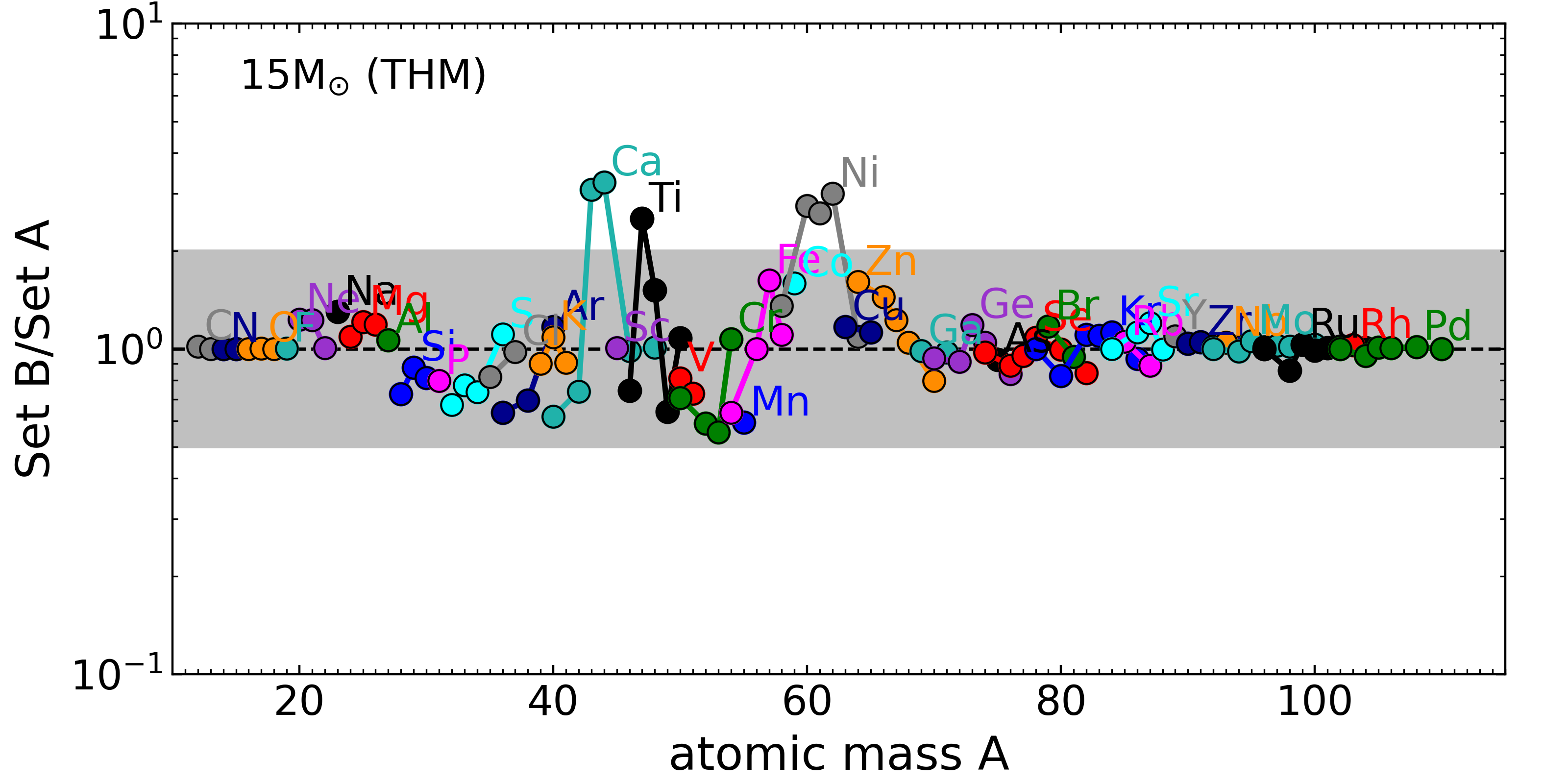}
            \includegraphics[width=.49\linewidth]{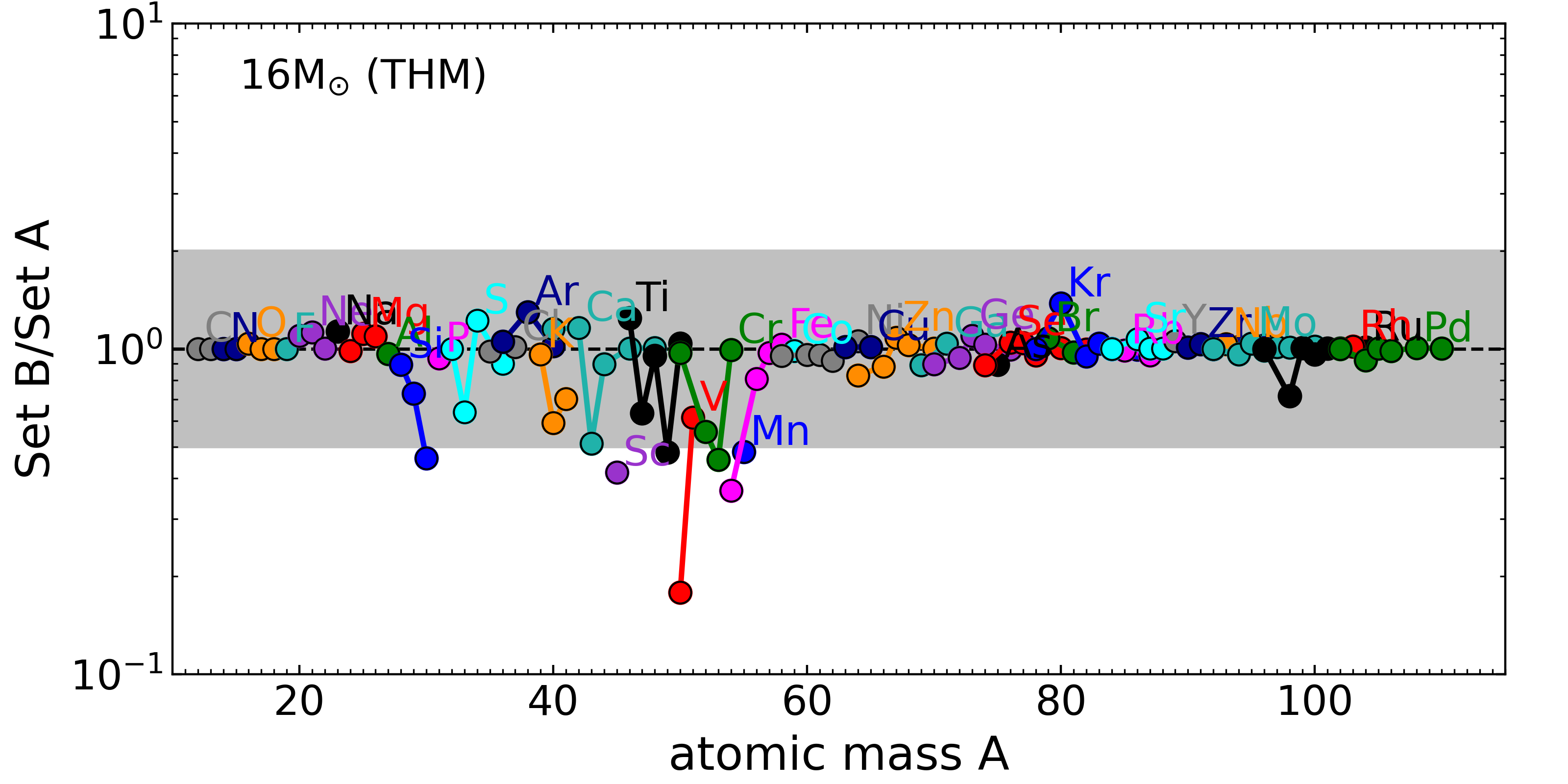}
            \includegraphics[width=.49\linewidth]{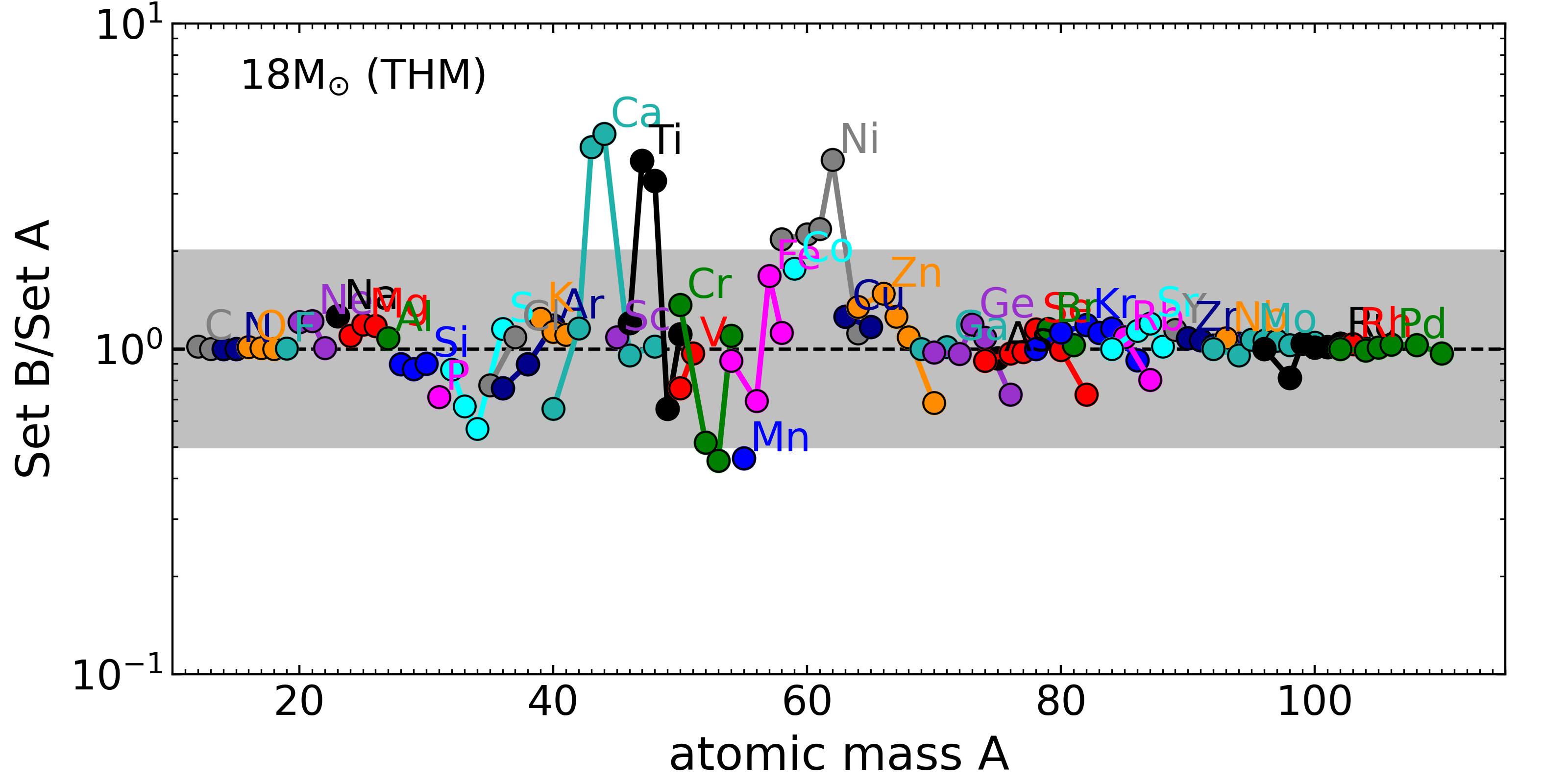}
            \includegraphics[width=.49\linewidth]{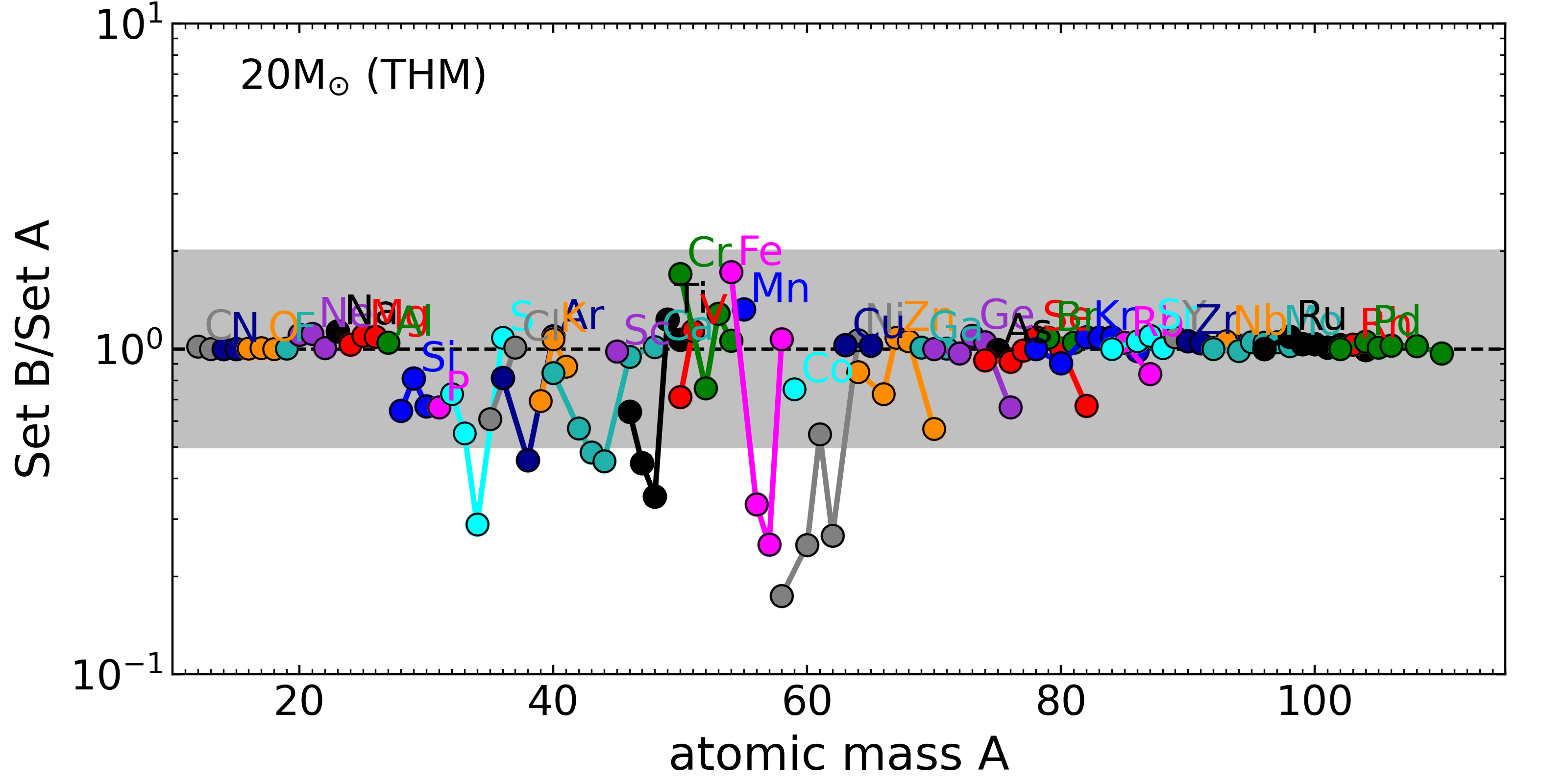}
            \includegraphics[width=.49\linewidth]{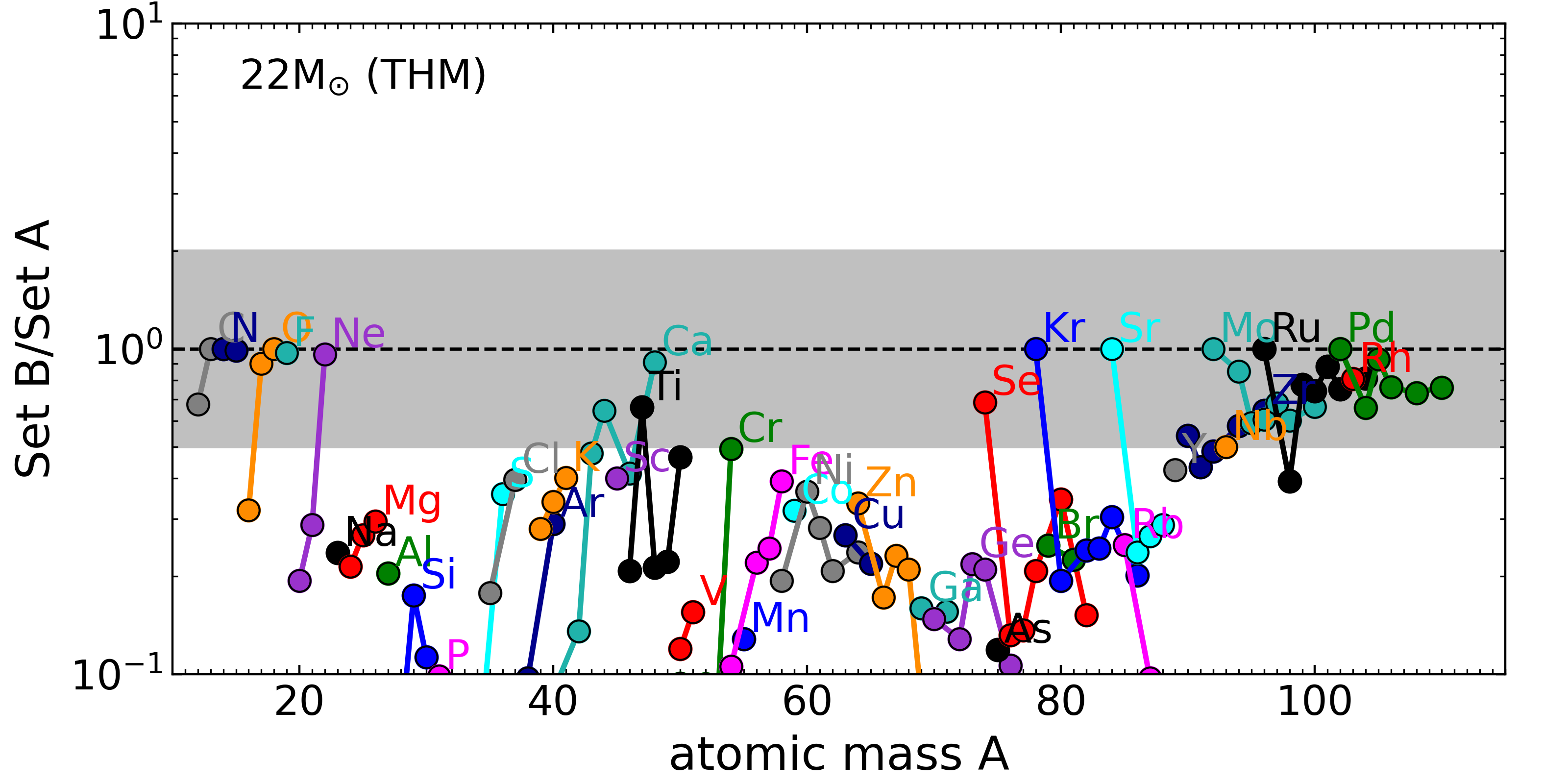}
            \includegraphics[width=.49\linewidth]{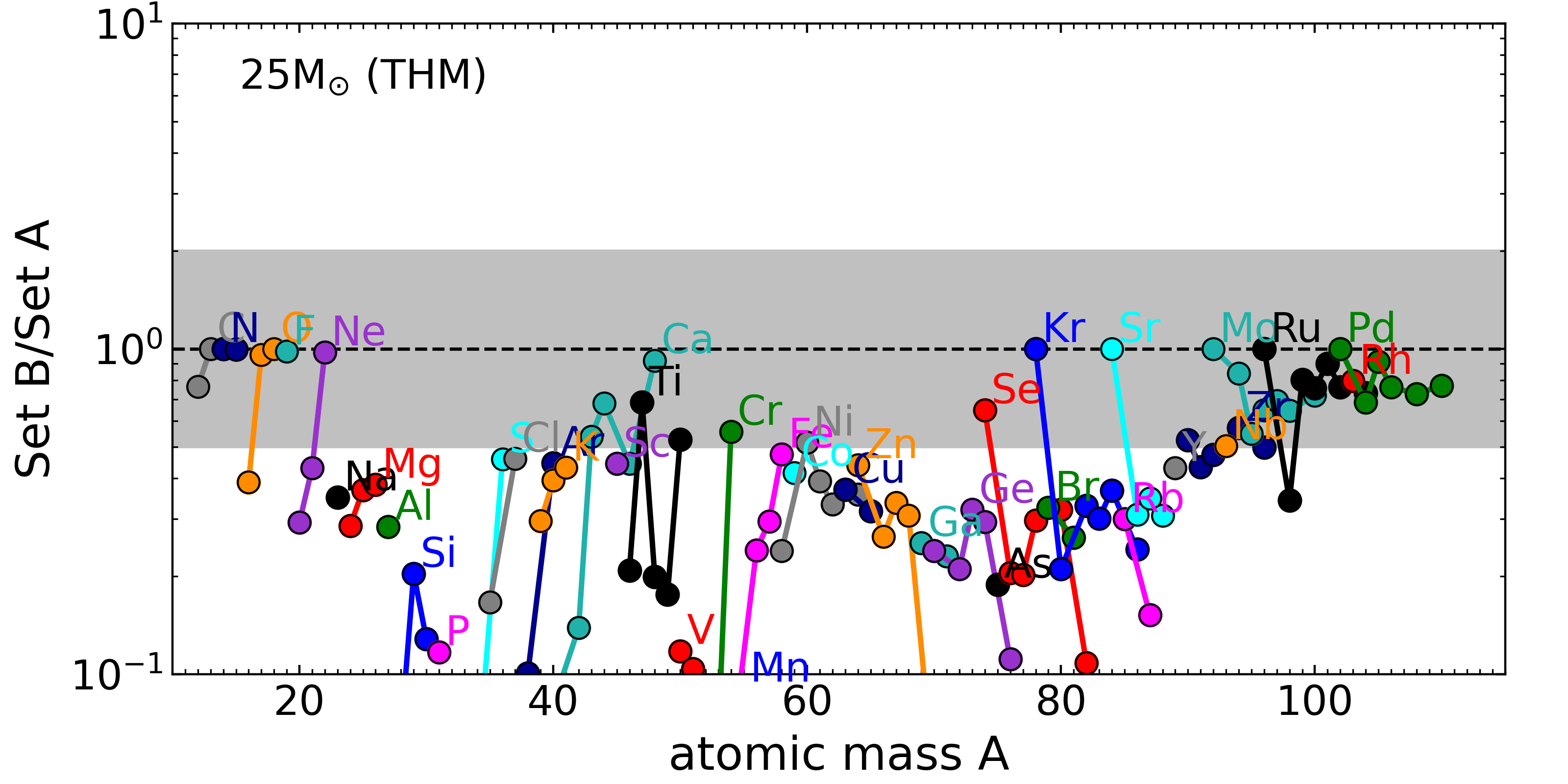}
            \includegraphics[width=.49\linewidth]{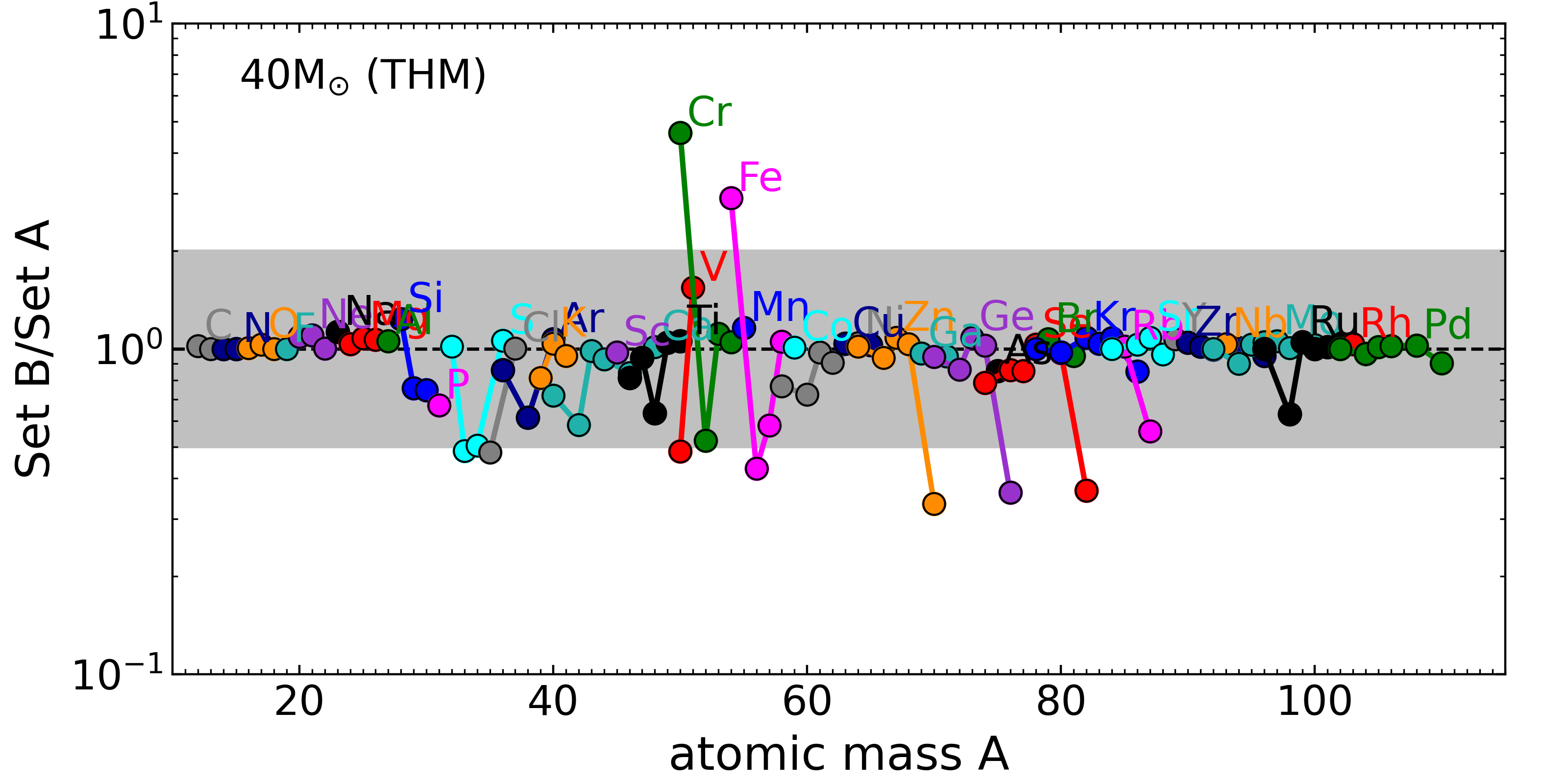}
            \caption{Same as \figurename~\ref{fig:setabcf88} but for THM models.}
            \label{fig:setabthm}
        \end{figure}

    To have a meaningful comparison with the Set A yields, wherever possible, we calculate the ejected masses from Set B imposing the same amount of \isotope[56]{Ni} as in Set A (0.07 \msun) and report the results in \figurename~\ref{fig:setabcf88} and \ref{fig:setabthm}. Larger $\rm Si_c$ regions means enhancing the production of nuclei produced under the full nuclear statistical equilibrium (NSE) conditions, such as \isotope[60,61,62]{Ni}, \isotope[43]{Ca}, and \isotope[44]{Ti} which decays to \isotope[44]{Ca} in $\sim60$ years. The smaller $\rm Si_i$ and $\rm O_x$ regions imply both a lower production of typical nuclei synthesized in these regions (such as \isotope[47]{Ti} and \isotope[51]{Cr}) and the preservation of hydrostatic products in the O shell, as \isotope[39]{K}, \isotope[42]{Ca}, and \isotope[46]{Ti}. When the shock enters the C shell, if $T_{\text{peak}}$ is high enough, the combination of \isotope[20]{Ne} photodisintegrations and the activation of the C fusion release $\alpha$ particles that can interact with the remaining \isotope[22]{Ne} and triggers a production of neutrons, which does not produce any \s process nucleosynthesis, but instead it may produce neutron richer isotopes of several elements, such as for example \isotope[60]{Fe} from \isotope[58]{Fe}. Moreover, the presence of \s process and pristine $r-$process material act as seeds for the $\gamma-$process nucleosynthesis \citep[][and references therein]{pignatari:16,roberti:23}. The $\gamma-$process nucleosynthesis mostly produces $p-$nuclei \citep[neutron deficient isotopes of elements beyond Fe, see, e.g.,][]{arnould:03}, but it can even produce some \s only nuclei, that are shielded by the neutron richer isotopes on the stability valley \citep[see Tab. 4 from][]{kaeppeler:89}. A more narrow $\rm Ne_x$ region would partially suppress these productions, which is particularly evident in the cases of the \s only nuclei \isotope[70,72]{Ge}, \isotope[76]{Se}, \isotope[80]{Kr}, and \isotope[98]{Ru} and the $r-$process nuclei \isotope[70]{Zn}, \isotope[76]{Ge} and \isotope[82]{Se}. Although \figurename~\ref{fig:expstrut} shows less destruction of the \s process nuclei, this is not significant on the total yield.

\section{Yield tables}

    In the following (\tablename~\ref{tab:seta_cf88}, \ref{tab:setb_cf88}, \ref{tab:seta_thm}, and \ref{tab:setb_thm}), we report the supernova yields from Set A and Set B for the CF88 and THM models (see also Sect. \ref{subsec:exp}).

    \startlongtable
\floattable
\begin{deluxetable}{rllllllllll}
\centering
\tablecolumns{11}
\tablecaption{Set A, CF88 ejected masses (\msun) \label{tab:seta_cf88}}
\tablehead{$\rm Isotope$ &    Z    & A   &   $\rm X_{sol}$ & $\rm 15~M_\odot$ & $\rm 16~M_\odot$ & $\rm 18~M_\odot$ & $\rm 20~M_\odot$ & $\rm 22~M_\odot$ & $\rm 25~M_\odot$ & $\rm 40~M_\odot$}
    \startdata
              n   &  0  &   1 &  0.0000E+00  & 2.4031E-36  & 7.4623E-30  & 1.1379E-27  & 3.6242E-27  & 1.1523E-25  & 6.1669E-25  & 4.7941E-23\\
              H   &  1  &   1 &  7.2155E-01  & 6.8469E+00  & 7.1834E+00  & 7.8525E+00  & 8.5245E+00  & 9.1456E+00  & 1.0071E+01  & 1.4711E+01\\
        $^{2}$H   &  1  &   2 &  2.9439E-05  & 3.7036E-06  & 4.7163E-06  & 5.2990E-06  & 5.8878E-06  & 6.4767E-06  & 7.3599E-06  & 1.1777E-05\\
        $^{3}$H   &  1  &   3 &  0.0000E+00  & 2.7799E-25  & 9.0524E-19  & 1.0801E-16  & 2.9266E-16  & 7.9295E-15  & 3.5169E-14  & 1.5602E-12\\
       $^{3}$He   &  2  &   3 &  3.1196E-05  & 3.9424E-04  & 4.0795E-04  & 4.3808E-04  & 4.6859E-04  & 4.9984E-04  & 5.4673E-04  & 8.0572E-04\\
       $^{4}$He   &  2  &   4 &  2.6500E-01  & 4.8387E+00  & 5.1357E+00  & 5.6926E+00  & 6.2021E+00  & 6.7887E+00  & 7.7972E+00  & 1.4338E+01\\
       $^{6}$Li   &  3  &   6 &  5.9602E-10  & 7.4980E-11  & 9.5695E-11  & 1.0763E-10  & 1.1973E-10  & 1.3188E-10  & 1.5020E-10  & 2.4326E-10\\
       $^{7}$Li   &  3  &   7 &  8.4661E-09  & 1.0651E-09  & 1.4201E-09  & 1.6376E-09  & 1.8273E-09  & 2.0223E-09  & 2.3142E-09  & 3.7947E-09\\
\enddata
\end{deluxetable} 

    \startlongtable
\floattable
\begin{deluxetable}{lllllllllll}
\centering
\tablecolumns{11}
\tablecaption{Set B, CF88 ejected masses (\msun) \label{tab:setb_cf88}}
\tablehead{$\rm Isotope$ &    Z    & A   &   $\rm X_{sol}$ & $\rm 15~M_\odot$ & $\rm 16~M_\odot$ & $\rm 18~M_\odot$ & $\rm 20~M_\odot$ & $\rm 22~M_\odot$ & $\rm 25~M_\odot$ & $\rm 40~M_\odot$}
    \startdata
          n  &  0 &   1 & 0.0000E+00 & 2.2982E-36 & 7.4537E-30 & 1.1379E-27 & 3.6242E-27 & 1.6058E-25 & 1.1844E-23 & 4.7941E-23 \\
          H  &  1 &   1 & 7.2155E-01 & 6.8469E+00 & 7.1834E+00 & 7.8525E+00 & 8.5245E+00 & 9.1455E+00 & 1.0071E+01 & 1.4711E+01 \\
    $^{2}$H  &  1 &   2 & 2.9439E-05 & 3.7036E-06 & 4.7109E-06 & 5.2990E-06 & 5.8878E-06 & 6.4767E-06 & 7.3599E-06 & 1.1777E-05 \\
    $^{3}$H  &  1 &   3 & 0.0000E+00 & 2.7799E-25 & 9.0419E-19 & 1.0801E-16 & 2.9266E-16 & 7.9295E-15 & 3.5169E-14 & 1.5602E-12 \\
   $^{3}$He  &  2 &   3 & 3.1196E-05 & 3.9424E-04 & 4.0795E-04 & 4.3808E-04 & 4.6859E-04 & 4.9984E-04 & 5.4673E-04 & 8.0572E-04 \\
   $^{4}$He  &  2 &   4 & 2.6500E-01 & 4.8548E+00 & 5.1671E+00 & 5.7345E+00 & 6.2429E+00 & 6.8217E+00 & 7.8345E+00 & 1.4370E+01 \\
   $^{6}$Li  &  3 &   6 & 5.9602E-10 & 7.4980E-11 & 9.5584E-11 & 1.0763E-10 & 1.1973E-10 & 1.3188E-10 & 1.5020E-10 & 2.4326E-10 \\
   $^{7}$Li  &  3 &   7 & 8.4661E-09 & 1.0651E-09 & 1.4185E-09 & 1.6376E-09 & 1.8273E-09 & 2.0223E-09 & 2.3142E-09 & 3.7947E-09 \\
\enddata
\end{deluxetable} 

    \startlongtable
\floattable
\begin{deluxetable}{rllllllllll}
\centering
\tablecolumns{11}
\tablecaption{Set A, THM ejected masses (\msun) \label{tab:seta_thm}}
\tablehead{$\rm Isotope$ &    Z    & A   &   $\rm X_{sol}$ & $\rm 15~M_\odot$ & $\rm 16~M_\odot$ & $\rm 18~M_\odot$ & $\rm 20~M_\odot$ & $\rm 22~M_\odot$ & $\rm 25~M_\odot$ & $\rm 40~M_\odot$}
    \startdata
          n  &  0 &   1 & 0.0000E+00 & 3.2212E-36 & 2.3119E-38 & 1.1379E-27 & 3.6242E-27 & 1.1523E-25 & 6.1668E-25 & 4.7941E-23 \\
          H  &  1 &   1 & 7.2155E-01 & 6.8463E+00 & 6.1277E+00 & 7.8509E+00 & 8.5233E+00 & 9.1461E+00 & 1.0071E+01 & 1.4711E+01 \\
    $^{2}$H  &  1 &   2 & 2.9439E-05 & 3.7030E-06 & 1.1223E-16 & 5.2991E-06 & 5.8878E-06 & 6.4767E-06 & 7.3598E-06 & 1.1777E-05 \\
    $^{3}$H  &  1 &   3 & 0.0000E+00 & 2.7794E-25 & 5.8517E-33 & 1.0801E-16 & 2.9266E-16 & 7.9294E-15 & 3.5169E-14 & 1.5602E-12 \\
   $^{3}$He  &  2 &   3 & 3.1196E-05 & 3.9417E-04 & 3.4788E-04 & 4.3809E-04 & 4.6859E-04 & 4.9984E-04 & 5.4673E-04 & 8.0572E-04 \\
   $^{4}$He  &  2 &   4 & 2.6500E-01 & 4.8146E+00 & 4.6875E+00 & 5.6838E+00 & 6.2065E+00 & 6.7853E+00 & 7.7905E+00 & 1.4342E+01 \\
   $^{6}$Li  &  3 &   6 & 5.9602E-10 & 7.4970E-11 & 4.5809E-17 & 1.0763E-10 & 1.1973E-10 & 1.3188E-10 & 1.5020E-10 & 2.4326E-10 \\
   $^{7}$Li  &  3 &   7 & 8.4661E-09 & 1.0649E-09 & 4.7745E-15 & 1.6376E-09 & 1.8273E-09 & 2.0223E-09 & 2.3142E-09 & 3.7946E-09 \\
\enddata
\end{deluxetable} 
    \startlongtable
\floattable
\begin{deluxetable}{rllllllllll}
\centering
\tablecolumns{11}
\tablecaption{Set B, THM ejected masses (\msun) \label{tab:setb_thm}}
\tablehead{$\rm Isotope$ &    Z    & A   &   $\rm X_{sol}$ & $\rm 15~M_\odot$ & $\rm 16~M_\odot$ & $\rm 18~M_\odot$ & $\rm 20~M_\odot$ & $\rm 22~M_\odot$ & $\rm 25~M_\odot$ & $\rm 40~M_\odot$}
    \startdata
          n  &  0 &   1 & 0.0000E+00 & 2.3930E-36 & 5.9803E-40 & 1.1379E-27 & 3.6242E-27 & 1.1523E-25 & 6.1668E-25 & 4.7941E-23 \\
          H  &  1 &   1 & 7.2155E-01 & 6.8463E+00 & 6.1277E+00 & 7.8509E+00 & 8.5233E+00 & 9.1461E+00 & 1.0071E+01 & 1.4711E+01 \\
    $^{2}$H  &  1 &   2 & 2.9439E-05 & 3.7030E-06 & 1.1223E-16 & 5.2991E-06 & 5.8878E-06 & 6.4767E-06 & 7.3598E-06 & 1.1777E-05 \\
    $^{3}$H  &  1 &   3 & 0.0000E+00 & 2.7794E-25 & 5.8517E-33 & 1.0801E-16 & 2.9266E-16 & 7.9294E-15 & 3.5169E-14 & 1.5602E-12 \\
   $^{3}$He  &  2 &   3 & 3.1196E-05 & 3.9417E-04 & 3.4788E-04 & 4.3809E-04 & 4.6859E-04 & 4.9984E-04 & 5.4673E-04 & 8.0572E-04 \\
   $^{4}$He  &  2 &   4 & 2.6500E-01 & 4.8310E+00 & 4.6940E+00 & 5.7249E+00 & 6.1917E+00 & 6.7836E+00 & 7.7903E+00 & 1.4330E+01 \\
   $^{6}$Li  &  3 &   6 & 5.9602E-10 & 7.4970E-11 & 4.5808E-17 & 1.0763E-10 & 1.1973E-10 & 1.3188E-10 & 1.5020E-10 & 2.4326E-10 \\
   $^{7}$Li  &  3 &   7 & 8.4661E-09 & 1.0649E-09 & 4.7745E-15 & 1.6376E-09 & 1.8273E-09 & 2.0223E-09 & 2.3142E-09 & 3.7946E-09 \\
\enddata
\end{deluxetable} 


\end{document}